\documentclass[fleqn,useAMS,usenatbib]{mn2e}
\usepackage{times,graphicx,verbatim,amsmath,amssymb,wasysym}
\usepackage[colorlinks=true, linkcolor=blue, citecolor=blue, urlcolor=blue]{hyperref}
\usepackage{etoolbox}
\newtoggle{referee}            %
\togglefalse{referee}          %
\iftoggle{referee}{
    \usepackage{color,soul}    %
}{
    \newcommand{\hl}[1]{{#1}}  %
}                              %
\newcommand{\hlold}[1]{{#1}}   %
\newcommand{\hlmold}[1]{{#1}}  

\makeatletter 
  \patchcmd{\NAT@citex}
    {\@citea\NAT@hyper@{%
      \NAT@nmfmt{\NAT@nm}%
      \hyper@natlinkbreak{\NAT@aysep\NAT@spacechar}{\@citeb\@extra@b@citeb}%
      \NAT@date}}
    {\@citea\NAT@nmfmt{\NAT@nm}%
    \NAT@aysep\NAT@spacechar\NAT@hyper@{\NAT@date}}{}{}

  \patchcmd{\NAT@citex}
    {\@citea\NAT@hyper@{%
      \NAT@nmfmt{\NAT@nm}%
      \hyper@natlinkbreak{\NAT@spacechar\NAT@@open\if*#1*\else#1\NAT@spacechar\fi}%
        {\@citeb\@extra@b@citeb}%
      \NAT@date}}
    {\@citea\NAT@nmfmt{\NAT@nm}%
    \NAT@spacechar\NAT@@open\if*#1*\else#1\NAT@spacechar\fi\NAT@hyper@{\NAT@date}}
    {}{}
\makeatother

\title[The Ly$\alpha$ signature of the first galaxies]{The Lyman-$\balpha$ signature of the first galaxies}
\author[A. Smith et al.]
{
  Aaron~Smith,$^1$\thanks{E-mail: \href{mailto:asmith@astro.as.utexas.edu}{asmith@astro.as.utexas.edu} }
  Chalence~Safranek-Shrader,$^{1,2}$
  Volker~Bromm$^1$ and
  Milo\v{s} Milosavljevi\'{c}$^1$ \\
  $^1$Department of Astronomy, The University of Texas at Austin, Austin, TX 78712, USA \\
  $^2$Department of Astronomy and Astrophysics, University of California, Santa Cruz, CA 95064, USA
}
\date{\today}

\pagerange{\pageref{firstpage}--\pageref{lastpage}} \pubyear{2014}

\newcommand\Msun{\text{M}_{\astrosun}} 
\newcommand\Lsun{\text{L}_{\astrosun}} 
\newcommand\colt{\mathbf{COLT}} 
\newcommand\HI{{H\,\textsc{i}}} 
\newcommand\HII{{H\,\textsc{ii}}} 

\def\app#1#2{%
  \mathrel{%
    \setbox0=\hbox{$#1\sim$}%
    \setbox2=\hbox{%
      \rlap{\hbox{$#1\propto$}}%
      \lower1.2\ht0\box0%
    }%
    \raise0.25\ht2\box2%
  }%
}
\def\approxprop{\mathpalette\app\relax}

\begin{document} \label{firstpage}

\maketitle

\begin{abstract}
  We present the $\mathbf{Co}$smic $\mathbf{L}$yman-$\alpha$ $\mathbf{T}$ransfer code ($\colt$), a massively parallel Monte-Carlo radiative transfer code, to simulate Lyman-$\alpha$ (Ly$\alpha$) resonant scattering through neutral hydrogen as a probe of the first galaxies. We explore the interaction of centrally produced Ly$\alpha$ radiation with the host galactic environment. Ly$\alpha$ photons emitted from the luminous starburst region escape with characteristic features in the line profile depending on the density distribution, ionization structure, and bulk velocity fields. For example, anisotropic \hlold{ionization} exhibits a tall peak close to line centre with a skewed tail that drops off gradually. Idealized models of first galaxies explore the effect of mass, anisotropic \HII\ regions, and radiation pressure driven winds on Ly$\alpha$ observables. We employ mesh refinement to resolve critical structures\hlold{.} We also post-process an \textit{ab initio} cosmological simulation and examine images captured at various distances within the 1~Mpc$^3$ comoving volume. Finally, we discuss the emergent spectra and surface brightness profiles of these objects in the context of high-$z$ observations. The first galaxies will likely be observed through the red damping wing of the Ly$\alpha$ line. Observations will be biased toward galaxies with an intrinsic red peak located far from line centre that reside in extensive \HII\ super bubbles, which allows Hubble flow to sufficiently redshift photons away from line centre and facilitate transmission through the intergalactic medium. Even with gravitational lensing to boost the luminosity \hlold{this preliminary work indicates} that Ly$\alpha$ emission from stellar clusters within haloes of $M_{\rm vir} < 10^9~\Msun$ is generally too faint to be detected by the \textit{James Webb Space Telescope}~(\textit{JWST}).
\end{abstract}


\section{Introduction}
  \label{sec:introduction}
  Observations of Lyman-$\alpha$~(Ly$\alpha$) sources are a powerful probe of the high-redshift Universe \citep[e.g.][]{Hu:1996Nature,Rhoads:2000,Taniguchi,Finkelstein:2009}. In particular, the prominence of the Ly$\alpha$ line at $\lambda_{\rm Ly\alpha} = 1216~\text{\AA}~(1+z)$ allows for spectroscopic confirmation of redshift measurements of individual distant galaxies. Ly$\alpha$ sources are also a compelling probe of the cosmic dark ages leading up to reionization -- see \citet{Dunlop:book} for a perspective on high-$z$ observations. Historically, \citet{1967ApJ...147..868P} determined that galaxies from the first billion years after the Big Bang would be powerful emitters of Ly$\alpha$ photons, though observations of these sources eluded us for longer than expected. However, robust detections are becoming more regular, especially if the stellar mass is comparable to the Milky Way or the star formation rate~(SFR) is elevated \citep[e.g. SFR~$\ga 100~\Msun~\text{yr}^{-1}$;][]{Pritchet:1994}.

  Within the earliest galaxies hard UV radiation from massive stars is reprocessed into Ly$\alpha$ photons; however, because neutral hydrogen~(\HI) is opaque to the Ly$\alpha$ line, many of these photons may be resonantly trapped, and consequently suffer significant dust absorption. Despite these effects, observations have determined that the Ly$\alpha$ escape fraction,~$f_{\rm e}$, actually increases at higher redshifts~\citep{Hayes,Curtis-Lake:2012}. At some point, although the photons are no longer destroyed by dust, they are scattered out of the line of sight and some fraction of the Ly$\alpha$ emission is lost to the background as their sources become spatially extended Ly$\alpha$ haloes \citep{Loeb:1999er}. Various mechanisms have been explored to explain the unusually high $f_{\rm e}$ of high-$z$ galaxies. In all likelihood this is a result of the complicated resonant line transfer, galactic structure, and peculiar dust properties. For example, multiple scatterings that facilitate excursions to the wings of the frequency profile; large-scale flows that induce Doppler shifts; and the geometry of dense, dusty clouds within a clumpy interstellar medium that provide pathways for escape~\citep[e.g.][]{Hansen:2005pm,Dijkstra:2008xz,Zheng:2009ax}. In this work, we push these questions to the very first galaxies~\citep[for a review see][]{Bromm:2011cw}.

  Assessing the observability of such early Ly$\alpha$ sources is nontrivial. Indeed, going to higher redshifts introduces physical effects that compete in either strengthening or attenuating the Ly$\alpha$ signal~\citep{Latif:2011ge,Latif:2011is,Dunlop:book}. On one hand, the intergalactic medium~(IGM) becomes increasingly neutral at higher $z$, resulting in a more difficult escape for Ly$\alpha$ photons \citep{Ono:2011ew}. On the other hand, the same IGM also becomes increasingly devoid of dust \citep{Pentericci:2011}. Furthermore, Population~III (so-called Pop~III) stellar sources are predicted to have been more efficient ionizers, boosting the Ly$\alpha$ luminosity \citep{Bromm:2013,Glover:2013ASSL}. The fact remains that high-redshift Ly$\alpha$ sources are being observed out to $z \sim 7.5$. Still, many details regarding the epoch of reionization~(EoR), or the inhomogeneous phase transition around $z \sim 6.5-15$, are uncertain and may greatly affect interpretations of Ly$\alpha$ transfer through the IGM \citep{Barkana:2007Review,Meiksin:2009,Zaroubi:2013Review}.

  Some of the most effective methods for identifying high-redshift objects involve the Ly$\alpha$ line. In particular, Lyman-break galaxies~(LBGs) are generally massive galaxies for which neutral hydrogen produces a sharp drop in the spectra due to absorption \citep{Meier:1976a,Meier:1976b,Steidel:1992,Steidel:1993}. Lyman-$\alpha$ emitters~(LAEs) are young, less-massive galaxies with active star formation and strong Ly$\alpha$ emission \citep{Charlot:1993}. It is an important frontier to push Ly$\alpha$ selection methods towards the highest possible redshifts. For galaxies at $z \ga 6$ the neutral fraction of the intervening IGM increases enough for their spectra to yield complete absorption of photons blueward of the Ly$\alpha$ line. This is the well-known ``Gunn-Peterson trough'' \citep{GunnPeterson} which is characteristic of LBGs. However, these massive, evolved galaxies become increasingly rare at high redshifts. The LAE luminosity function also declines as redshift increases, and the observed trend is robustly established for $4 \la z \la 7$ \citep[e.g.][]{Bowens:2007,Oesch:2012} and expected to continue beyond $z \sim 7$ \citep{Ellis:2013}. Because a strong detection of the highly-redshifted Ly$\alpha$ line requires the emitter to be young and relatively dust free -- conditions which are naturally expected for the first galaxies -- LAEs are likely their typical manifestation.

  High-redshift Ly$\alpha$ candidates must be followed up by spectroscopy in order to guard against false positives from foreground contaminants. Fortunately, moderate- to high-$z$ surveys are underway that will dramatically increase the sample size of Ly$\alpha$ galaxies and better characterize their statistical properties. For example, the Hobby-Eberly Telescope Dark Energy Experiment (HETDEX) is a large integral-field spectroscopic survey expected to detect a million LAEs \citep{Hill:2008HETDEX,Adams:2010un,Finkelstein:2010ut,Chonis:2013HETDEX}. Currently, there are several candidates at $z \ga 7$~\citep[e.g.][]{Ellis:2013}, with the highest spectroscopically-confirmed source announced at $z = 7.51$ by \citet{Finkelstein:2013}. Other records have been found using gamma-ray bursts~(GRBs), active galactic nuclei~(AGN), or (sub-)mm observations of redshifted thermal dust emission \citep{Dunlop:book}. However, it is unclear how these other selection methods relate to Ly$\alpha$ predictions. Such connections may complement Ly$\alpha$ observations, even if the phenomena originate from the luminous deaths of individual massive stars (i.e. GRBs) or are not associated with a `normal' activity of the first galaxies (e.g. AGN or high amounts of dust).

  Ly$\alpha$ radiative transfer within the first galaxies is a timely problem because next-generation facilities will provide high resolution data by the end of the decade. The \textit{James Webb Space Telescope}~\citep[\textit{JWST};][]{Gardner:2006ky} and large-aperture ground-based observatories\footnote{Infrared telescopes with integral field spectrographs and adaptive optics imaging \hl{will} include the Giant Magellan Telescope~(GMT; \href{http://www.gmto.org/}{www.gmto.org}), Thirty Meter Telescope~(TMT; \href{http://www.tmt.org/}{www.tmt.org}), and the European Extremely Large Telescope~(E-ELT; \href{http://www.eso.org/sci/facilities/eelt/}{www.eso.org/sci/facilities/eelt}).} offer the prime avenue for observing Ly$\alpha$ emission at the high-$z$ frontier and will significantly contribute to our understanding of the ionization history at the end of the dark ages \citep{Stiavelli:2009book}. However, significant progress has also been made on a number of complementary probes of the high-$z$ Universe. Several 21-cm array experiments\footnote{For example, the Low Frequency Array (LOFAR; \href{http://www.lofar.org/}{www.lofar.org}), the Murchison Wide-Field Array (MWA; \href{http://www.mwatelescope.org/}{www.mwatelescope.org}), the Precision Array to Probe the Epoch of Reionization (PAPER; \href{http://eor.berkeley.edu/}{eor.berkeley.edu}), and ultimately the Square Kilometer Array (SKA; \href{http://www.skatelescope.org/}{www.skatelescope.org}).} are coming online to map the distribution of \HI\ over the course of reionization. The Ly$\alpha$ and 21-cm lines are related through the Wouthuysen-Field mechanism for which Ly$\alpha$ scatterings pump electrons into the excited hyperfine state, thereby coupling the spin and kinetic temperatures~\citep*{Wouthuysen,Field,Furlanetto:2006jb}. Finally, an ideal complementary Ly$\alpha$ probe is encoded in the cosmic infrared background~(CIB) because the integrated radiation from all background stars and galaxies has been redshifted to IR wavelengths \citep*{1967ApJ...148..377P,Santos:2002}. The Ly$\alpha$ contribution is seen through the correlation of sources across characteristic length scales \citep[for a review see][]{Kashlinsky:2005Review}.

  A number of authors have studied Ly$\alpha$ radiative transfer within different contexts. We have greatly benefited from and hope to add to the body of work in this area. A partial list of references include: \citet*{Ahn:2001pz,Zheng:2002qc,Dijkstra:2005ny,Tasitsiomi:2006,Verhamme:2006tc,Semelin:2007,Laursen:2008aq,Forero-Romero:2011,Yajima:2012}.
  The state of the art is to apply post-processing radiative transfer to realistic hydrodynamical simulations, which is justified for many large scale systems. We use this method in conjunction with semi-analytic models to ascertain the feedback of Ly$\alpha$ radiation on the galactic assembly process. Our focus on the very first galaxies, in their proper cosmological context, is different from previous research that has targeted more massive systems at redshifts close to, or after, reionization \citep[e.g.][]{Yajima:2014}. Such systems require a statistical description of Ly$\alpha$ transmission through the IGM as described by \citet*{Dijkstra:2011} and \citet*{Laursen:2011}. Photons that scatter out of the line of sight due to the neutral fraction of the IGM are effectively lost to the background. Therefore, the observability of Ly$\alpha$ emitters will provide independent constraints on reionization \citep*{Fan:2006,Jeeson-Daniel:2012,Jensen:2013,Jensen:2014,Dijkstra:2014}. 

  Measurements of the Ly$\alpha$ flux from first galaxies depend heavily on the observed line of sight in addition to the properties of the host system and IGM. Therefore, rather than solving a potentially intractable transfer equation with complex angular dependence we take advantage of Monte-Carlo Radiative Transfer techniques to accurately build emergent spectral energy distributions~(SEDs). In order to perform Ly$\alpha$ simulations we have developed a new massively parallel code called $\colt$ -- the $\mathbf{Co}$smic $\mathbf{L}$yman-$\alpha$ $\mathbf{T}$ransfer code. In Section~\ref{subsec:RS}, the basic physics of Ly$\alpha$ transport is presented. In Section~\ref{sec:methodology}, we discuss the general methodology behind $\colt$ and provide further algorithmic details in Section~\ref{sec:schemes}. In Section~\ref{sec:tests}, the code is tested against both static and dynamic setups. In Section~\ref{sec:toy}, we construct idealized analytical models to explore how fundamental parameters, including halo mass, virialization redshift, bulk velocity, and ionization structure, affect Ly$\alpha$ transport in the first galaxies. These well-motivated models help test our methods and sensitivity. Section~\ref{sec:sim} describes our implementation of the \textit{ab initio} cosmological simulation of \citet{SafranekShrader:2012qn} as post-processing conditions for $\colt$. In Section~\ref{sec:results} we analyze and discuss the emergent line of sight flux distributions and surface brightness profiles for both the idealized analytic cases and the cosmological simulation. Finally, in Section~\ref{sec:conc} we reflect on the implications of this study with regard to \hlold{future} Ly$\alpha$ observations with the \textit{JWST}.

\section{Basic physics of Lyman-$\balpha$ transport}
  \label{subsec:RS}
  Photons with frequencies close to the Ly$\alpha$ resonance line, corresponding to the transition from the first excited state~(2p) to the ground state~(1s), may be absorbed and quickly re-emitted by neutral hydrogen. Thus, in optically thick environments ($\tau \gg 1$) 
  \hlold{the main mechanism of spatial diffusion is by the rare excursion to the Lorentz wing of the spectral line.}
  The cross section~$\sigma_\nu$ and number density $n_{\rm \HI}$ describe the optical depth~$\tau_\nu$ along a photon's path:
  \begin{equation} \label{eq:tau}
    \tau_\nu = \int_{\rm path} n_{\rm \HI}\,\sigma_\nu\,d\ell \, ,
  \end{equation}
  where $\nu$ specifies the frequency dependence.

  \hlold{It is often convenient to express frequency in a dimensionless manner as the number of Doppler widths from line centre}
  \begin{equation}
    x \equiv \frac{\nu-\nu_0}{\Delta\nu_{\rm D}} \, ,
  \end{equation}
  where $\nu_0 = 2.466 \times 10^{15}$~Hz is the frequency of Ly$\alpha$ and the thermal Doppler width of the profile is $\Delta\nu_{\rm D} \equiv (v_{\rm th}/c)\,\nu_0$. The thermal velocity in terms of $T_4 \equiv T/(10^4~\text{K})$ is
  \begin{equation}
    v_{\rm th} = \sqrt{\frac{2 k_{\rm B} T}{m_{\rm H}}} = 12.85~T^{1/2}_4~\text{km~s}^{-1} \, .
  \end{equation}
  Furthermore, if the natural Ly$\alpha$ line width is $\Delta \nu_{\rm L} = 9.936 \times 10^7$~Hz then the `damping parameter' represents the relative broadening of the natural line width:
  \begin{equation}
    a \equiv \frac{\Delta\nu_{\rm L}}{2\Delta\nu_{\rm D}} = 4.702 \times 10^{-4}~T^{-1/2}_4 \, .
  \end{equation}
  Therefore, the final cross section is
  \begin{align}
    \sigma_\nu = f_{12} \, \frac{\pi e^2}{m_{\rm e} c} \, \phi_{\rm Voigt}
             = f_{12} \, \frac{\sqrt{\pi} e^2}{m_{\rm e} c \Delta\nu_{\rm D}} \, H(a, x) \, ,
  \end{align}
  where $f_{12} = 0.4162$ is the oscillator strength of the Ly$\alpha$ transition and the Hjerting-Voigt function~$H(a,x)$ is the dimensionless convolution of Lorentzian and Maxwellian distributions,
  \begin{equation}
    H(a,x)  = \sqrt{\pi} \Delta\nu_{\rm D} \, \phi_{\rm Voigt} = \frac{a}{\pi} \int_{-\infty}^\infty \frac{e^{-y^2}dy}{a^2+(y-x)^2} \, .
  \end{equation}
  For reference, we define the cross section at line centre as $\sigma_0 \equiv \sigma_x / H(a, x) = 5.898 \times 10^{-14}~T^{-1/2}_4$~cm$^2$. For a region of constant density~$n_{\rm \HI}$ -- such as a cell in a computational domain -- the integral in Equation~(\ref{eq:tau}) simplifies to
  \begin{align} \label{eq:tauH}
    \tau_x &= n_{\rm \HI} \, \sigma_0 \, \ell \, H(a,x) \notag \\
           &= 1.820 \times 10^5 \, \frac{H(a,x)}{T^{1/2}_4} \left(\frac{n_{\rm \HI}}{\rm cm^{-3}}\right) \left(\frac{\ell}{\rm pc}\right) \, .
  \end{align}

  Typically, the parameter~$a$ is much less than unity so the Hjerting-Voigt function is dominated in the centre by a resonant scattering Doppler core, $\phi_{\rm D}$, and the wings are dominated by the Lorentzian component, $\phi_{\rm L}$. If the approximate frequency marking the crossover from core to wing is denoted by $x_{\rm cw}$, i.e. where $\phi_{\rm D}(x_{\rm cw}) \simeq \phi_{\rm L}(x_{\rm cw})$, then the Hjerting-Voigt function is roughly
  \begin{equation}
    H(a,x) \approx
      \begin{cases}
        e^{-x^2} & \quad |x| < x_{\rm cw} \qquad \text{`core'} \\
        {\displaystyle \frac{a}{\sqrt{\pi} x^2} } & \quad |x| > x_{\rm cw} \qquad \text{`wing'}
      \end{cases} \, .
  \end{equation}
  See Sections~\ref{subsub:H_approx}~and~\ref{subsub:x_cw_approx} for a more rigorous discussion as well as numerical approximations for $H(a,x)$ and $x_{\rm cw}$.
  For most conditions with significant neutral hydrogen density the gas is optically thick to core photons. However, in the wing, $H(a,x)$ can be quite small allowing a photon to escape with greater ease. The approximate optical depth for a wing photon is then
  \begin{align} \label{eq:tau_wing}
    \tau_{\rm wing} & \approx \frac{48.28}{x^2 \, T_4} \left(\frac{n_{\rm \HI}}{\rm cm^{-3}}\right) \left(\frac{\ell}{\rm pc}\right) \notag \\
                    & \hlmold{ \approx 1.0_{\,} \left( \frac{\Delta v}{500~\text{km~s}^{-1}} \right)^{-2} \left(\frac{N_{\rm \HI}}{10^{20}~{\rm cm^{-2}}}\right) } \, ,
  \end{align}
  \hlold{which is independent of temperature as is immediately apparent when written in terms of the Doppler velocity $\Delta v \equiv c \Delta \lambda / \lambda$. Note the relation between Doppler frequency and velocity: $x = \Delta v / v_{\rm th}$.} Regions of high column density, $N_{\rm \HI} \gg 1$~pc~cm$^{-3} \sim 10^{18}$~cm$^{-2}$,
  may produce very high opacities. The trapped Ly$\alpha$ photons are therefore exposed to greater extinction from dust. This may be best measured by the optical depth at line centre, which \hlold{has moderate temperature dependence and} can be read from Equation~(\ref{eq:tauH}) as $\tau_0 \equiv \tau|_{x=0} \approx 5.9 \times 10^6 \, [N_{\rm \HI}/(10^{20}~\text{cm}^{-2})]~T_4^{-1/2}$.

\section{Numerical Methodology}
  \label{sec:code}
  The $\colt$ code is based on previous Monte-Carlo radiative transfer (MCRT) algorithms \citep[See e.g. ][]{Ahn:2001pz,Zheng:2002qc,Dijkstra:2005ny,Verhamme:2006tc,Laursen:2008aq}. The code reads initial conditions of velocity, density, and temperature for each cell of a three-dimensional grid employing adaptive mesh refinement~(AMR). Sampling the Ly$\alpha$ emission profile of high opacity systems is possible because acceleration schemes avoid unnecessary computations, e.g. frequent core scatterings. Additionally, $\colt$ is massively parallel allowing a greater number of photon packets and therefore less statistical error. Section~\ref{sec:methodology} outlines the general methodology for a Ly$\alpha$ transport code while Section~\ref{sec:schemes} describes the specific implementations used in $\colt$. \hlold{While the Monte-Carlo method for Ly$\alpha$ radiative transfer is fairly standard, each portion of Section}~\ref{sec:schemes} \hlold{contains significant discussion or numerical schemes unique to this work.}

  \subsection{Basic Methodology}
    \label{sec:methodology}
    \hlold{In order to sample Ly$\alpha$ observables we apply the method described below to individual photon packets. Depending on the desired resolution and physical setup a large number of packets may be necessary to obtain statistical convergence. Unless otherwise specified the number of photon packets in each simulation is $N_{\rm ph} = 10^7$.}

    \subsubsection{Photon emission}
      \label{subsub:photon_emission}
      The initial spatial distribution of Ly$\alpha$ photons is based on the physical setup, i.e. initial conditions. The first mechanism for producing Ly$\alpha$ line emission is interstellar recombination as a result of ionizing radiation from hot stars. The second mechanism is collisional excitation of neutral hydrogen, usually resulting from shocks caused by accretion or supernovae. An initialization criterion accounts for photoionizing sources and diffuse emission, both of which may be significant for atomic cooling haloes. However, throughout this work we choose to initialize photons from the central location $\mathbf{r} = \mathbf{0}$. The initial direction~$\mathbf{k}_i$ of each photon is drawn from an isotropic distribution in the rest frame of the embedded source. For convenience, velocities are expressed in terms of the thermal velocity:
      \begin{equation}
        \mathbf{u} \equiv \frac{\mathbf{v}}{v_{\rm th}} \, .
      \end{equation}
      The photon is emitted at the natural frequency of the Ly$\alpha$ photon~$x_{\rm nat}$ in the rest frame of the atom. To obtain the initial frequency~$x_i$ in the moving frame of ambient gas we apply a Doppler shift appropriate to the velocity of the atom~$\mathbf{u}_{\rm atom} = \mathbf{v}_{\rm atom} / v_{\rm th}$ \citep{Laursen:2008aq}
      \begin{equation} \label{eq:init_freq}
        x_i = x_{\rm nat} + \mathbf{k}_i \cdot \mathbf{u}_{\rm atom} \, .
      \end{equation}
      To be explicit, $x_{\rm nat}$ is drawn from a Lorentzian distribution and the components of $\mathbf{u}_{\rm atom}$ are each taken from a Maxwellian distribution describing the thermal motion of the ambient gas. Although we use the expression in Equation~(\ref{eq:init_freq}), the memory of $x_i$ is quickly lost by multiple scattering events, so for optically thick environments one may simply inject the photon at line centre, or $x_i = 0$.

    \subsubsection{Ray tracing}
      \label{subsub:ray_tracing}
      The propagation distance of any photon is determined by the optical depth~$\tau_\nu$ drawn from an exponential distribution. This is because the mean optical depth $\langle \tau_\nu \rangle \equiv \int_0^\infty \tau_\nu e^{-\tau_\nu} d\tau_\nu = 1$ defines the mean free path~$\lambda_{\rm mfp}$, or the average distance a photon can travel without being absorbed (or scattered) by the intervening medium~\citep{Rybicki:book}. Therefore, travel beyond each mean free path becomes less and less probable. Formally, the mean free path is then $\lambda_{\rm mfp} \equiv \ell |_{\tau_\nu = 1} = 1/n_{\rm \HI} \sigma_\nu$.

      From Equation~(\ref{eq:tau}) the respective optical depths of infinitesimal paths are $d\tau_{\nu, i} = n_{\rm \HI}\,\sigma_\nu\,d\ell_i = d\ell_i/\lambda_{\rm mfp}$ so the probability of \textit{no} interaction is $1 - d\ell_i/\lambda_{\rm mfp} = 1 - d\tau_{\nu, i}$. With $N$ partitions of $\Delta\tau_\nu$, the numerical representation of discrete intervals, the probability distribution function over the total path is the product
      \begin{equation}
        P(\tau_\nu) = \lim_{N \rightarrow \infty} \left( 1 - \frac{\Delta\tau_\nu}{N} \right)^N = e^{-\tau_\nu} \, ,
      \end{equation}
      which has been normalized so that $\int_0^\infty P(\tau_\nu) d\tau_\nu = 1$. The cumulative distribution function is the integrated distribution defined by $F(\tau_\nu) \equiv P(\leq \tau_\nu) = \int_0^{\tau_\nu} P(\tau'_\nu) d\tau'_\nu$, which can be inverted to give the optical depth at which an interaction event occurs,
      \begin{equation} \label{eq:lnR}
        \tau_{\rm event} = -\ln R \, ,
      \end{equation}
      where $R$ is drawn from a univariate distribution.

      To perform Cartesian-like ray tracing, the Monte-Carlo method selects a photon with an optical depth~$\tau_{\rm event}$ according to Equation~(\ref{eq:lnR}). Because individual cells are regions of uniform density we may equate this with the calculated optical depth~$\tau_x$ from Equation~(\ref{eq:tauH}) to find the propagation distance, i.e. $\tau_x = \tau_{\rm event}$ provides $\ell(\tau_{\rm event})$. However, this often leads to a scattering which is outside the original cell. Therefore, we first calculate the optical depth required to travel through the cell: $\tau_{x, \rm cell} = n_{\rm \HI, cell} \, \sigma_0 \, \ell_{\rm cell} \, H(a_{\rm cell},x)$, where $\ell_{\rm cell}$ is the distance from the current position to the point where the photon exits the cell. If $\tau_{\rm event} > \tau_{x, \rm cell}$ the `spent' optical depth is subtracted from the current optical depth, i.e. $\tau_{\rm event} = \tau_{\rm event} - \tau_{x, \rm cell}$. Likewise, the current position is updated to $\mathbf{r} = \mathbf{r} + \ell_{\rm cell} \mathbf{k}_i$ and ray tracing continues through the next cell.

      As photons traverse from cell to cell the temperature~$T$ and ambient bulk velocity~$\mathbf{u}_{\rm bulk}$ may change. Therefore, Doppler shifting induces terms of $\pm \mathbf{k}_i \cdot \mathbf{u}_{\rm bulk}$ corresponding to a nonrelativistic Lorentz transformation, i.e. one where $\| \mathbf{u}_{\rm bulk} \| \ll c$. Additionally, the frequency differs between cells due to the $x \propto \Delta \nu_{\rm D}^{-1} \propto T^{-1/2}$ scaling relation. The frequency in the new (primed) cell is then
      \begin{equation}
        x' = ( x + \mathbf{k}_i \cdot \mathbf{u}_{\rm bulk} ) \sqrt{T/T'} - \mathbf{k}_i \cdot \mathbf{u}'_{\rm bulk} \, .
      \end{equation}

      If $\tau_{\rm event} \leq \tau_{x, \rm cell}$ the optical depth is `exhausted' in the current cell and the photon undergoes a scattering event. The propagation distance is recalculated according to the remaining optical depth, i.e. $\ell_{\rm event} = \tau_{\rm event} / [n_{\rm \HI, cell} \, \sigma_0 \, H(a_{\rm cell},x)]$. The position is advanced by $\ell_{\rm event}$ in the $\mathbf{k}_i$ direction and the algorithm proceeds to update the frequency and direction as described in Section~\ref{sec:scatter}. The photon ray traces along a new direction with each subsequent scattering until it ultimately escapes the computational domain.

    \subsubsection{Scattering events}
      \label{sec:scatter}
      A scattering event changes the photon's frequency according to the atom's velocity~$\mathbf{u}_{\rm atom}$, the final~$\mathbf{k}_f$ and initial~$\mathbf{k}_i$ directions, and a small recoil effect to satisfy conservation of momentum
      \begin{equation} \label{eq:x_f}
        x_f = x_i + ( \mathbf{k}_f - \mathbf{k}_i ) \cdot \mathbf{u}_{\rm atom} + g \, ( \mathbf{k}_i \cdot \mathbf{k}_f - 1 ) \, .
      \end{equation}
      The recoil parameter~$g$ is defined as \citep[see e.g.][]{Adams:1971}
      \begin{equation} \label{eq:recoil_parameter}
        g \equiv \frac{h \Delta\nu_{\rm D}}{2 k_{\rm B} T} = 2.536 \times 10^{-4}~T^{-1/2}_4 \approx 0.54 a \, ,
      \end{equation}
      which is included but negligible for the applications of this paper.

      The velocity of the scattering atom is most conveniently expressed in terms of its parallel~$\mathbf{u}_\|$ and perpendicular~$\mathbf{u}_\perp$ components. The perpendicular magnitudes are unbiased by the photon's frequency and are therefore drawn from a Gaussian, i.e. $\exp(-u^2_\perp)/\sqrt{\pi}$. However, the magnitude of the parallel velocity $u_\| = \mathbf{k}_i \cdot \mathbf{u}_{\rm atom}$ is affected by the presence of the resonance line. The distribution function for $u_\|$ depends on frequency as the convolution of a Gaussian with a Doppler-shifted Lorentzian peak:
      \begin{equation} \label{eq:uparf}
        f(u_\|) = \frac{a}{\pi H(a,x)} \frac{e^{-u^2_\|}}{a^2 + (x-u_\|)^2} \, ,
      \end{equation}
      which highly favors velocities with $u_\| = x$ for core photons. For wing photons ($x \gg x_{\rm cw}$), however, the probability that an atom has a high enough velocity to Doppler shift into resonance becomes vanishingly small, so $f(u_\|)$ becomes increasingly Gaussian.

      The angle~$\theta$ between the incident and outgoing scattering directions is governed by the phase (probability) function
      \begin{equation} \label{eq:phaseW}
        W(\theta) \propto 1 + \frac{R}{Q} \cos^2 \theta \, ,
      \end{equation}
      where $R/Q$ is the degree of polarization for $90\degr$ scattering. Due to the physical symmetry the phase function is independent of the azimuthal angle $\phi$. For core photons with $x < x_{\rm cw}$ the electron transition to the $2p_{1/2}$ state results in isotropic scattering, i.e. $R/Q = 0$, while the $2p_{3/2}$ transition results in polarization with $R/Q = 3/7$ \citep{Hamilton:1940}. Since a quantum state with angular momentum~$j$ has a spin multiplicity of $2j+1$, Ly$\alpha$ photons in the core are excited to the $2p_{1/2}$ state with $1/3$ probability and the $2p_{3/2}$ state with $2/3$ probability. Nonresonant wing photons on the other hand are dominated by Rayleigh scattering because their wavelength is much larger than the Bohr radius, $\lambda_{\rm Ly\alpha} \gg a_0$, therefore the resultant polarization is maximal, i.e. $R/Q = 1$ \citep{Stenflo:1980}. Anisotropic scattering is used throughout $\colt$.

  \subsection{Computational optimization schemes}
    \label{sec:schemes}
    The previous section was written with little regard to computational efficiency, something we now seek to remedy. The details are presented in order of introduction, rather than importance.

    \begin{figure}
      \centering
      \includegraphics[width=\columnwidth]{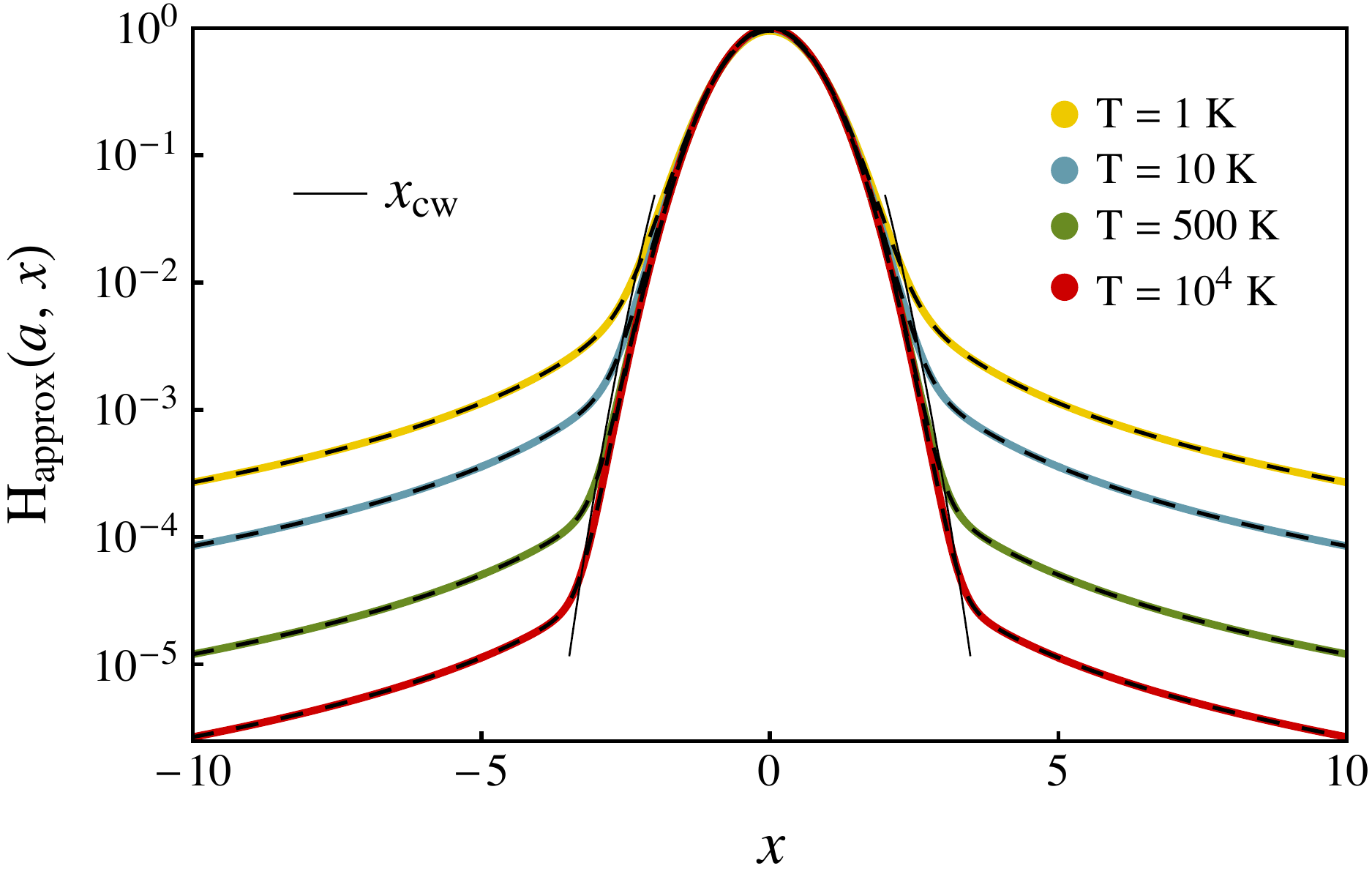}
      \includegraphics[width=\columnwidth]{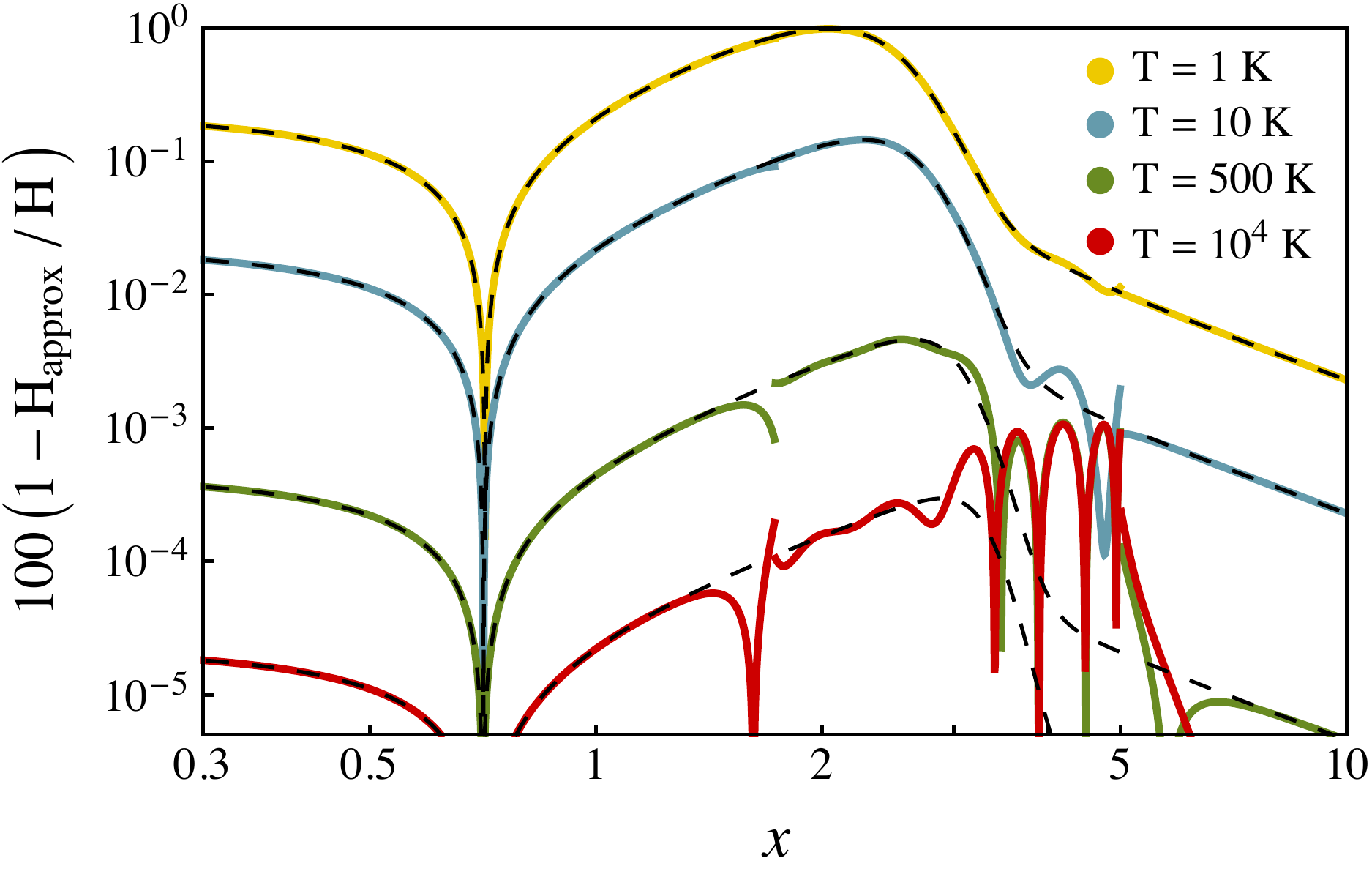}
      \vspace{-.25cm}
      \caption{Top panel: The Hjerting-Voigt function~$H(a,x)$ and our approximation for different values of temperature. The crossover from core to wing~$x_{\rm cw}$ as given by Equ.~(\ref{eq:xcw}) is also shown. Bottom panel: A demonstration of the relative per cent error $100\,[1-H_{\rm approx}(a,x)/H(a,x)]$ where $H_{\rm approx}(a,x)$ is  given by Equ.~(\ref{eq:AB}). The yellow, blue, green, and red curves correspond to temperatures of $1$~K, $10$~K, $500$~K and $10^4$~K, respectively. The overlying dashed lines represent the best case scenario when only keeping first order terms in $a$, i.e. using the exact Dawson integral.}
      \label{fig:Herror}
    \end{figure}

    \subsubsection{Approximation for H(a,x)}
      \label{subsub:H_approx}
      An approximation for the Voigt profile is important because $H(a,x)$ is evaluated after every scattering. A substantial effort has gone into studying this profile and implementing efficient algorithms with double precision accuracy -- see e.g. \citet{Schreier:Happrox} and references therein. However, the Ly$\alpha$ resonance line is a unique application with a specific parameter range, so the approximations used in this work are entirely our own. We require our algorithm to provide better than one per cent accuracy for all frequencies and all realistic temperatures. To do this we first evaluate the integral $H(a,x)$ with special functions and expand to second order in $a$:
      \begin{align}
        H(a,x) &= \frac{a}{\pi} \int_{-\infty}^\infty \frac{e^{-y^2}dy}{a^2+(y-x)^2} \notag \\
               &= Re \left( e^{(a-ix)^2} \text{erfc} (a-ix) \right) \notag \\
               &= e^{-x^2} + \frac{2 a}{\sqrt{\pi}} (2 x F(x) - 1) \notag \\
               & \quad + a^2 e^{-x^2} \left( 1 - 2 x^2 \right) + \mathcal{O} \left( a^3 \right) \, ,
      \end{align}
      where the (complex) complementary error function is related to the area under a Gaussian by $\text{erfc}(z) \equiv 1 - 2 \int_0^z e^{-y^2} dy / \sqrt{\pi}$ and the associated Dawson integral is $F(x) \equiv \int_0^x e^{y^2-x^2} dy$.

      $\colt$ takes advantage of the fact that $H(a,x)$ is symmetric by first evaluating $z = x^2$. Then because the behavior of the profile differs for small and large $z$ the domain is decomposed into three regions. The `core' region is derived by expanding around $z = 0$, while the `wing' region is an asymptotic expansion. An intermediate region acts as a smooth transition between the two. The approximations utilize continued fractions in order to maximize the efficiency of a small number of operations, e.g. $8$ additions and $4$ divisions. See Appendix~\ref{appendix:H_approx} for the implementation of $H_{\rm approx}$.

      Second and higher order terms in $a$ are required to achieve one per cent accuracy for sub-Kelvin temperatures. However, since the CMB temperature floor prevents gas from reaching such low temperatures only first order terms are used in $\colt$. Figure~\ref{fig:Herror} demonstrates the relative error, i.e. $100\,[1-H_{\rm approx}(a,x)/H(a,x)]$, for temperatures ranging from $1-10^4$~K. As can be seen, the error is well controlled, improving even further for higher temperatures.

    \subsubsection{Approximation for $x_{\rm cw}$}
      \label{subsub:x_cw_approx}
      \begin{figure}
        \centering
        \includegraphics[width=\columnwidth]{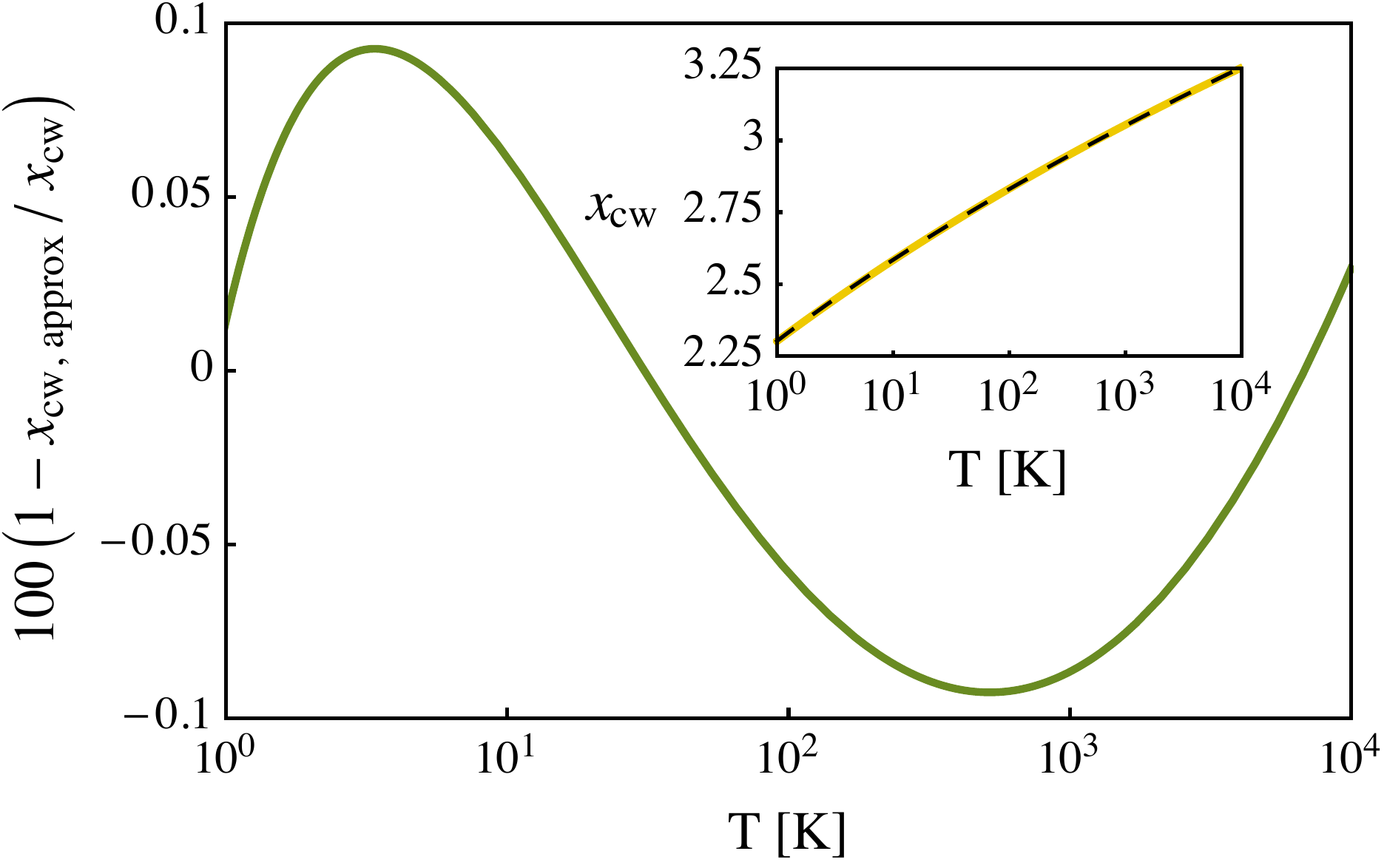}
        \vspace{-.25cm}
        \caption{The relative per cent error, $100\,[1-x_{\rm cw, approx}/x_{\rm cw}]$, is shown in green. The yellow curve in the insert is $x_{\rm cw, approx}$ as given by Equation~(\ref{eq:xcw}). All curves cover a temperature range of $T \in [1, 10^4]$~K. }
        \label{fig:xcwerror}
      \end{figure}
      The crossover from core to wing~$x_{\rm cw}$ determines when $H(a,x)$ changes from a Gaussian to a Lorentzian. This is important because $x_{\rm cw}$ marks the point where difficulties arise in generating $u_\|$ and also identify whether core skipping is necessary (see Sections \ref{sec:upar}~and~\ref{sec:coreskip}). Our calculation assumes that the core and wing limits of $H(a,x)$ compete in their contribution to the profile:
      \begin{equation} \label{eq:xcwA}
        e^{-x_{\rm cw}^2} \approx \frac{a}{\sqrt{\pi} x_{\rm cw}} \, .
      \end{equation}
      Equation~(\ref{eq:xcwA}) provides a conservative approximation for $x_{\rm cw}$, providing a sharp boundary with the core~(see Fig.~\ref{fig:Herror}). The exact solution can be written in terms of the lower branch of the Lambert~$W$ function, which in turn may be approximated by a low-order rational function with a relative accuracy of less than $0.1$ per cent for the same parameter range as $H_{\rm approx}(a,x)$. In summary, our optimal calculation of the core-to-wing crossover frequency is
      \begin{align} \label{eq:xcw}
        x_{\rm cw} &= \sqrt{ -W_{-1} \left( -\frac{a}{\sqrt{\pi}} \right) } \notag \\
                   &\approx \sqrt{ L_1 - L_2 + \frac{L_2}{L_1} + \mathcal{O}(L_1^{-2}) } \notag \\
                   &\approx 6.9184721 + \frac{81.766279}{\log a - 14.651253} \, ,
      \end{align}
      where $L_1 = \log (a/\sqrt{\pi})$ and $L_2 = \log [ - \log (a/\sqrt{\pi}) ]$. The relative per cent error is shown in Fig.~\ref{fig:xcwerror}.

    \subsubsection{Generating the scattering velocity $u_\|$}
      \label{sec:upar}
\begin{figure}
        \centering
        \includegraphics[width=\columnwidth]{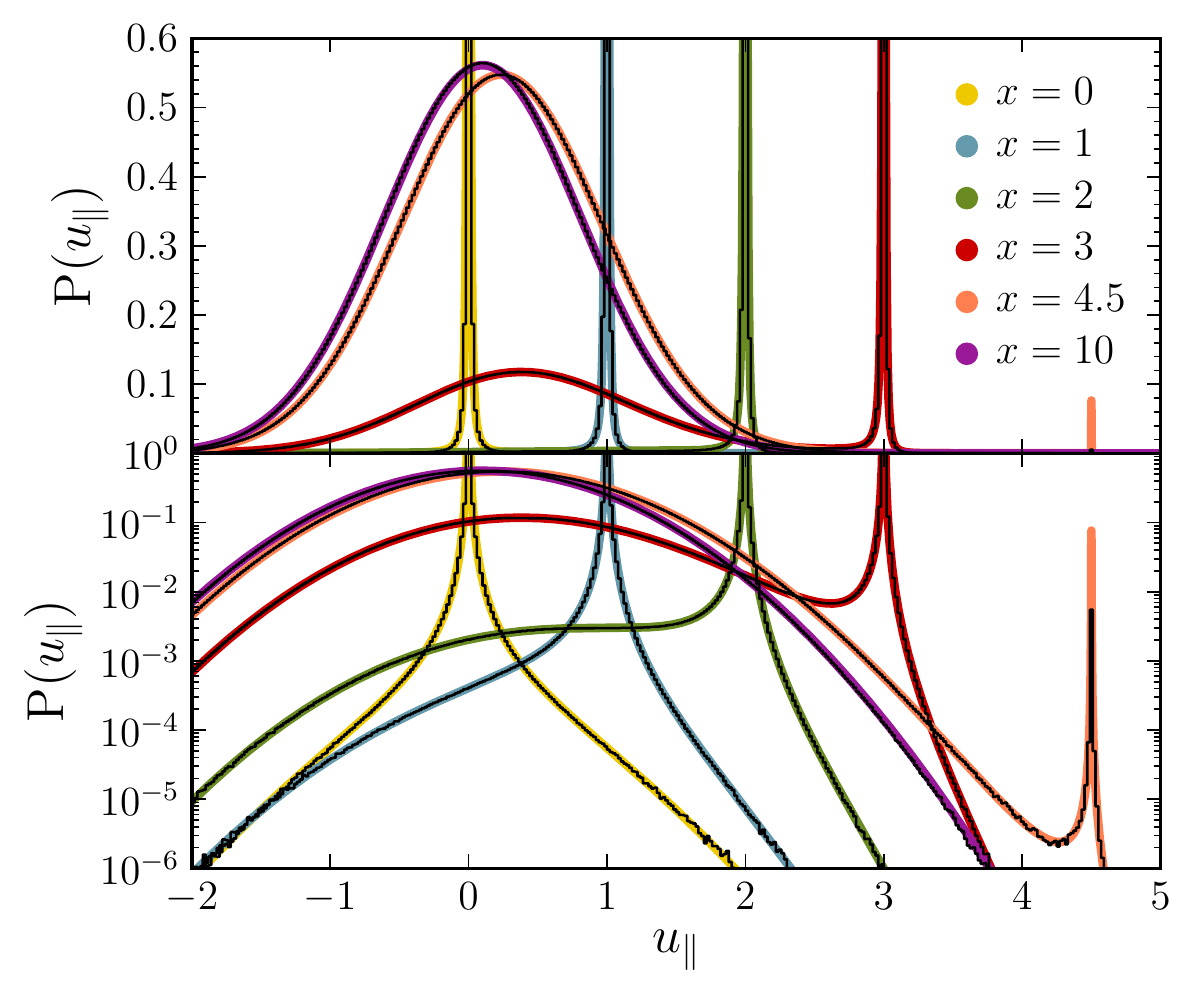}
        \vspace{-.4cm}
        \caption{The distribution of parallel velocities~$f(u_\|)$ as given by Equ.~(\ref{eq:uparf}) for different values of incoming frequency~$x$. The profile resembles a Gaussian with a sharp peak around the point $u_\| = x$. For large $x$ it becomes too improbable for atoms to have velocities high enough to Doppler shift into resonance so the peak at $u_\| = x$ disappears and a shifted Gaussian is a good approximation. The frequencies sampled are $x = \{0, 1, 2, 3, 4.5, 10\}$.}
        \label{fig:upar}
      \end{figure}

      As described in Section~\ref{sec:scatter} (cf. Equation~\ref{eq:uparf}) the distribution for the parallel velocity $u_\|$ with respect to the incoming photon is
      \begin{equation}
        f(u_\|) \propto \frac{e^{-u^2_\|}}{a^2 + (x-u_\|)^2} \, .
      \end{equation}
      To good approximation this profile resembles a Gaussian with a sharp peak around the point $u_\| = x$ (see Fig.~\ref{fig:upar}).

      Unfortunately, $f(u_\|)$ is not integrable so we instead use the inverted cumulative distribution function method on a related distribution and employ the rejection method to accept each draw. In this case the comparison function is chosen to be
      \begin{equation}
        g(u_\|) \propto \frac{1}{a^2 + (x-u_\|)^2} \, ,
      \end{equation}
      and draws are accepted with a probability of $f/g = \exp(-u_\|^2)$. However, we only employ this algorithm for small frequencies, i.e. $x \leq 1$, because as $x$ increases the method becomes quite inefficient.

      We learn more about $f(u_\|)$ by examining its behavior when $x\rightarrow\infty$. Here the peak at $u_\| = x$ is pushed so far into the wing that there are essentially no atoms with speeds fast enough to absorb at the Doppler shifted resonance line, i.e. $f \approx \exp{-u_\|^2}/(a^2+x^2)$. Therefore, we may simply draw from a proper Gaussian with a slight modification -- the peak is shifted by $x^{-1}$. We demonstrate this by finding the local extrema in the core:
      \begin{align}
        \frac{df(u_\|)}{du_\|} & \propto u_\| \left( 1 + a^2 + (x-u_\|)^2 \right) - x \approx u_\| x^2 - x = 0 \notag \\
                               & \Rightarrow \quad u_{\rm \|, max} = \frac{1}{x} \, .
      \end{align}
      Here we have assumed $|x| \gg 1 \gg u_\|$. It is straightforward to show $f''(u_\|) \approx -2 x^{-2} < 0$ so this is a local maximum as expected. Therefore, for $x \ga 9$ we draw from a Gaussian with mean $x^{-1}$.

      The intermediate region is problematic for either of these methods. Therefore, we follow \citet{Zheng:2002qc} and use the piecewise comparison function:
      \begin{equation}
        g(u_\|) \propto
          \begin{cases}
            g_1 = 1 / \left[ a^2 + (x-u_\|)^2 \right] & \; u_\| \leq u_0 \\
            g_2 = e^{-u_0^2} / \left[ a^2 + (x-u_\|)^2 \right] & \; u_\| > u_0
          \end{cases}
      \end{equation}
      where $u_0$ is a separation parameter and the corresponding acceptance fraction is $\exp(-u_\|^2)$ for $g_1$ and $\exp(u_0^2-u_\|^2)$ for $g_2$. We now restrict the discussion to positive $x$, which is possible because $f(-x,u_\|) = f(x,-u_\|)$ allows us to recover velocities drawn from negative $x$. The probability that a velocity is less than $u_0$ is
      \begin{equation}
        p = \frac{\int_{-\infty}^{u_0} g(u_\|) du_\|}{\int_{-\infty}^{\infty} g(u_\|) du_\|}
          = \frac{\theta_0+\frac{\pi}{2}}{\left(1-e^{-u_0^2} \right)\,\theta_0 + \left(1+e^{-u_0^2} \right)\,\frac{\pi}{2}} \, ,
      \end{equation}
      where
      \begin{equation}
        \theta_0 = \tan^{-1} \left( \frac{u_0 - x}{a} \right) \, .
      \end{equation}
      If $p < R$, a univariate, then $\theta$ is drawn uniformly from the interval $[\theta_0, \pi/2]$, otherwise $\theta \in [-\pi/2, \theta_0]$. Finally, a velocity candidate,
      \begin{equation}
        u_\| = a\,\tan\theta + x \, ,
      \end{equation}
      is accepted if another univariate, $R'$, is less than the acceptance fraction, i.e. $\exp(-u_\|^2)$ or $\exp(u_0^2-u_\|^2)$ for each respective region.

      The algorithm from \citet{Zheng:2002qc} works well as long as the probability of the two regions are balanced, i.e. in $\colt$ we attempt to maintain $p \sim \frac{1}{2}$. The reason for this is that $u_0$ controls the acceptance fraction, so if $u_0$ is too small then we do not gain much for larger $x$ and conversely if $u_0$ is too large then we defeat the purpose for smaller $x$. The exact value of $p$ is very sensitive to both $x$ and $a$, so we can only hope for average acceptance rates to be reasonable given the variance in $u_0(a, x)$. If we assume $u_0 < x$ then to first order in $a$ we have:
      \begin{equation} \label{eq:papprox}
        p \approx \frac{a}{\pi} \frac{e^{u_0^2}}{x-u_0} \, .
      \end{equation}
      The behavior of this function is different for core and wing photons. In the core it is reasonable to assume a perturbation from the natural peak location and apply the transformation $u_0 \rightarrow x - a u'_0$ where $u'_0 = u'_0(a,x)$ admits an analytic solution to the approximate balance of $p \sim \frac{1}{2}$. Again, to first order in $a$ Equation~(\ref{eq:papprox}) becomes
      \begin{equation}
        p \approx \frac{e^{(x - a u'_0)^2}}{\pi u'_0} \approx \frac{1 - 2ax\,u'_0}{\pi u'_0} e^{x^2} \approx \frac{1}{2} \, .
      \end{equation}
      The solution in terms of $u_0 = x - u'_0$ is
      \begin{equation}
        u_{0{\rm , core}} \approx x - \frac{1/2}{x + \frac{\pi}{4 a} e^{-x^2}} \approx x - \frac{1}{x + e^{1-x^2}/a} \, ,
      \end{equation}
      where the final equality allows the approximation to extend to larger $x$ and is used in $\colt$ for $1 < x < x_{\rm cw}$.

\begin{figure}
        \centering
        \includegraphics[width=\columnwidth]{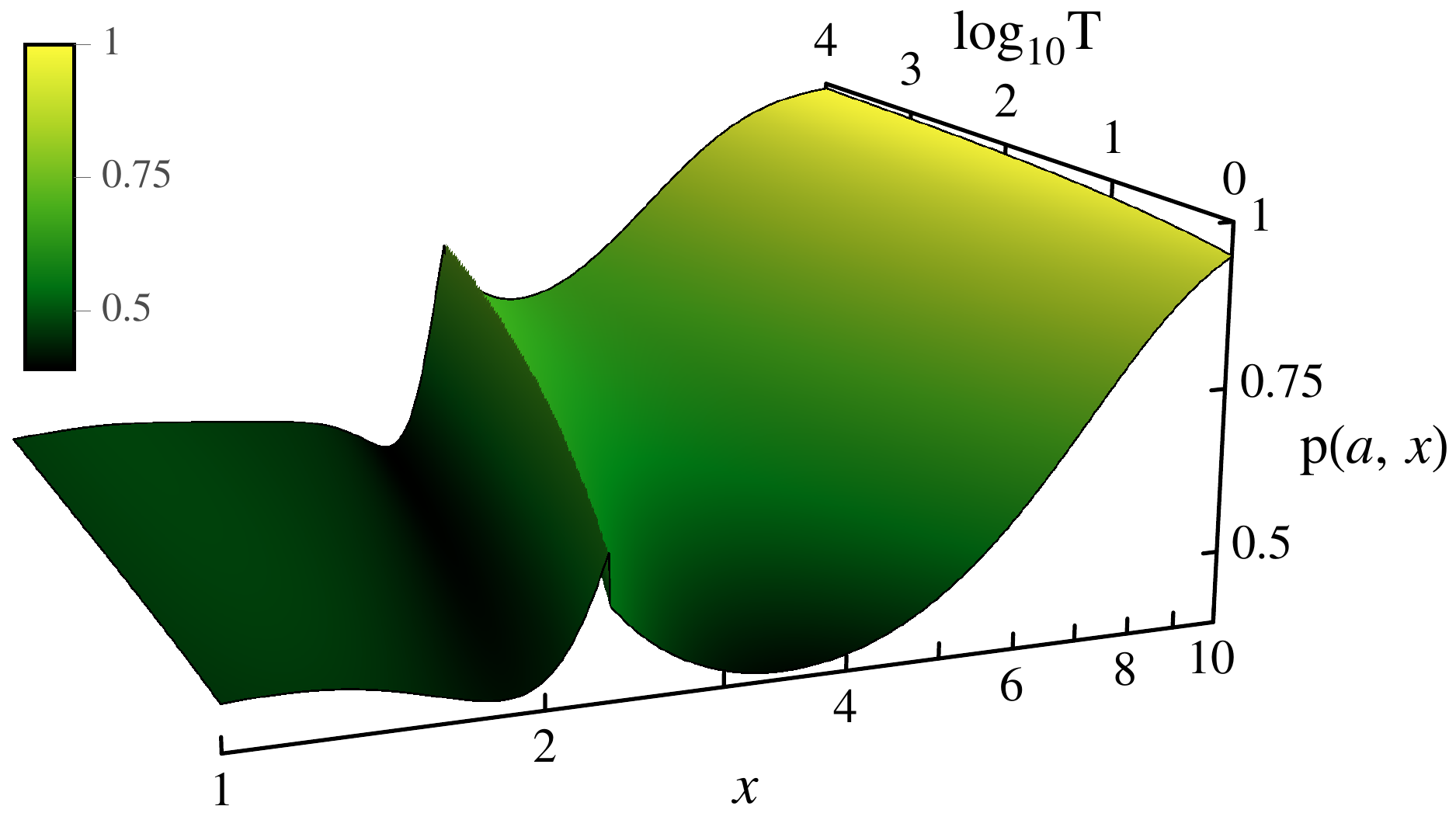}
        \vspace{-.25cm}
        \caption{The fractional probability~$p$ that $u_\| < u_0$. This separation captures the behavior of $g$ for frequencies above and below the transitional frequency~$x_{\rm cw}$. The axes are frequency~$x \in (1, 10)$ and (log$_{10}$) temperature~$T \in (1,10^4)$~K.}
        \label{fig:u0p}
      \end{figure}

      For the region in which the wing dominates, i.e. $x > x_{\rm cw}$, the algorithm suffers an identity crisis. As discussed earlier, the high-$x$ behavior of $f$ approaches a Gaussian distribution though each of the candidate $u_\|$ samples are from a Lorentzian distribution. Therefore, the exact value of $u_0$ is less important as long as it is greater than $x_{\rm cw}$. We test various prescriptions for $u_0(a,x)$, also varying initial frequency~$x$ and temperature~$T$, in order to minimize the average number of draws in a simulation. A linear function, matched to the previous region
      provides sufficient acceptance of candidate random numbers. We use the following separation constant for $x_{\rm cw} < x < 9$, denoted the `wing' region:
      \begin{equation}
        u_{0{\rm , wing}} \approx x_{\rm cw} - x_{\rm cw}^{-1} + 0.15 \left( x - x_{\rm cw} \right) \, .
      \end{equation}
      Figure~\ref{fig:u0p} demonstrates the fractional probability~$p(a,x)$ for drawing $u_\| < u_0$ given our piecewise prescription of $u_0 = u_{0{\rm , core}}$ for $1 < x < x_{\rm cw}$ and $u_0 = u_{0{\rm , wing}}$ for $x_{\rm cw} \leq x < 9$. The efficiency of the algorithm is especially sensitive to frequency.

    \subsubsection{Core-skipping}
      \label{sec:coreskip}
      In optically thick regimes photons spend much of their time undergoing core scatterings with negligible diffusion in physical or frequency space. These scatterings can be avoided by only selecting atoms with perpendicular velocity components greater than a critical frequency, i.e. photons have zero mean free path if $u_\perp < x_{\rm crit}$. Following \citet{Ahn:2001pz} we employ the Box-Muller method to generate two independent draws according to:
      \begin{align}
        u_{\perp,1} & = \sqrt{x_{\rm crit}^2 - \ln R_1} \cos 2 \pi R_2 \notag \\
        u_{\perp,2} & = \sqrt{x_{\rm crit}^2 - \ln R_1} \sin 2 \pi R_2 \, ,
      \end{align}
      where $R_1$ and $R_2$ are univariates. The speedup time achieved here is significant since the probability of drawing a wing photon is roughly $P_{\rm wing} \sim \int^\infty_{x_{\rm cw}} \exp (-x^2) \, dx \sim 10^{-5}$ corresponding to skipping roughly $10^5$ core scatterings.

      The crucial problem is to find an appropriate value for the critical frequency~$x_{\rm crit}$. $\colt$ introduces an algorithm with core-skipping based on both local and nonlocal criteria. This is especially important for high resolution, adaptively structured grids where the range of densities may cover several orders of magnitude. We desire a local determination of $x_{\rm crit}$ that accelerates the code but does not artificially push photons too far into the wings. As noted by \citet{Laursen:2008aq} the important parameter is the product~$a\tau_0$ so we seek a relation of the form $x_{\rm crit}(a\tau_0)$. The model we consider for the near zone environment is that of an optically thick static uniform sphere. This is motivated by the idealized geometry and an analytic solution for the angular averaged intensity~$J$ at the surface first given by \citet{Dijkstra:2005ny},
      \begin{equation} \label{eq:sphere}
        J(\tau_0, a, x) = \frac{1}{8} \sqrt{\frac{\pi}{6}} \frac{x^2}{a \tau_0} \text{sech}^2 \left( \sqrt{\frac{\pi^3}{54}} \frac{x^3}{a \tau_0} \right) \, ,
      \end{equation}
      which has been normalized to $1/4 \pi$, reflecting an integration over solid angle. The peaks are located at $x_{\rm p} = \pm 0.931 \, (a \tau_0)^{1/3}$, which is derived by solving the equation $\partial J / \partial x = 0$, or equivalently $\bar{x} \tanh \bar{x} = \frac{1}{3}$ with $\bar{x} = \sqrt{\pi^3/54} \, x^3/a \tau_0$. Therefore, the peak heights correspond to $J_{\rm p} \equiv J(x_{\rm p}) = 0.0551 \, (a \tau_0)^{-1/3}$. We next expand Equation~(\ref{eq:sphere}) around $x = 0$ and define $x_{\rm crit}$ for large $a\tau_0$ as the frequency where $J \approx \frac{1}{8} \sqrt{\frac{\pi}{6}} x^2 / a \tau_0 + \mathcal{O}(x^8)$ reaches a small fraction of $J_{\rm p}$, giving $x_{\rm crit} \approx \frac{1}{5} ( a\tau_0 )^{1/3} ( {}^{\rm percentage}_{\rm calibration} / 6.57 \% )^{-1/2}$. Thorough tests demonstrate this expression for $x_{\rm crit}$ is valid for all $a\tau_0 > 1$ (see Appendix~\ref{appendix:x_crit}). Furthermore, it has a negligible effect on the emergent spectrum but greatly reduces the computation time. In summary, $\colt$ uses the following approximation:
      \begin{equation} \label{eq:xcrit}
        x_{\rm crit} =
          \begin{cases}
            0 & \quad \text{for} \;\; a\tau_0 < 1 \\
            \frac{1}{5} \left( a\tau_0 \right)^{1/3} & \quad \text{for} \;\; a\tau_0 \geq 1
          \end{cases} \, .
      \end{equation}
      If the photon is already in the wing we do not use a cutoff because there are no core scatterings to skip.

      The final ingredient in Equation~(\ref{eq:xcrit}) to be explained is how to calculate the product $a\tau_0$. $\colt$ employs a combination of local and nonlocal estimates of how aggressive to be with core-skipping. As we cannot possibly predetermine the escape path of a given photon we instead place a conservative limit on $a\tau_0$ for different lines of sight. The local criterion is computed on-the-fly, using the minimum optical depth to the edge of the current cell~$a\tau_{\rm cell}$. However, in highly refined regions (perhaps with partial ionization) and for scattering events near cell edges this can be quite small, i.e. if $\ell_{\rm cell}$ is the minimum distance to the cell boundary 
      Equation~(\ref{eq:tauH}) gives
      \begin{equation} \label{eq:atau0min}
        a \tau_{\rm cell} = 85.56~T_4^{-1}~\left( \frac{n_{\rm \HI}}{\text{cm}^{-3}} \right) \left( \frac{\ell_{\rm cell}}{\text{pc}} \right) \, .
      \end{equation}
      It is apparent that if $n_{\rm \HI}$ is roughly constant then the only nonlocal quantity needed is the physical size of the system. Therefore, the minimum integrated column density along rays emanating from the scattering event sets up an effective sphere with the intensity of Equation~(\ref{eq:sphere}). Rather than calculate this at every scattering we combine the cell-based determination with a nonlocal~(\textsc{nl}) estimate:
      \begin{equation} \label{eq:atauNL}
        a \tau_\textsc{nl} \equiv \min \sum_{\rm path} a\tau_0 \, ,
      \end{equation}
      where the path is followed as long as the relative change in neutral hydrogen density remains less than a prescribed threshold, i.e. $| \Delta n_{\rm \HI} / n_{\rm \HI} | < f_\textsc{nl} \sim \frac{1}{2}$. A value of $f_\textsc{nl} = 0$ ignores the nonlocal scheme completely and setting $f_\textsc{nl} \sim 1$ is too aggressive, failing to even detect sharp ionization fronts. To avoid double counting the local contribution, the paths originate at the edge of each cell and proceed outward. In practice the paths include at least the six directions of the coordinate axes and possibly more to achieve greater angular coverage. The nonlocal estimate can be computed once for each cell as the initial conditions are read in. $\colt$ currently uses the sum of the local determination $a\tau_{\rm cell}$ and the nonlocal estimate $a\tau_\textsc{nl}$ as the value of $a\tau_0$ used in Equation~(\ref{eq:xcrit}).

      We note that other criteria could be used for estimating $a\tau_0$. For example, the nonlocal integration might stop if the velocity gradient exceeds the threshold for Sobolev escape; however, in most cases this is a secondary factor with little affect on core-skipping. Finally, additional local distances may be used in certain cases. Of special interest is to use the Jeans length to estimate core-skipping in idealized galactic setups. The Jeans length~$\lambda_{\rm J} \equiv \sqrt{15 k_{\rm B} T / 4 \pi G \mu \rho}$ acts as a physical upper limit for the size of a system with uniform density $\rho$. A primordial gas with mean molecular weight $\mu \approx 1.23~m_{\rm H}$, mass fraction of hydrogen $X \approx 0.75$, and density $\rho \approx m_{\rm H} n_{\rm \HI}/X$ corresponds to a Jeans length of $\lambda_{\rm J} = 0.75~\text{kpc}~T_4^{1/2}~(n_{\rm \HI}/\text{cm}^{-3})^{-1/2}$ and a local $a\tau_0$ of
      \begin{equation} \label{eq:atau0J}
        a \tau_{\rm J} = 6.428 \times 10^4~T_4^{-1/2}~\left( \frac{n_{\rm \HI}}{\text{cm}^{-3}} \right)^{1/2} \, ,
      \end{equation}
      so that $x_{\rm crit, J} \approx 8$ for $T = 10^4$~K and $n_{\rm \HI} = 1~\text{cm}^{-3}$.

    \subsubsection{Parallel implementation}
      Monte Carlo codes benefit greatly from parallel computation because each photon packet is an independent event. Therefore, the scalability is nearly linear. $\colt$ uses the Message Passing Interface~(MPI) libraries to implement dynamic load balancing between different processors. This is done because photons that undergo many scatterings take longer to escape. Therefore, the master process allocates a reasonable amount of work to each slave, e.g. ($1-10$\% of the photons)/(number of processors), until the required number of photons are assigned. The nonlocal determination of $\int d(a\tau)$ for each cell, see Equation~(\ref{eq:atauNL}), is also performed with an efficient parallel computation.

\section{Test Cases}
  \label{sec:tests}
  At this stage it is important to verify the code against known solutions. We choose tests that are complementary to each other in order to isolate certain key aspects of Ly$\alpha$ transport. The static test is the well-known Neufeld analytical solution for an optically thick homogeneous slab \citep{Harrington,Neufeld}. The dynamic test is that of an isotropically expanding sphere with different maximal velocities as described by \citet{Laursen:2008aq}.

  \begin{figure}
    \centering
    \includegraphics[width=\columnwidth]{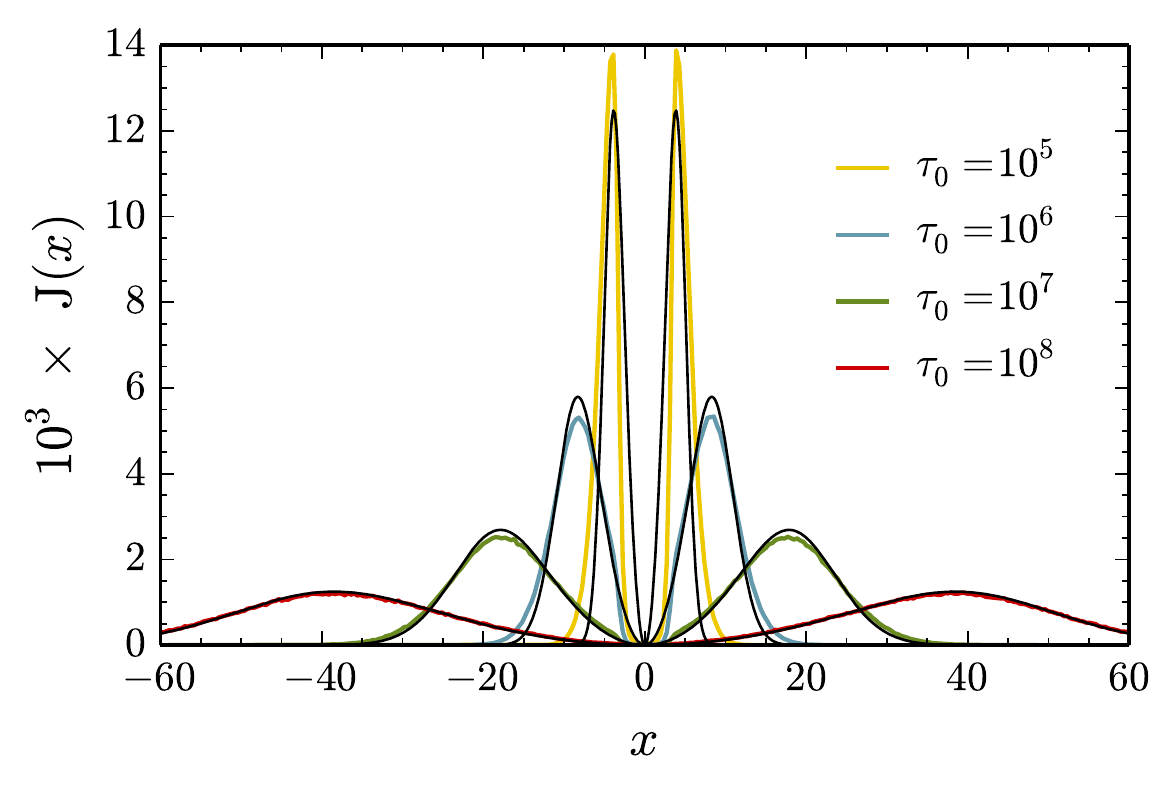}
    \includegraphics[width=\columnwidth]{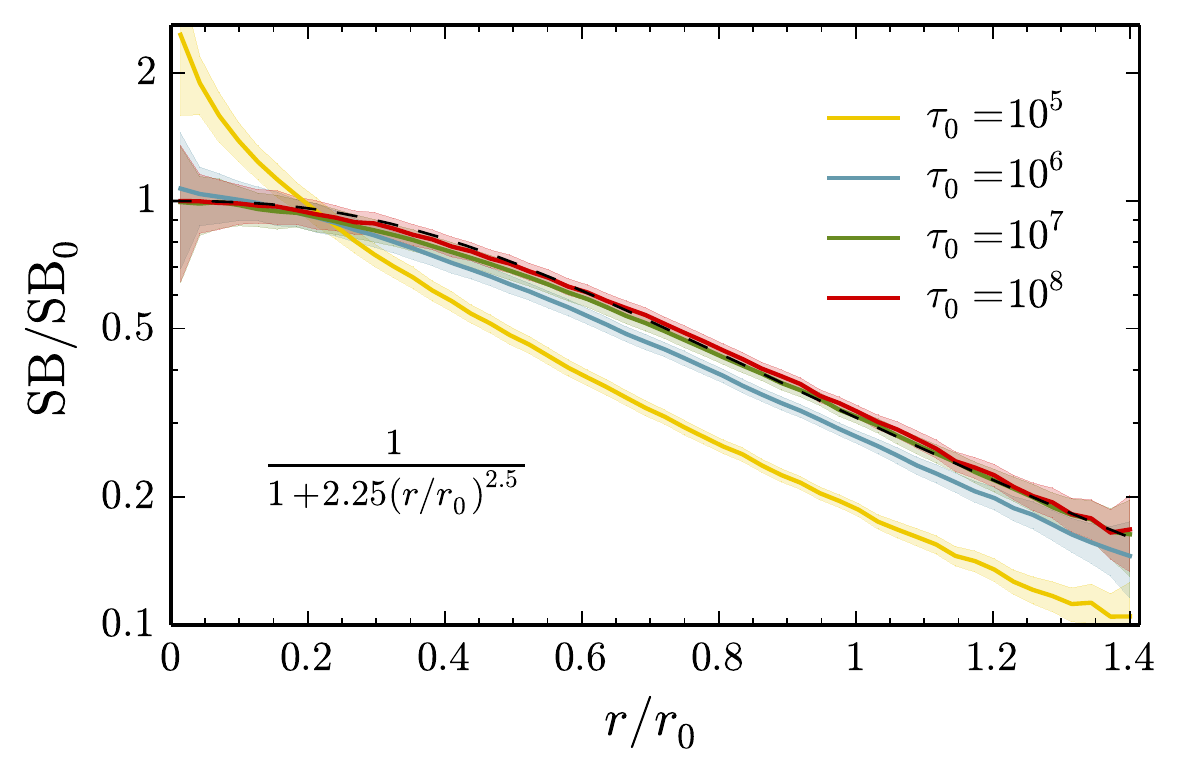}
    \vspace{-.4cm}
    \caption{Top panel: The Neufeld solution for the angular averaged intensity~$J(\tau_0, x)$ of a static, homogeneous slab at $T = 10^4$~K. The central optical depth~$\tau_0$ covers a range of moderate to extreme values, where the yellow, blue, green, and red curves represent $\colt$ simulations using $\sim 10^6$ photon packets for $\tau_0 = 10^5$, $10^6$, $10^7$, and $10^8$, respectively. The agreement with the approximate analytic solutions (thin black lines) is quite good, especially for higher values of $\tau_0$. At lower optical depths the underlying assumptions used to derive Equation~(\ref{eq:neufeld}) break down, therefore the true measure of this test is the limiting behavior as $\tau_0$ tends to infinity. Bottom panel: Face-on line of sight radial surface brightness profiles for the same slabs as calculated by Equation~(\ref{eq:SB}) of Section~\ref{sub:surface_brightness_construction}. The scaling is such that the highest optical depth system is unity at the centre, i.e. $\text{SB}_0 \equiv \text{SB}(r=0)$. The dashed line is a rough analytic fit to guide the eye.
    }
    \label{fig:neufeld}
  \end{figure}

  \subsection{Static test -- the Neufeld profile}
    The angular averaged intensity~$J(\tau_0, a, x)$ for a static one-dimensional uniform slab was derived by~\citet{Harrington} and \citet{Neufeld}. Photons are injected at line centre at the origin and continuously scatter until escape occurs at a `centre-to-edge' optical depth of $\tau_0$. When the bulk motion of the gas is set to zero, $\mathbf{u}_{\rm bulk} = \mathbf{0}$, and the recoil effect is ignored, $g = 0$, the emergent spectra are double-peaked and symmetric around the line-central frequency~$x=0$. The intensity at the surface is given by
    \begin{equation} \label{eq:neufeld}
      J(\tau_0, a, x) = \frac{1}{4\sqrt{6 \pi}} \frac{x^2}{a \tau_0} \text{sech} \left( \sqrt{\frac{\pi^3}{54}} \frac{x^3}{a \tau_0} \right) \, ,
    \end{equation}
    which has been normalized to $1/4 \pi$, reflecting an integration over solid angle. The peaks are located at $x_{\rm p} = \pm 1.06642 \, (a \tau_0)^{1/3}$, which is derived by solving the equation $\partial J / \partial x = 0$, or equivalently $\bar{x} \tanh \bar{x} = \frac{2}{3}$ with $\bar{x} = \sqrt{\pi^3/54} \, x^3/a \tau_0$. Therefore, the peak heights correspond to $10^3 \, J(\tau_0, a, x_{\rm p}) = 45.074 \, (a \tau_0)^{-1/3}$.

    As can be seen in Fig.~\ref{fig:neufeld} the agreement \hlold{between numerical and analytic solutions} is quite good for high optical depths. \hlold{For example, at an optical depth of $\tau_0 = 10^8$ when comparing the simulated profile (red line) with the approximate analytic solution (thin black line) there is a maximum difference of $\sim 5$ per cent.} The Neufeld approximation of Equation~(\ref{eq:neufeld}) fails to capture the correct spectra at lower optical depths because core scatterings also contribute to the spatial diffusion of Ly$\alpha$ photons. \hlold{Therefore, the true measure of this test is the limiting behavior as $\tau_0$ tends to infinity. The bottom panel of Fig.}~\ref{fig:neufeld} \hlold{demonstrates the face-on radial surface brightness profiles for the same slabs as calculated by Equation}~(\ref{eq:SB}) \hlold{of Section}~\ref{sub:surface_brightness_construction}. \hlold{The scaling is such that the highest optical depth system is unity at the centre, i.e. $\text{SB}_0 \equiv \text{SB}(r=0)$. The dashed line is a rough analytic fit to guide the eye and is given in the figure. The shaded regions represent one standard deviation of uncertainty on $\text{SB}$ due to radial binning.} \hlold{Finally,} this test has also been performed at other temperatures with similar results, \hlold{cf. Appendix}~\ref{appendix:temperature_tests}. \hlold{Such} simulations of Ly$\alpha$ scatterings through optically thick slabs help measure the effectiveness of the acceleration schemes, \hlold{cf. Appendix}~\ref{appendix:x_crit}.

    \begin{figure}
      \centering
      \includegraphics[width=\columnwidth,trim=0in 0.1in 0in 0in,clip=true]{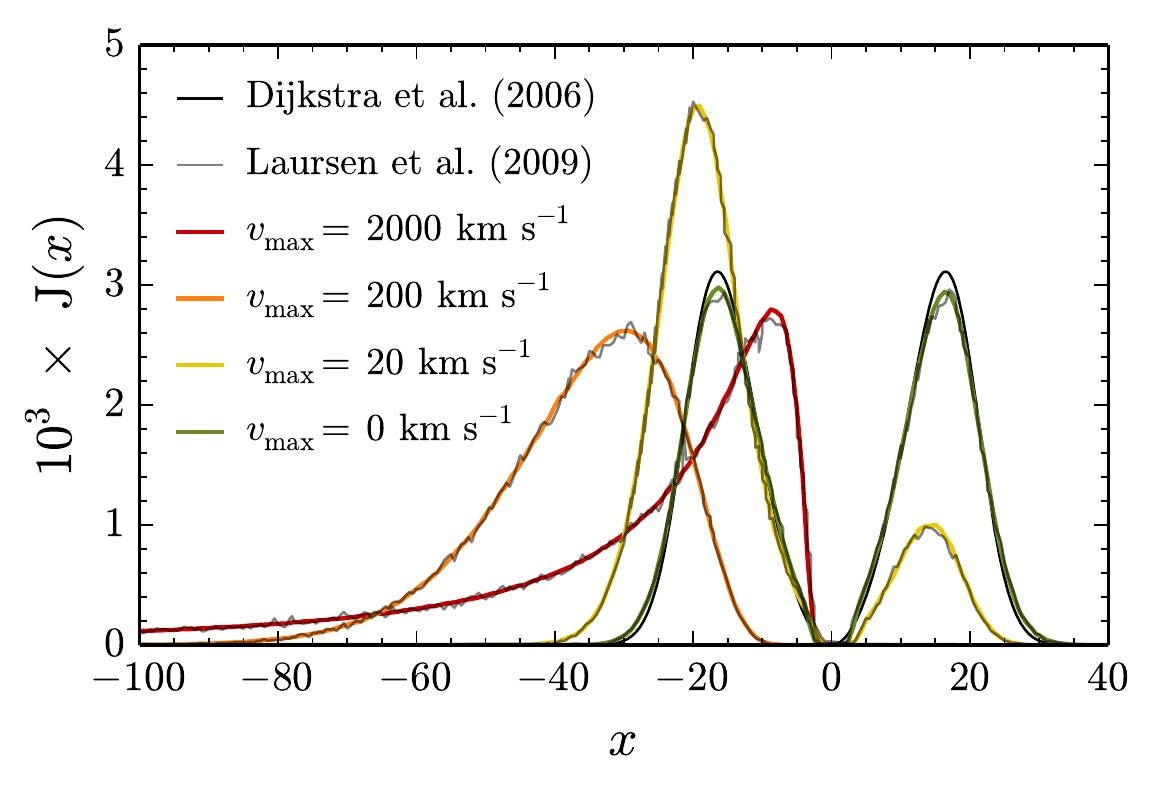}
      \includegraphics[width=\columnwidth,trim=.1in 0.04in 0in 0in,clip=true]{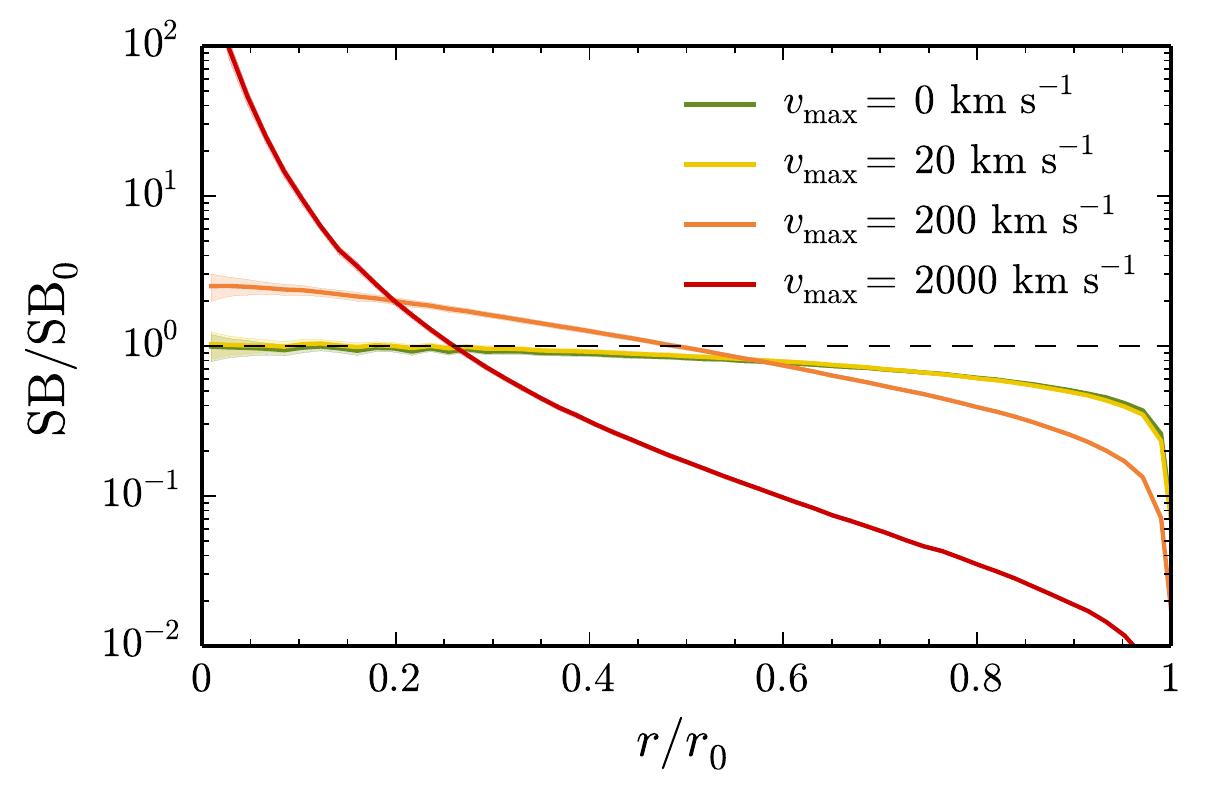}
      \vspace{-.4cm}
      \caption{Top panel: The dynamic test case gives the angular averaged intensity $J(v_{\rm max}, x)$ of an isothermal ($T = 10^4$~K), homogeneous sphere of column density $N_{\rm \HI} = 2 \times 10^{20}~\text{cm}^{-2}$ experiencing isotropic outflow. The Hubble-like expansion is parametrized by the maximum velocity $v_{\rm max}$ at the edge of the sphere. The static case agrees with the analytic solution of \citet{Dijkstra:2005ny} while the colored curves reproduce to high precision the same cases presented by \citet{Laursen:2008aq}. The velocity field applies an overall redshift to the profile, suppressing the blue peak until it disappears entirely. At first the red peak is pushed further from line centre, however, past a critical $v_{\rm max}$ the peak approaches the centre again because the velocity gradient facilitates escape from Doppler shifting. Bottom panel: The (averaged) line of sight radial surface brightness profile for each sphere calculated according to Equation~(\ref{eq:SB}) of Section~\ref{sub:surface_brightness_construction}. The scaling is such that the static sphere is unity at the centre, i.e. $\text{SB}_0 \equiv \text{SB}(r=0)$. The dashed line is a reference for a flat profile to guide the eye. Only the largest velocity gradients substantially alter the profile so that the apparent image is much more concentrated toward the centre.}
      \label{fig:velocity}
    \end{figure}

  \subsection{Dynamic test -- isotropic expansion of a uniform sphere}
    To test the code for the case of a nonzero gas bulk velocity we compare our results to previous simulations with isotropic expansion of a uniform sphere -- see \citet{Zheng:2002qc,Dijkstra:2005ny,Verhamme:2006tc,Tasitsiomi:2006,Semelin:2007,Laursen:2008aq,Yajima:2012}. In general, velocity gradients affect Ly$\alpha$ escape because photons are Doppler shifted out of line centre thereby reducing the effective optical depth. In general, even velocity fields on the order of the thermal velocity in photoionized gas ($v_{\rm th} \sim 10~\text{km~s}^{-1}$) can change the emergent spectrum of Ly$\alpha$ photons.

    Following \citet{Laursen:2008aq} we consider the intensity of an isothermal ($T = 10^4$~K), homogeneous sphere of column density $N_{\rm \HI} = 2 \times 10^{20}~\text{cm}^{-2}$ experiencing isotropic expansion. The Hubble relation gives the velocity $\mathbf{v_{\rm bulk}}(\mathbf{r})$ of the gas at the position $\mathbf{r}$ from the centre of a sphere with radius~$R$:
    \begin{equation}
      \mathbf{v_{\rm bulk}}(\mathbf{r}) = \mathcal{H} \mathbf{r} = \frac{v_{\rm max}}{R} \mathbf{r} \,
    \end{equation}
    where the Hubble-like parameter $\mathcal{H}$ sets the maximal velocity $v_{\rm max}$ at the edge of the sphere, i.e. $v_{\rm max} \equiv v_{\rm bulk}(R)$. Figure~\ref{fig:velocity} shows the result of this test which demonstrates excellent agreement with \citet{Laursen:2008aq}. The static case ($v_{\rm max} = 0$) also agrees with the analytical solution of \citet{Dijkstra:2005ny} for a static, optically thick spherical ``slab''. This test is also similar to that of \citet{Loeb:1999er} who calculated the zero-temperature spectrum of a Ly$\alpha$ source embedded in a neutral, homogeneous IGM undergoing Hubble expansion. Whereas a static solution produces two distinct peaks for blue and red modes of escape, the blue photons are continuously redshifted back to the core and the red mode becomes the only means of escape. This inevitably leads to free streaming on cosmological~($\gtrsim$Mpc) scales. Unfortunately, there is no analytical solution for a medium possessing both thermal and bulk motions. Figure~\ref{fig:velocity} illustrates the effect of increasing $v_{\rm max}$, which acts to suppress the blue peak until it disappears entirely by $v_{\rm max} \sim 200~\text{km~s}^{-1}$. At first the red peak is pushed further from line centre, however, past a critical $v_{\rm max}$ the peak approaches the centre again because the velocity gradient facilitates escape.
    \hlold{The bottom panel of Fig.}~\ref{fig:velocity} \hlold{shows the (averaged) line of sight radial surface brightness profile for each sphere. The scaling is such that the static sphere is unity at the centre, i.e. $\text{SB}_0 \equiv \text{SB}(r=0)$. The dashed line is a reference for a flat profile to guide the eye. Only the largest velocity gradients substantially alter the profile so that the apparent image is much more concentrated toward the centre.}

\section{First galaxy models}
  \label{sec:first_galaxy_models}
  Galaxy formation is a highly complex process; however, studying the formation and radiative transport in the first, comparatively simple, systems provides an ideal laboratory for the physics involved \citep{Bromm:2011cw}. We here employ two complementary methodologies to represent the structure and dynamics of a first galaxy for $\colt$ -- first, in Section~\ref{sec:toy} we construct idealized analytic models of these galaxies, and second, in Section~\ref{sec:sim} we extract a virialized halo from an \textit{ab initio} cosmological simulation for post-processing. These first galaxy models will then be the input for the radiative transfer calculations discussed in Section~\ref{sec:results}. Our idealized models allow us to explore the basic physics by directly adjusting parameters in order-of-magnitude fashion. Our cutout from the cosmological simulation, on the other hand, provides us with one representative example of a realistic first galaxy. We can thus gauge the validity of our exploratory toy models.

  \subsection{Idealized models}
    \label{sec:toy}
    We now explore idealized models for the first galaxies at redshift $z \sim 10$ in preparation for extractions from a cosmological simulation. The number density of hydrogen nuclei~$n_{\rm H}$ is built up from the following assumptions: 
    First, we require spherical symmetry so that $n_{\rm H} = n_{\rm H}(r)$. Second, we adopt a power law density profile within the galaxy, i.e. $n_{\rm H} \propto r^{-\beta}$ out to the edge of the galaxy~$r_{\rm edge}$ defined as the point where $n_{\rm H}$ equals the background IGM, $n_{\rm H, IGM}$. An isothermal law with $\beta = 2$ provides a good description of a virialized system~\citep[e.g.][]{BT}. Finally, we prefer a non-cuspy `core' in the centre of the galaxy, meaning the density profile flattens off to a constant density~$n_{\rm H, 0}$ within a core radius of~$r_{\rm core} \approx 10$~pc. This is inspired by observations of low surface brightness galaxies that suggest a softening in the centre~\citep{Burkert:1995,deBlok:2001,Kormendy:2009}. In summary, the model is piecewise in the radial coordinate~$r$ according to:
    \begin{equation} \label{eq:nHprofile}
      n_{\rm H}(r) =
        \begin{cases}
          n_{\rm H, 0}     & \text{for} \quad r \leq r_{\rm core} = 10~{\rm pc} \\
          n_{\rm H, 0} \left( \displaystyle \frac{r}{r_{\rm core}} \right)^{-2}  & \text{for} \quad r_{\rm core} < r < r_{\rm edge} \\
          n_{\rm H,IGM} & \text{for} \quad r \geq r_{\rm edge}
        \end{cases} \, .
    \end{equation}

    Here $n_{\rm H, IGM}$ is the background atomic hydrogen number density derived from a $\Lambda$CDM model with cosmological parameters taken from the \citet{Planck2013}, and incorporating constraints from the \textit{Wilkinson Microwave Anisotropy Probe} \citep[WMAP;][]{WMAP9}, the \textit{South Pole Telescope} \citep[SPT;][]{Keisler:2011aw}, and the \textit{Atacama Cosmology Telescope} \citep[ACT;][]{Das:2014}. Specifically, the Hubble constant is taken to be $H_0 = 67.8$~km~s$^{-1}$~Mpc$^{-1}$ while the fractional energy contributions of baryons, matter, and dark energy are $\Omega_{\rm b} = 0.0485$, $\Omega_{\rm m} = 0.307$, and $\Omega_\Lambda = 0.693$, respectively. Therefore, $n_{\rm H, IGM}$ is $\rho_{\rm cr,0} X \Omega_{\rm b} (1+ z)^3 / m_{\rm H}$, or
    \begin{equation}
      n_{\rm H, IGM} \approx \hlmold{2 \times 10^{-4}~\text{cm}^{-3} \left( \frac{1+z}{10} \right)^3} \, , 
    \end{equation}
    where $X \approx 0.75$ is the mass fraction of hydrogen and $\rho_{\rm cr,0} \equiv 3 H_0^2 / (8 \pi G)$ is the critical energy density at present.

    Furthermore, the edge of the galaxy is given by solving the equation $n_{\rm H}(r_{\rm edge}) = n_{\rm H, IGM}$, which yields a radius of $r_{\rm edge} = r_{\rm core} ( n_{\rm H, 0} / n_{\rm H, IGM} )^{1/2}$. The density parameter~$n_{\rm H, 0}$ is then found by normalizing the overall mass of hydrogen in the galaxy~$M_{\rm H, tot}$ to some value, e.g. $\sim 10^6-10^8~\Msun$ for an atomic cooling halo. The total mass is given by the integral
    \begin{align} \label{eq:MHtot}
      M_{\rm H, tot} &= 4 \pi m_{\rm H} \int_0^{r_{\rm edge}} n_{\rm H}(r) \, r^2 dr \notag \\
                     &= 4 \pi m_{\rm H} n_{\rm H, 0} r_{\rm core}^3 \left( \sqrt{ \frac{n_{\rm H, 0}}{n_{\rm H, IGM}} } - \frac{2}{3} \right) \, .
    \end{align}
    Equation~(\ref{eq:MHtot}) is a cubic polynomial in $n_{\rm H, 0}$ whose solution is not particularly insightful. However, we may expand about large masses~$M_{\rm H, tot}$ to consolidate the leading order terms. These terms provide a relative accuracy in $n_{\rm H, 0}$ of better than 1~part per billion for masses larger than $10^6~\Msun$. For the parameters chosen above this implies a central density of
    \begin{equation} \label{eq:nH0}
      n_{\rm H, 0} \approx n_{\rm H, IGM}~\left( \chi^2 + \frac{4}{9} \chi + \frac{4}{27} \right) \, ,
    \end{equation}
    where we have introduced the dimensionless parameter:
    \begin{align} \label{eq:chi}
      \chi &\equiv \frac{n_{\rm H, IGM}^{-1/3}}{r_{\rm core}} \left( \frac{M_{\rm H, tot}}{4 \pi m_{\rm H}} \right)^{1/3} \notag \\
           &= \hlmold{560~\left(\frac{M_{\rm H, tot}}{10^7~\Msun} \right)^{1/3} \left( \frac{r_{\rm core}}{10~\text{pc}} \right)^{-1} \left( \frac{ 1 + z }{ 10 } \right)^{-1}} \, .
    \end{align}
    This term may also be found by dropping the $\frac{2}{3}$ term in Equation~(\ref{eq:MHtot}). The $\chi$ parameter relates the (leading order) ratio of core and background densities via $n_{\rm H, 0} \approx \chi^2 n_{\rm H, IGM}$ and the ratio of core and edge radii via $r_{\rm edge} \approx \chi r_{\rm core}$. With this central density the optical depth to Ly$\alpha$ scattering of the core region is at least $\tau_{\rm core} \sim n_{\rm H, 0} \sigma_0 r_{\rm core} \sim 10^8$, a value that would increase with a larger core radius, a lower temperature, or a more massive system.

    We are primarily interested in the mass and density of hydrogen, therefore we decided to use $M_{\rm H, tot}$ in the comparison. The total or virial mass of the galaxy is larger for two main reasons: $(i)$ The mass fraction of hydrogen to baryons is less than unity, i.e. $X \approx 0.75$. $(ii)$ The contribution of baryonic mass is considerably less than the contribution of dark matter, which is usually more extreme in smaller galaxies. In fact, if a substantial amount of gas is lost through ram pressure stripping or supernova blowout, for example, the baryonic mass may be significantly below the cosmological baryon fraction of $\Omega_{\rm b} / \Omega_{\rm m} \sim 16$ per cent \citep*{Allen:2002sr}. At any rate we can reasonably relate the virial and total hydrogen masses by $M_{\rm vir} \sim 10 \, M_{\rm H, tot}$.

    \subsubsection{Structure and evolution of the ionized region}
      The model above did not include an ionized region around a central star cluster. This is important because Ly$\alpha$ photons can easily escape such regions. Furthermore, a Str\"{o}mgren analysis suggests that the ionized region may be on the order of $r_{\rm core}$ for a reasonable set of parameters:
      \begin{equation} \label{eq:stromgren}
        R_{\rm S} = \left( \frac{3}{4 \pi} \frac{\dot{N}_{\rm ion}}{n_0^2 \alpha_{\textsc B}} \right)^{1/3}
                  \approx 15~T_4^{1/4}~\dot{N}_{\rm ion, 51}^{1/3}~n_{100}^{-2/3}~\text{pc} \, ,
      \end{equation}
      where we have defined normalized values for ionizing photon rate $\dot{N}_{\rm ion, 51} \equiv \dot{N}_{\rm ion}/(10^{51}~\text{s}^{-1})$ and density $n_{100} \equiv n_0/(100~\text{cm}^{-3})$. The normalization used for $\dot{N}_{\rm ion}$ is a plausible guess at the rate expected from the starbursts residing in the first galaxies (see Section~\ref{sec:central_starburst} for further discussion). The expression was simplified by approximating the total Case~B recombination rate by $\alpha_{\textsc B} \approx 2.5 \times 10^{-13}~T_4^{-3/4}$~cm$^3$~s$^{-1}$ \citep{Osterbrock:2006}. In order to understand how the ionized region affects Ly$\alpha$ transfer we need to consider how the full three-dimensional ionization structure evolves in time. The Str\"{o}mgren radius is a good approximation early on when the radiation is bottled up and the ionization rate balances the recombination rate, i.e. $\dot{N}_{\rm ion} \approx \dot{N}_{\rm rec}$. However, as the ionization front moves out, the \HII\ regions blow pockets through which Ly$\alpha$ photons may escape. Therefore, we distinguish between \textit{early} ionization scenarios when the \HII\ region is ultra-compact as in Equation~(\ref{eq:stromgren}) and \textit{late} scenarios when the \HII\ starburst regions have overtaken the entire halo. These cases bracket the entire evolution.

\begin{figure}
      \centering
      \includegraphics[width=.825\columnwidth]{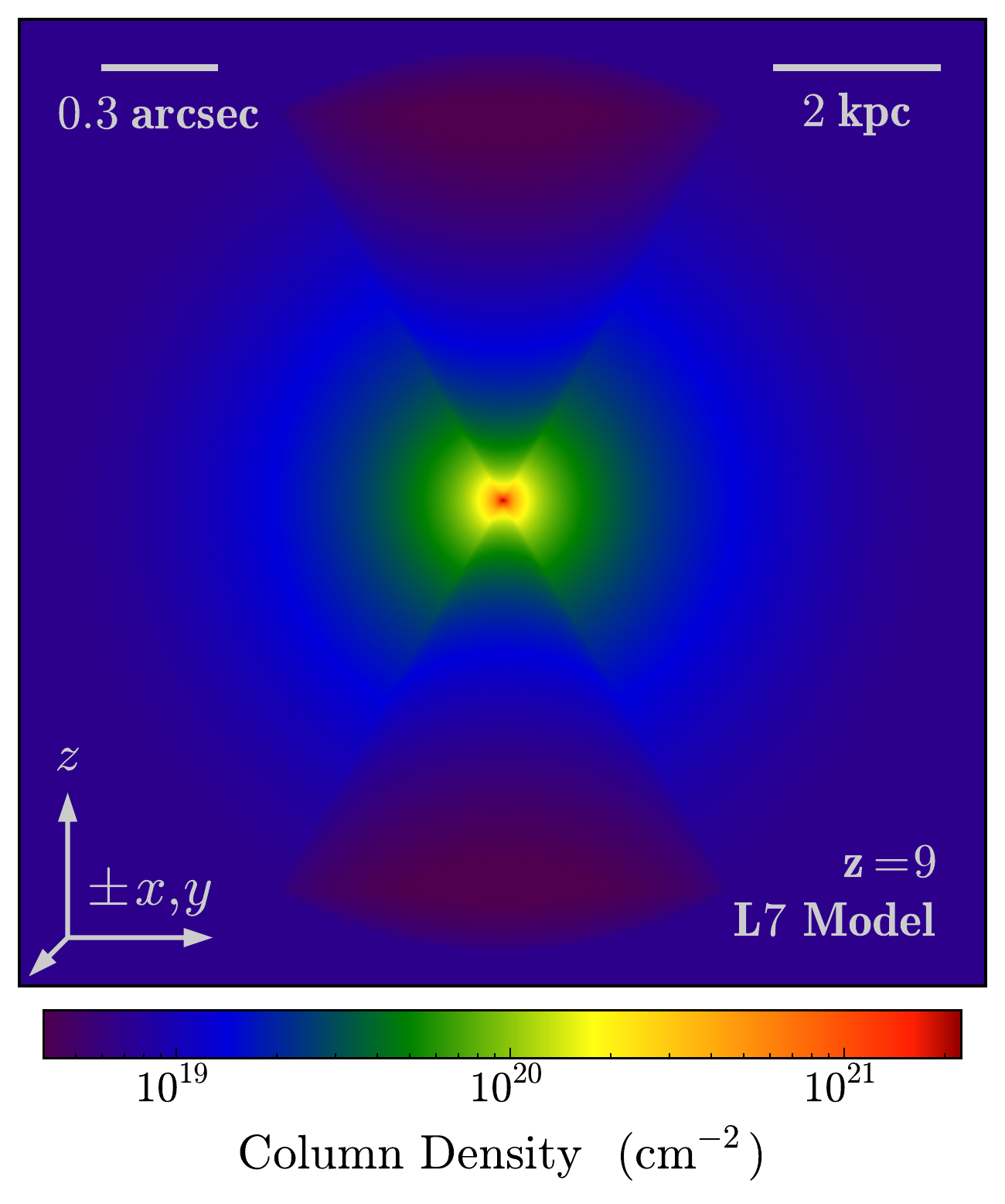}
      \vspace{-.2cm}
      \caption{Neutral hydrogen column density~$N_{\rm \HI}$ for a ``Late'' model at a source redshift of $z = 9$ with $M_{\rm H, tot} = 10^7~\Msun$. The spatial scale, in physical kiloparsecs and arcseconds, are provided. The edge-on view of the halo demonstrates the \hlold{butterfly-shaped} 
      ionization structure of the models. \hlold{For comparison with Eq.}~(\ref{eq:t_late}) \hlold{the corresponding ionizing photon rate is $\dot{N}_{\rm ion} \approx 10^{53}~\text{s}^{-1}$, 
      giving a late time lower limit of $t_{\rm late} \gtrsim 10^3$~yr.}
      }
      \label{fig:idealized_density}
    \end{figure}

      The late stages of the ionization structure must be anisotropic because the cosmological filaments guide the ionized bubbles into \hlold{a butterfly-shaped morphology} around the centre. For simplicity we model this as a bipolar cavity. Specifically, in addition to the Str\"{o}mgren sphere, our \textit{late} scenario also has a biconical region out to $r_{\rm edge}$. The size of the \HII\ region is determined by the total number of ionizing photons, i.e. $N_{\rm ion} \hlmold{>} N_{\rm H}$. If the photons are produced at a constant rate and the volume of the cone is described by the opening angle~$\theta_{\rm open}$ from the axis of symmetry then the time to evacuate the cavity is \hlold{at least} of order 
      \begin{align} \label{eq:t_late}
        t_{\rm late} &\hlmold{\gtrsim} \frac{N_{\rm H}}{\dot{N}_{\rm ion}}
                      \approx \frac{\theta_{\rm open}^2 M_{\rm H, tot}}{2 m_{\rm H} \dot{N}_{\rm ion}} \notag \\
                     &\approx \hlmold{0.1~\text{Myr}~\dot{N}_{\rm ion, 51}^{-1} \left(\frac{M_{\rm H, tot}}{10^7~\Msun}\right)} \left(\frac{\theta_{\rm open}}{30\degr}\right)^2 \, .
      \end{align}
      Therefore, the \textit{late} scenario corresponds to a time when the biconic region out to $r_{\rm edge}$ is fully ionized -- see Fig.~\ref{fig:idealized_density} for an edge-on view of the column density through one such halo.

    \subsubsection{Radiation pressure driven wind}
      The central starburst provides a feedback mechanism on the host galaxy. 
      In order to motivate a spherically symmetric velocity law we consider the relative strength of the gravitational force~$F_{\rm grav}$ to the radiation force~$F_{\rm rad}$. \hlold{As a first estimate, if we attribute the radiation pressure to Thompson scattering in the single-scattering limit, then because of the $1/r^2$ scaling of both forces,} their ratio is 
      of order
      \begin{equation} \label{eq:rad_grav}
        \frac{F_{\rm rad}}{F_{\rm grav}} \approx \frac{L_\ast \sigma_{\rm T}}{4 \pi c m_{\rm H} G M_\ast}
                                         \approx 0.3~\left(\frac{\Upsilon}{10^4}\right)^{-1} \, ,
      \end{equation}
      where $L_\ast$ and $M_\ast$ are respectively the total luminosity and mass of the entire cluster. The normalization is chosen to correspond to the mass to light ratio for Pop~III stars, i.e. $\Upsilon \approx 10^4$ (see Section~\ref{sec:central_starburst}). For simplicity Equation~(\ref{eq:rad_grav}) assumes a point source. In reality, the ratio decreases radially from the centre of the gravitational potential as the dark matter halo mass dominates the gravitational force. \hlold{Furthermore, Equation}~(\ref{eq:rad_grav}) \hlold{is only applicable to ionized gas, whereas Ly$\alpha$ scattering occurs in neutral gas so other sources of pressure must also be present for our idealized law.} \hlold{Therefore, we consider a force multiplier in analogy to that of} \citet[CAK;][]{CAK} 
      originating from many resonant and optically thin lines. 
      The multiplier can amplify the scattering efficiency by many orders of magnitude. Thus, the luminosities considered here ($\gtrsim 10^7~\Lsun$) may be capable of sustaining radiation-driven winds \citep{Wise:2012a}. Indeed, the Ly$\alpha$ line alone may be responsible for much of the opacity these winds require \citep{Dijkstra:2008xz,Dijkstra:2009}. The CAK theory uses mass conservation and momentum balance to arrive at a `beta law' \hlold{with an exponent of $1/2$} for the radial velocity profile
      \begin{equation} \label{eq:wind}
        v(r) = v_\infty \left( 1 - \frac{R_\ast}{r} \right)^{1/2} \, ,
      \end{equation}
      where $R_\ast$ is the size of the source. The terminal velocity $v_\infty$ is typically larger than the escape velocity $v_{\rm esc}$ by a factor of a couple. Plausible values for our models are obtained by requiring that the mass loss rate $\dot{M} = 4 \pi r^2 \rho v$ be less than the maximally efficient mass loss for single scattering $\dot{M}_{\rm max} v_\infty = L_\ast/c$ for which the radiation momentum is imparted entirely to the gas. In the halo's isothermal region, i.e. $r_{\rm core} < r < r_{\rm edge}$, the velocity approaches $v_\infty$ and $\rho \propto r^{-2}$ so
      \begin{align}
        v_\infty &\lesssim \left( 4 \pi m_{\rm H} M_{\rm H, tot}^2 n_{\rm H,IGM} \right)^{-1/6} \sqrt{\frac{L_\ast}{c}} \notag \\
                 &\approx \hlmold{10~\text{km~s}^{-1}\left(\frac{M_{\rm H, tot}}{10^7~\Msun}\right)^{-1/3}\!\left(\frac{L_\ast}{10^7~\Lsun} \frac{10}{1+z}\right)^{1/2} .}
      \end{align}
      In principle, multiple scattering in the optically thick environment could boost $v_\infty$ by a factor of the square root of the optical depth, or at least an order of magnitude. However, for simplicity our wind models all assume the velocity profile of Equation~(\ref{eq:wind}) with $R_\ast = 1$~pc, the approximate size of the central star cluster, and $v_\infty = 10~\text{km~s}^{-1}$, a value comparable to the thermal velocity~$v_{\rm th}$. \hlold{We note that our value of $v_\infty$ is also close to the escape velocity of the system, $v_{\rm esc} \approx ( 2 G M_{\rm vir} / R_{\rm vir} )^{1/2}$ $\approx 13~\text{km~s}^{-1}~[ M_{\rm vir} / ( 10^7~\Msun ) ]^{1/3}$ $[ (1+z) / 10 ]^{1/2}$. For completeness, the original CAK force multiplier $M(t) = kt^{-\alpha}$ would need to be reevaluated for H and He lines of gas with primordial composition. It is beyond the scope of this work to do so here but for reference the model of} \citet{CAK} \hlold{derives values of $k \approx 1/30$ and $\alpha \approx 0.7$. The parameter $t$ is either the electron-scattering optical depth $\int_r^\infty \sigma_{\rm T} \rho dr$ or the Sobolev optical depth $\sigma_{\rm T} \rho v_{\rm th} |dv/dr|^{-1}$ in static or expanding media, respectively.} \hl{Therefore, our terminal velocity is similar in magnitude to the 
      CAK theory formula of $v_\infty = v_{\rm esc} \big[(1-F_{\rm rad}/F_{\rm grav}) \alpha / (1-\alpha)\big]^{1/2}.$}

      The density and velocity profiles considered here are not self-consistent in the dynamical sense. However, our simplified $\beta$-model, calibrated in the above fashion, should give us a rough window into the otherwise extremely complex physics of galactic winds \citep*[e.g.][]{Veilleux:2005}. This also avoids overconstructing models that are already quite idealized. \hlold{Finally, the radiation pressure driven winds considered above may require some dust and metals to be present in the ISM. However, we have not included dust in these primordial environments. Further study on how dust affects Ly$\alpha$ transfer in anisotropic models should be carried out and is left for future work. Still, various mechanisms such as scattering in clumpy media have been proposed to explain why the Ly$\alpha$ escape fraction is so significant in environments where core depletion seems unavoidable. In fact,} \citet{Hayes} \hlold{suggest that dust attenuation is unlikely to affect Ly$\alpha$ escape at $z \gtrsim 11$. For our purposes, we expect dust to lower the bolometric line flux by at most a factor of a few. For an example of coupling large scale cosmological simulations with dust and radiative transfer models designed to constrain IGM ionization and the dust distribution in the ISM of early galaxies see} \citet*{Dayal:2011} \hlold{and} \citet{Hutter:2014}.

    \begin{table}
      \caption{Classification scheme for the idealized galactic models.}
      \label{tab:models}
      \begin{tabular}{@{} c ccc  @{}}
      \hline
      \;\;Classification\; & \;Velocity\; & \;\HII\ Scenario\; & \;$M_{\rm H, tot}$~$\left[\Msun\right]$\;\; \\
      \hline
      \, SE\,[$5 - 8$] & Static & Early & $10^5 - 10^8$ \\
      \, SL\,[$5 - 8$] & Static & Late  & $10^5 - 10^8$ \\
         WE\,[$5 - 8$] &  Wind  & Early & $10^5 - 10^8$ \\
         WL\,[$5 - 8$] &  Wind  & Late  & $10^5 - 10^8$ \\
      \hline
      \end{tabular}
    \end{table}

    \begin{figure*}
      \centering
      \includegraphics[width=2.12\columnwidth]{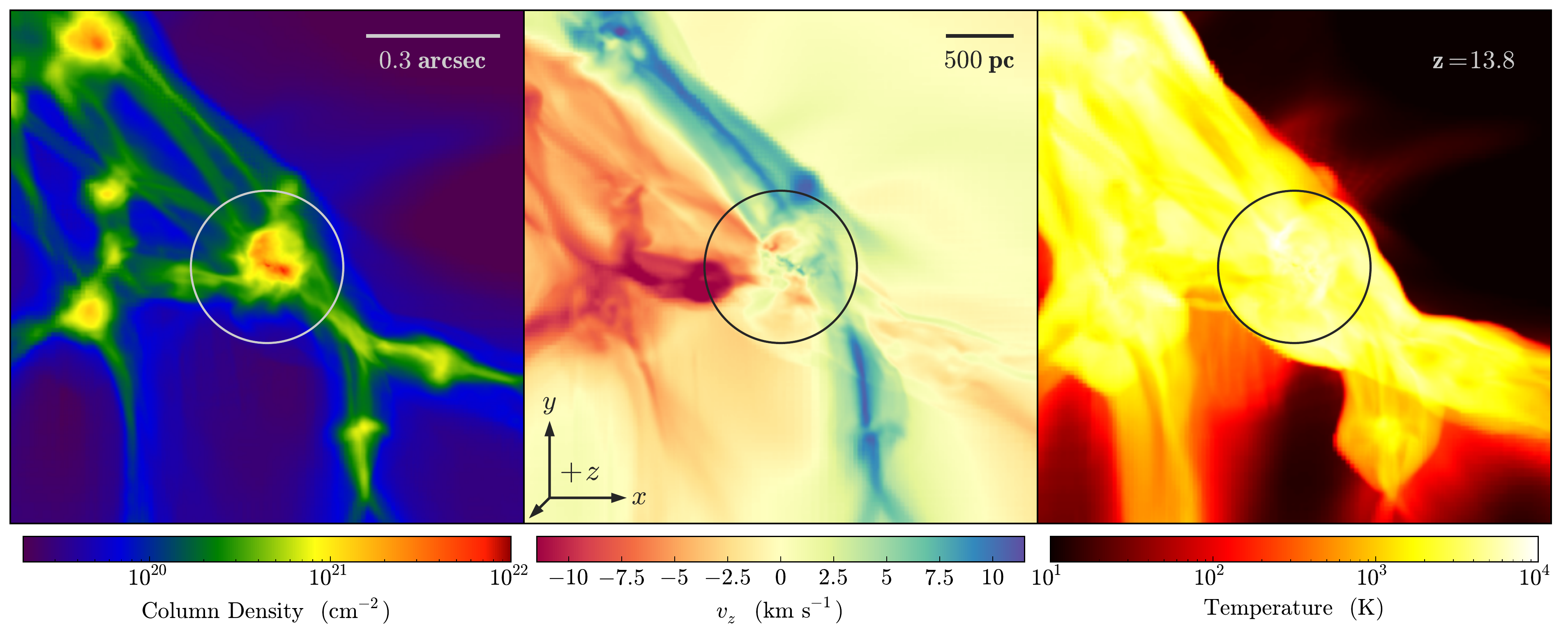}
      \vspace{-.2cm}
      \caption{Neutral hydrogen column density~$N_{\rm \HI}$ (left), line of sight velocity~$v_z$ (middle), and ambient gas temperature~$T$ (right) surrounding the atomic cooling halo of \citet{SafranekShrader:2012qn} at a source redshift of $z = 13.8$. The velocity and temperature projections are weighted by the gas density. The spatial scale, in physical parsecs and angular units of arcseconds, is provided along with circles to represent the halo's virial radius of $R_{\rm vir} \approx 600$~pc, corresponding to a virial mass of $M_{\rm vir} \approx 2 \times 10^7~\Msun$ or roughly $M_{\rm H, tot} \approx 2.5 \times 10^6~\Msun$ for comparison with our idealized models.}
      \label{fig:extractor_density}
    \end{figure*}

    \subsubsection{Model parameter space}
      The various galactic models we consider are based on the following physical quantities: $(i)$ The velocity structure as either static or with a radial wind. $(ii)$ The \HII\ structure as either an early or late ionization scenario. $(iii)$ The total mass of hydrogen in the halo~$M_{\rm H, tot}$ chosen as either $10^5$, $10^6$, $10^7$,~or~$10^8~\Msun$. For clarity we describe the model classifications summarized in Table~\ref{tab:models}:
      \begin{itemize}
        \item[{\bf S\,}] $\Rightarrow$~``Static''
          -- The bulk velocity of every cell is zero.
        \item[{\bf W}] $\Rightarrow$~``Wind''
          -- A radiation-driven wind assuming the velocity \indent\indent\indent\indent
             profile of Eq.\,(\ref{eq:wind}) with $R_\ast = 1$~pc and $v_\infty = 10~\text{km~s}^{-1}$. \indent\indent\indent\indent
             Note: For simplicity we chose the same $v_\infty$ for all models \indent\indent\indent\indent
             although it could very well be larger by a factor of a few.
        \item[{\bf E}] $\Rightarrow$~``Early''
          -- The density profile of Eq.\,(\ref{eq:nHprofile}) is modified to have \indent\indent\indent\indent
             zero neutral hydrogen density within a Str\"{o}mgren sphere \indent\indent\indent\indent
             of radius $R_{\rm S} = r_{\rm core}$ at the centre of the galaxy.
        \item[{\bf L}] $\Rightarrow$~``Late''
          -- The density profile of Eq.~(\ref{eq:nHprofile}) is also modified to \indent\indent\indent\indent
             have zero neutral hydrogen density within a bipolar cone \indent\indent\indent\indent
             of opening angle $\theta = 30\degr$ out to $r_{\rm edge}$.
        \item[{\bf 5--8}] $\Rightarrow~M_{\rm H, tot} = 10^5 - 10^8~\Msun$
          -- These models allow us to \indent\indent\indent\indent
             explore the effect of mass, or column density, on Ly$\alpha$ \indent\indent\indent\indent
             escape. $M_{\rm vir}$ is roughly an order of magnitude larger.
      \end{itemize}

      \noindent \hlold{For simplicity the temperature is set to a constant value of $T = 10^4$~K throughout the entire computational domain.} \hlold{We note that in some cases the residual {\HI} within such {\HII} regions may still be optically thick to the Ly$\alpha$ line centre, i.e. $a\tau_0 \gg 1$. However, as this is many orders of magnitude lower than the optical depth outside an ionized region we have assumed the neutral fraction is negligible in idealized models.}
      These models are intended to test the basic physics involved with Ly$\alpha$ transport under the conditions discussed above, and we consider more realistic conditions from a cosmological simulation in Section~\ref{sec:sim}.

    \subsubsection{Refinement criteria}
      The above models require accurate spatial discretization with a dynamic range of several orders of magnitude. Because we work primarily with cosmological simulations utilizing adaptive mesh refinement~(AMR) we also incorporate the AMR grid structure into these idealized models. This has the dual benefit of \textit{(i)} efficiently characterizing the field information even with ionization fronts or high density formations and \textit{(ii)} unifying the data structures and ray tracing algorithms for both idealized setups and extractions from hydrodynamical simulations. In order to map the analytic conditions onto an AMR grid we first construct a Cartesian grid with dimensions $\{x, y, z\} \in (-2 r_{\rm edge}, 2 r_{\rm edge})$. This choice for the box size is somewhat arbitrary but provides enough of a buffer from the IGM to redistribute any remnant Ly$\alpha$ core photons that might have escaped through the bipolar cavity. At this point we recursively refine the grid structure until the following criteria are all met: (\textit{i}) Density -- the cell dimensions must be smaller than the Jeans length~$\lambda_{\rm J}$ by a factor of $N_{\rm J}$, i.e. $\Delta \ell_{\rm cell} N_{\rm J} \leq \lambda_{\rm J}$. A choice of $N_{\rm J} = 64$ was implemented in the models. (\textit{ii}) Velocity gradient -- the cell must be smaller than the Sobolev length~$\lambda_{\rm S}$ by a factor of $N_{\rm S}$, i.e. $\Delta \ell_{\rm cell} N_{\rm S} \leq \lambda_{\rm S}$. A choice of $N_{\rm S} = 32$ was made to avoid unresolved Doppler shifting from continuous Sobolev escape. (\textit{iii}) Geometric -- refine based on whether a cell is within a specified volume or if the boundary of a geometric shape passes through the cell. For example, the shape of the edge of the galaxy was resolved by requiring that cells containing points where $r = r_{\rm edge}$ satisfy $\Delta \ell_{\rm cell} N_{\rm G} \leq r_{\rm edge}$. The chosen value was $N_{\rm G} = 64$. Similarly for $r_{\rm core}$. The refinement criteria for the boundary of the ionized cone in the ``Late'' scenario was chosen to be $512 \Delta \ell_{\rm cell} \leq r_{\rm edge}$.

    \begin{table*}
      \caption{The Ly$\alpha$ luminosity is related to the Pop~III star formation efficiency, $\eta_\ast \equiv M_\ast / M_{\rm gas}$, according to Equation~(\ref{eq:L_Lya}), where $M_\ast$ is the total mass in Pop~III stars and $M_{\rm gas}$ is the total baryonic mass of the halo. Specifically, a primordial gas satisfies $M_{\rm H, tot} \approx 0.75 \, M_{\rm gas}$. The ionizing photon rate for a Pop~III cluster is roughly $\dot{N}_{\rm ion} \sim 10^{48}~(M_\ast/\Msun)~\text{s}^{-1}$. If metal enriched Pop~II stars were present then $\dot{N}_{\rm ion}$, and therefore $L_{\rm Ly\alpha}$, would be an order of magnitude lower. For our purposes we assume the escape fraction of ionizing photons, $f^{\rm ion}_{\rm esc}$, is negligible. The virial mass~$M_{\rm vir}$ is also given for reference.}
      \label{tab:central_starburst}
      \begin{tabular}{@{} c cccccc @{}}
        \hline
        Model & $M_{\rm vir}$ & $M_{\rm H, tot}$ & $\eta_\ast$ & $M_\ast$ & $\dot{N}_{\rm ion}$ & $L_{\rm Ly\alpha}$ \\
        \hline
        \vspace{.05cm}
        \;Idealized\; & $10^{[6-9]}~\Msun$ & $10^{[5-8]}~\Msun$ & $0.01$ & $1.3 \times 10^{[3-6]}~\Msun$ & $1.3 \times 10^{[51-54]}~\text{s}^{-1}$ & $3.9 \times 10^{[6-9]}~\Lsun$ \\
            SS12      & $2.13 \times 10^7~\Msun$ & $2.65 \times 10^6~\Msun$ & $0.01$ & $3.5 \times 10^4~\Msun$ & $3.5 \times 10^{52}~\text{s}^{-1}$ & $10^8~\Lsun$ \\
        \hline
      \end{tabular}
    \end{table*}

  \subsection{First atomic cooling haloes}
    \label{sec:sim}
    The analytic models considered above inform us about key aspects of Ly$\alpha$ line transfer in the first galaxies, testing our methods and sensitivity. However, we can push these questions further by employing \textit{ab initio} cosmological simulations as post-processing initial conditions for $\colt$. We examine one cosmological simulation in this paper, and analyze additional cases in a follow-up study. In this section we summarize the simulation described by \citet[][hereafter SS12]{SafranekShrader:2012qn}, in preparation for the radiative transfer calculations of Section~\ref{sub:realistic_first_galaxy}. The simulation of SS12 uses the hydrodynamical/$N$-body code \textsc{flash} \citep{Fryxell:2000}, \hlold{version 3.3,} to evolve cosmological initial conditions through the nonlinear collapse of structure formation. \hlold{The cosmological initial conditions were generated with \textsc{mpgrafic}} \citep{Prunet:2008}, \hlold{which provides multi-scale Gaussian random fields at $z = 146$ in a $1$~Mpc$^3$ comoving volume. A hierarchical zoom-in procedure with three levels of dark matter refinement was employed to obtain a maximum effective resolution of $512^3$ and an effective dark matter particle mass of $230~\Msun$ in the target halo. The baryonic refinement strategy is based on (i) a Lagrangian refinement factor that implies a cell mass of $\approx 0.1~\Msun$ at the highest refinement level and (ii) a criterion that the Jeans length be resolved by at least 12 grid cells. Further details regarding the initial conditions, hydrodynamics, refinement strategies, chemistry, gas cooling, sink particles, H$_2$-dissociating radiation, and various other schemes may be found in SS12.} The first galaxy model we study with $\colt$ is an extraction of a virialized halo at redshift $z = 13.8$ with $R_{\rm vir} \approx 600$~pc and $M_{\rm vir} = 2.1 \times 10^7~\Msun$, corresponding to $M_{\rm H, tot} = 2.6 \times 10^6~\Msun$ for comparison with our idealized models. \hlold{With 22 levels of refinement and $8^3$ cells per block the effective spatial resolution of the halo is 0.004~pc, or 830~AU, considerably higher than previous Ly$\alpha$ radiative transfer simulations.}

    SS12 examine the formation and fragmentation conditions for a star cluster inside a cosmological atomic cooling halo, i.e. a system with virial temperature~$T_{\rm vir} \gtrsim 10^4$~K such that Ly$\alpha$ line cooling is enabled. These systems are important for the first galaxies because Ly$\alpha$ line cooling is much more efficient than cooling by molecular hydrogen or metal lines, and catalyzes the star formation process. Figure~\ref{fig:extractor_density} shows the neutral hydrogen column density~$N_{\rm \HI}$, line of sight velocity~$v_z$, and ambient gas temperature~$T$ of the cutout region. The filamentary, irregular nature of the gas is apparent, stressing the need for more realistic conditions than analytic models may allow. For simplicity we assume the stellar population of this galaxy has had no significant \hlold{impact} on its galactic surroundings, thereby isolating the radiative transfer effects as originating from a Ly$\alpha$ source within a cosmological environment harboring gas accretion inflow. \hlold{Indeed, the only feedback mechanism in the simulation is an external Lyman-Werner radiation field incident from the six faces of the computational domain. However, as the Ly$\alpha$ observability may be enhanced by feedback at later times, e.g. galactic outflows and ionization,} we will analyze additional haloes for which feedback is accounted for with greater sophistication in a follow-up paper. \hlold{In Section}~\ref{subsub:IGM_Trans} \hlold{we discuss the effect of the IGM on Ly$\alpha$ observations; 
    any reddening of the intrinsic line profile induces less attenuation.} Finally, we also test cutouts of varying sizes within the $(1$~Mpc$)^3$ comoving volume, or $(67.5$~kpc$)^3$ physical volume in Appendix~\ref{appendix:box_sizes}. We find minimal differences between the emergent flux densities~$f_\lambda$, especially for the larger cutouts.

  \subsection{Properties of the central starburst}
    \label{sec:central_starburst}
    We now clarify the assumptions regarding the central starburst luminosity and stellar properties. The initial mass function~(IMF) of a cluster depends on the metallicity of the population, where more massive Pop~III stars are distributed with a top-heavy IMF and Pop~II stars display a normal IMF biased toward low-mass stars. For simplicity, we assume a Pop~III starburst with a top-heavy IMF, although a significant fraction of the first galaxies may typically already be populated by metal enriched Pop~II stars, or a mixture of populations \citep{Johnson:2008,Greif:2010,Ritter:2012,Wise:2012b,Muratov:2013,Ritter:2014}. The Ly$\alpha$ luminosity,~$L_{\rm Ly\alpha}$, depends on the Pop~III star formation efficiency, $\eta_\ast \equiv M_\ast / M_{\rm gas}$, where $M_\ast$ is the mass in Pop~III stars and $M_{\rm gas}$ is the total baryonic mass in the host halo. For ease of comparison we assume a fixed star formation efficiency of $\eta_\ast = 0.01$ for both the idealized models and the cutout simulation of SS12. The assumption that the cluster consists of Pop~III stars with a top-heavy IMF sets the ionizing photon rate to $\dot{N}_{\rm ion} \sim 10^{48}~(M_\ast/\Msun)~\text{s}^{-1}$. However, if metal enriched Pop~II stars were present the rate would be an order of magnitude lower \citep*{Bromm:2001,Schaerer:2002}. The luminosity in Lyman-$\alpha$ is \citep{Dijkstra:2014}
    \begin{align} \label{eq:L_Lya}
      L_{\rm Ly\alpha} &= 0.68~h \nu_0 \left( 1 - f^{\rm ion}_{\rm esc} \right) \dot{N}_{\rm ion} \notag \\
                       &\approx 5 \times 10^8~\Lsun~\left( \frac{\eta_\ast}{0.01} \right)~\left( \frac{M_{\rm vir}}{10^8~\Msun} \right) \, ,
    \end{align}
    where $h \nu_0 = 10.2$~eV and $f^{\rm ion}_{\rm esc}$ is the fraction of ionizing photons escaping the central starburst region, which we assume to be zero. Because Equation~(\ref{eq:L_Lya}) scales with $\dot{N}_{\rm ion}$, if one assumes a Pop~II IMF the radiative transfer calculations of Section~\ref{sec:results}, including flux and surface brightness, scale down by roughly a factor of ten compared to the Pop~III case if $\eta_\ast$ would remain the same as before. Again, we emphasize that the Ly$\alpha$ flux and intensity profiles throughout this paper are scale free because the radiative transfer is decoupled from the hydrodynamics. The choice of a fixed star formation efficiency serves as the primary normalization for our profiles and fundamentally captures the basic idea that source luminosity should depend on halo mass. The one per cent star formation efficiency is admittedly an optimistic value that represents a likely upper limit on the prospects of detecting Ly$\alpha$ photons from Pop~III sources. A comprehensive list of the properties of the central starburst is given in Table~\ref{tab:central_starburst}.

    \hlold{In the models considered above, emission from Ly$\alpha$ line cooling is insignificant compared to the central luminosity of the starburst. This is primarily because the electron abundance is quite low at this stage of galaxy formation. To justify this we calculate the Ly$\alpha$ emissivity due to collisional excitation according to $\psi_{\rm H} = 7.5 \times 10^{-19}$~erg~cm$^3$~s$^{-1}~(1 + T_5^{1/2})^{-1} e^{-118348/T} n_e n_{\rm H}$} \citep[see equation~15a of][]{Cen:1992}, \hlold{where $T_5 \equiv T / (10^5~\text{K})$, and integrate over volume to obtain the Ly$\alpha$ cooling luminosity. Applying this method to each cell of the entire SS12 simulation results in a total computed luminosity of $\sim 250~\Lsun$, 
    several orders of magnitude below the stellar luminosity estimated by Equation}~(\ref{eq:L_Lya}). \hlold{However, this should be thought of as a lower bound on the cooling luminosity as additional radiative feedback would lead to greater ionization. To obtain an estimate of the upper bound on Ly$\alpha$ cooling emission we may assume an ionization scenario such that $n_e \approx n_{\rm H}$, which in this case leads to a maximal luminosity of $7.3 \times 10^5~\Lsun$, less than one per cent the Ly$\alpha$ stellar luminosity. Finally, even though the cooling production rate is inconsequential in this case, it may be important for more massive haloes as the electron abundance significantly increases for virial temperatures well above $10^4$~K, which is necessary to activate more efficient atomic cooling.}

    \hlold{We briefly mention that the assumed 0.68 conversion factor from ionizing to Ly$\alpha$ photons may underestimate the actual luminosity in high-density {\HII} regions.
    This is because metal poor stars are harder sources of ionization. In these environments the factor takes into account the mean ionizing photon energy above 13.6~eV} \citep{Raiter:2010}. \hlold{Secondary ionization effects may also play a role. However, as this would only boost the luminosity by a factor of $\sim 1.5$ it is not likely to change the conclusions in this study.}

\section{Radiative Transfer Calculations}
  \label{sec:results}
  The output from $\colt$ can be viewed as a redistribution of Ly$\alpha$ photons in both frequency and spatial position. In this section we describe the next-event estimator method for calculating surface brightness profiles (Section~\ref{sub:surface_brightness_construction}) and the results from each of the first galaxy models described above (Sections~\ref{sub:idealized_models}~and~\ref{sub:realistic_first_galaxy}).

    \begin{figure*}
      \centering
      \includegraphics[width=1.04598\columnwidth]{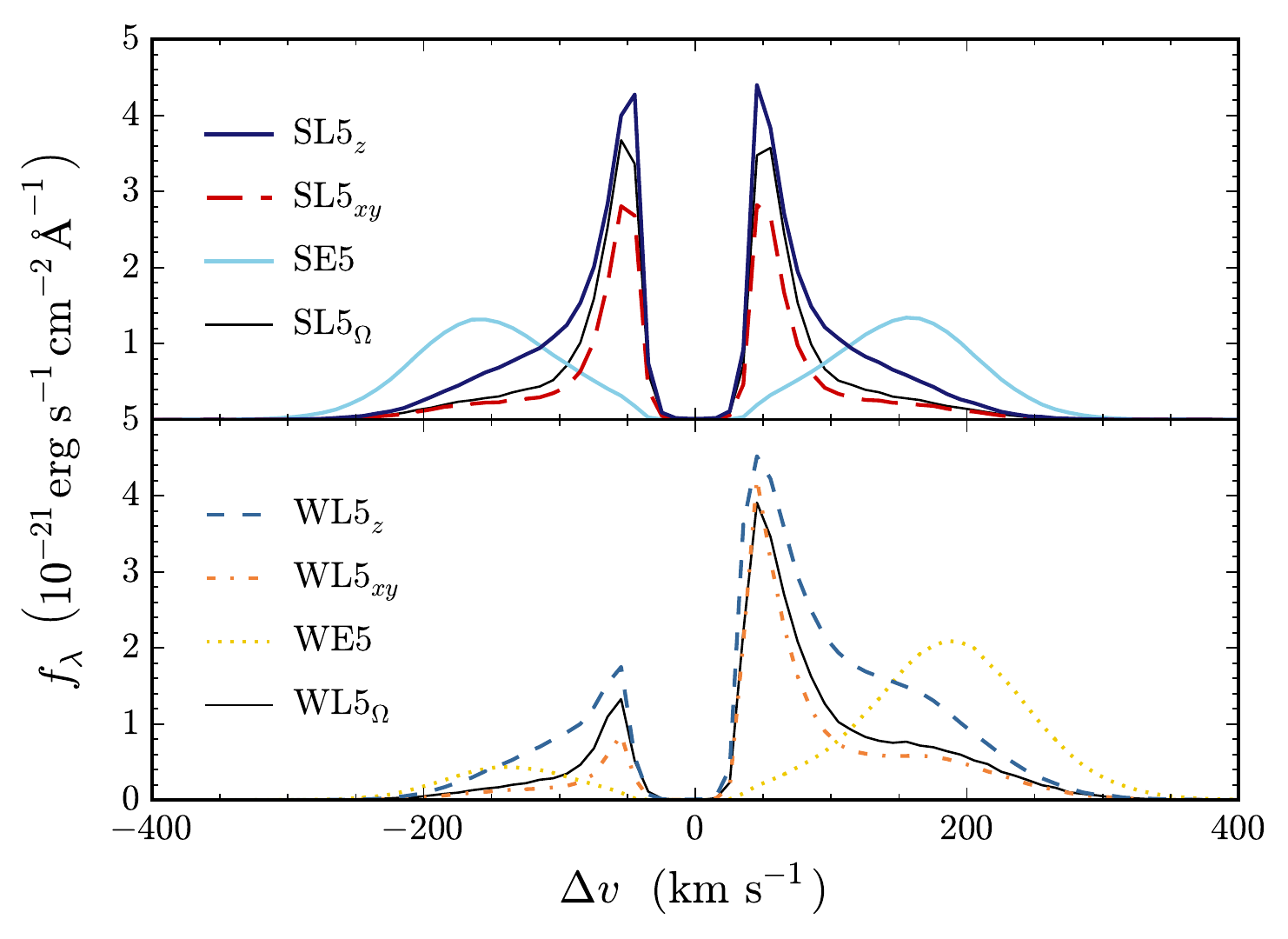}
      \includegraphics[width=1.04598\columnwidth]{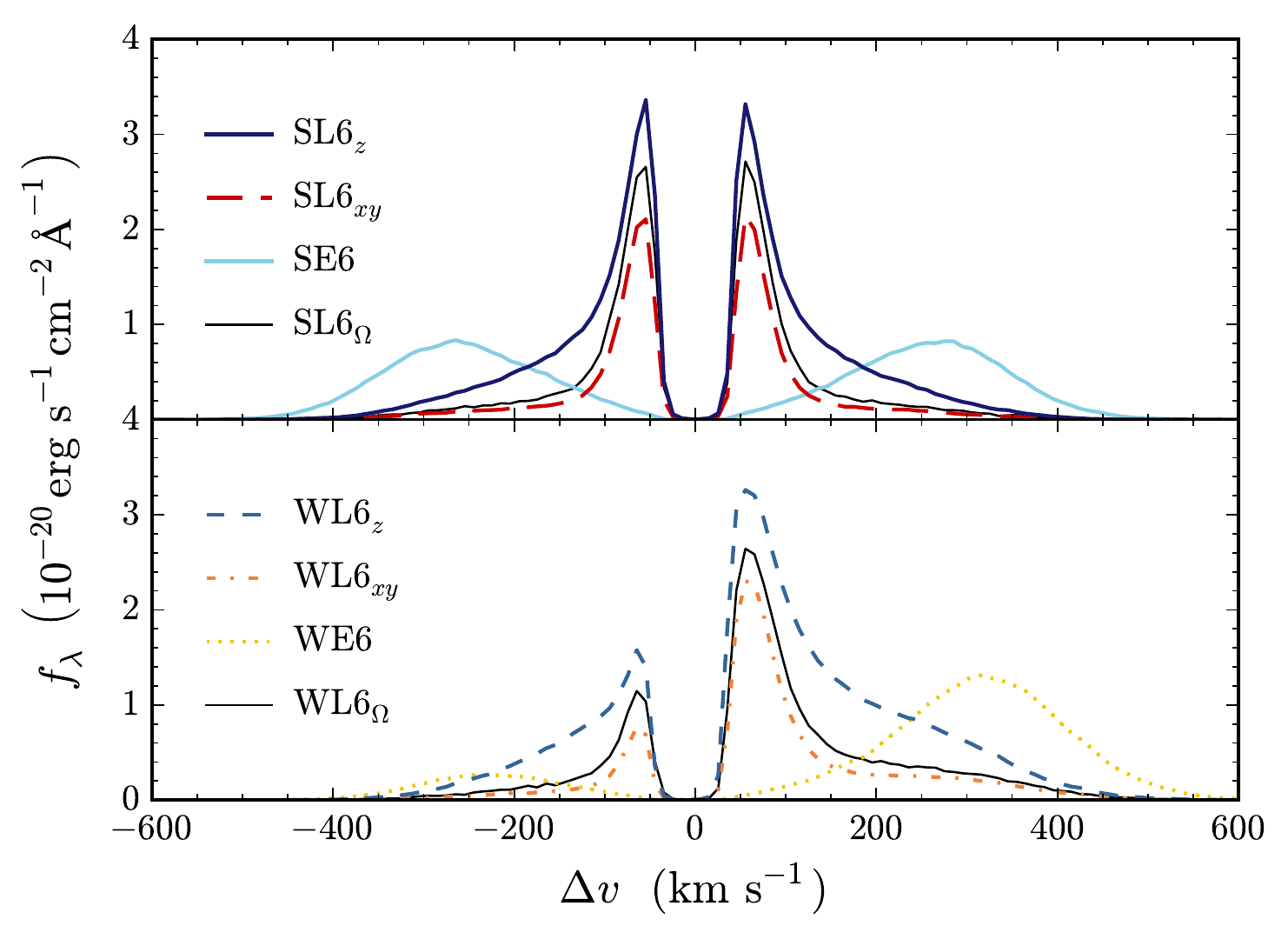}
      \includegraphics[width=1.04599\columnwidth]{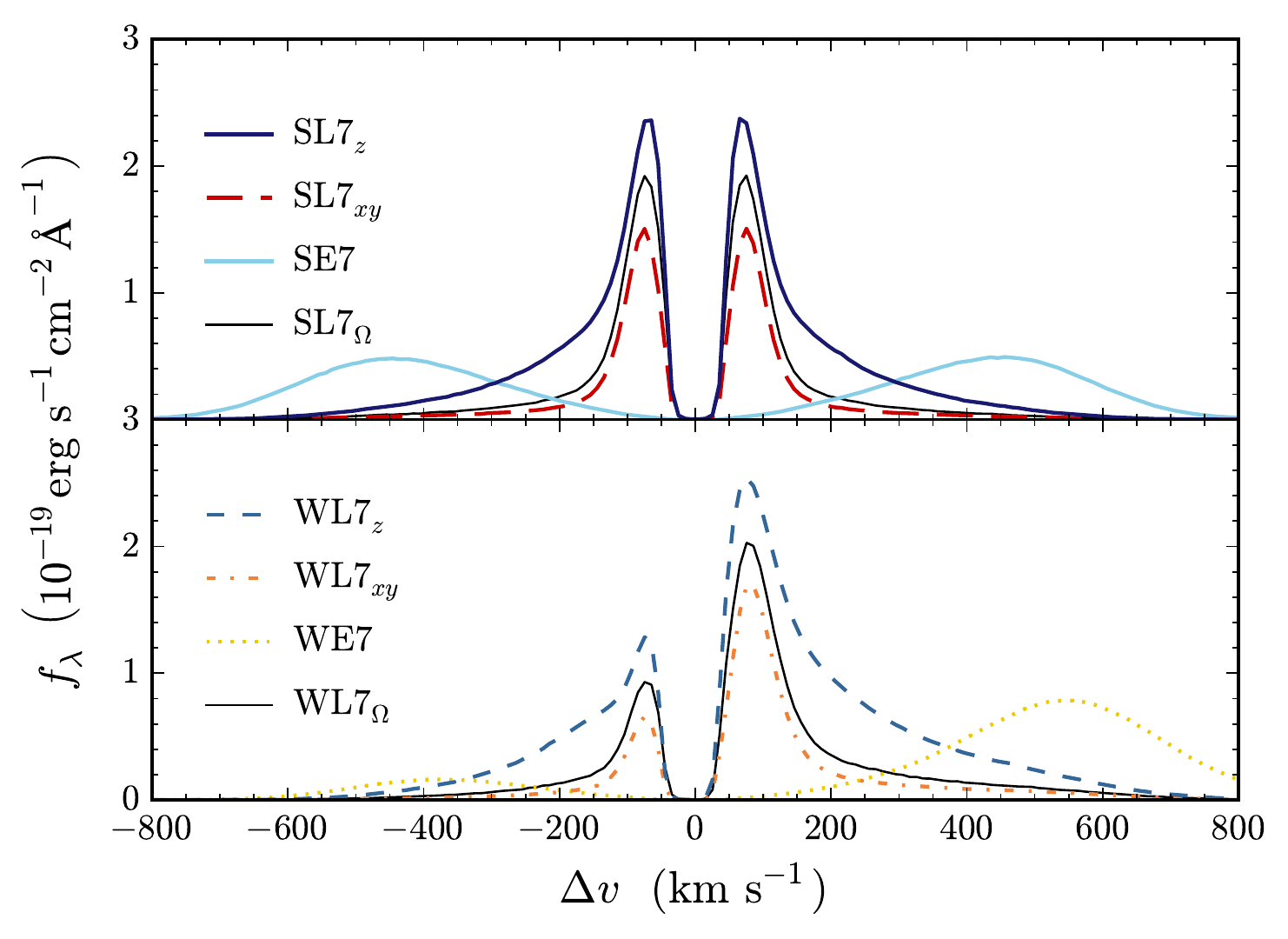}
      \includegraphics[width=1.04599\columnwidth]{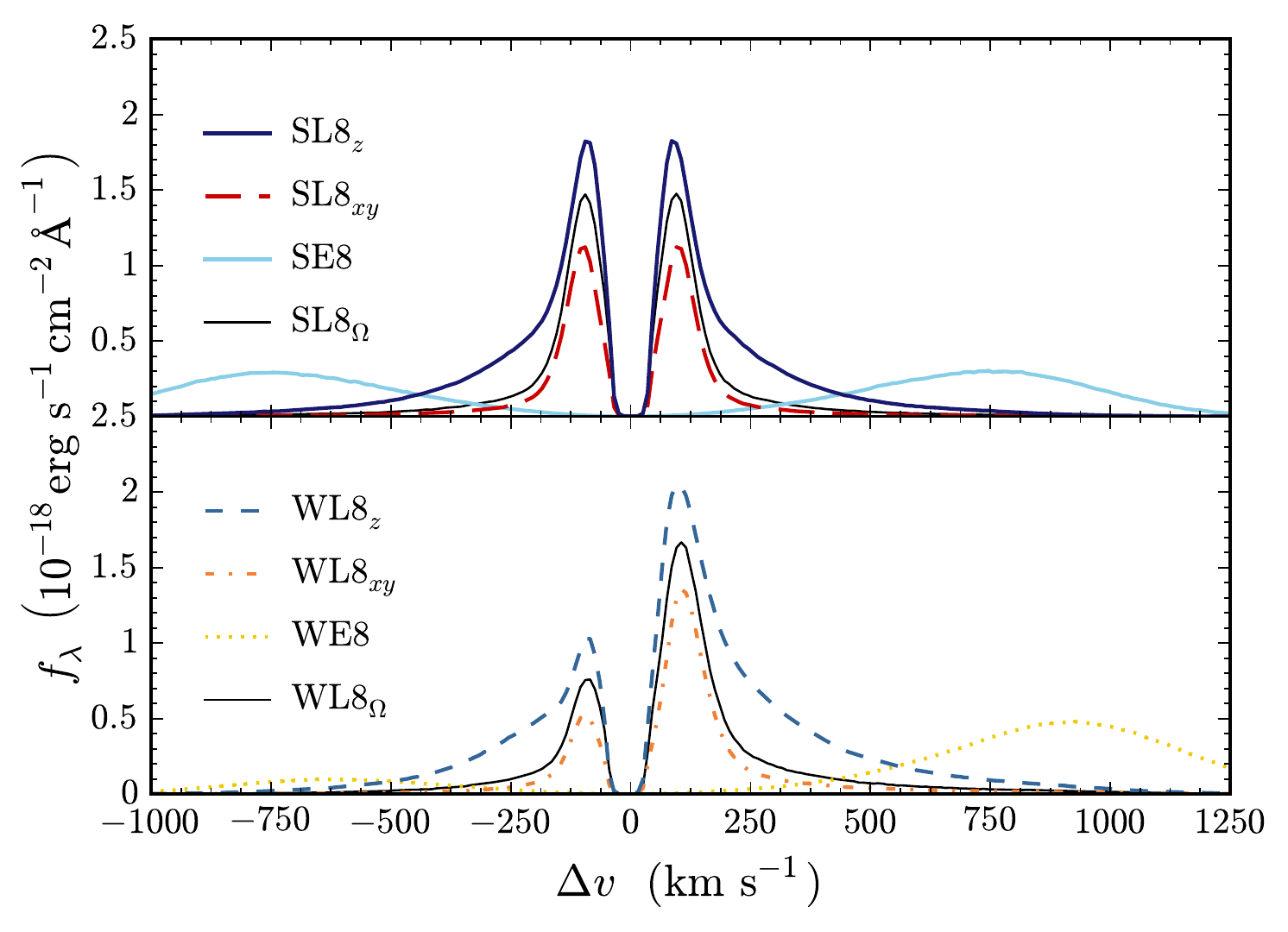}
      \vspace{-.4cm}
      \caption{Line of sight flux as a function of Doppler velocity $\Delta v = c \Delta \lambda / \lambda$ for each of the different models organized by mass (separate subfigures) and wind structure (panels within each subfigure). Anisotropic ionization and wind-driven outflows lower the effective line-of-sight opacity. The halo models are characterized by velocity structure (S for ``Static'' and W for ``Wind''), ionization scenario (E for ``Early'' and L for ``Late''), and the total mass of hydrogen in the halo $M_{\rm H, tot}~(10^5-10^8~\Msun$). Anisotropic ``Late'' profiles have considerably more flux along the face-on line of sight than the edge-on view. Also, asymmetric ``Wind'' models produce greater flux redward of the line centre. Both effects are summarized for quantitative comparison in Table~\ref{tab:model_bolometric_flux}. The units are set by having each halo reside at redshift $z=9$ with a Pop~III star formation efficiency of $\eta_\ast = 0.01$, corresponding to a central starburst of $L_{\rm Ly\alpha} = 3.9 \times 10^{[6-9]}~\Lsun$ (see Table~\ref{tab:central_starburst} for more information). For reference, the observed flux density~$f_\lambda$ increases with source luminosity~$L_{\rm Ly\alpha}$ and decreases with redshift so that $f_\lambda \approxprop L_{\rm Ly\alpha} (1+z)^{-3}$ at high redshifts. Transmission through the (neutral) IGM is not accounted for here, however, see Sections~\ref{sub:surface_brightness_construction} and \ref{subsub:IGM_Trans} for a discussion.}
      \label{fig:halo_Flux}
    \end{figure*}

  \begin{figure*}
    \centering
    \includegraphics[width=1.825\columnwidth]{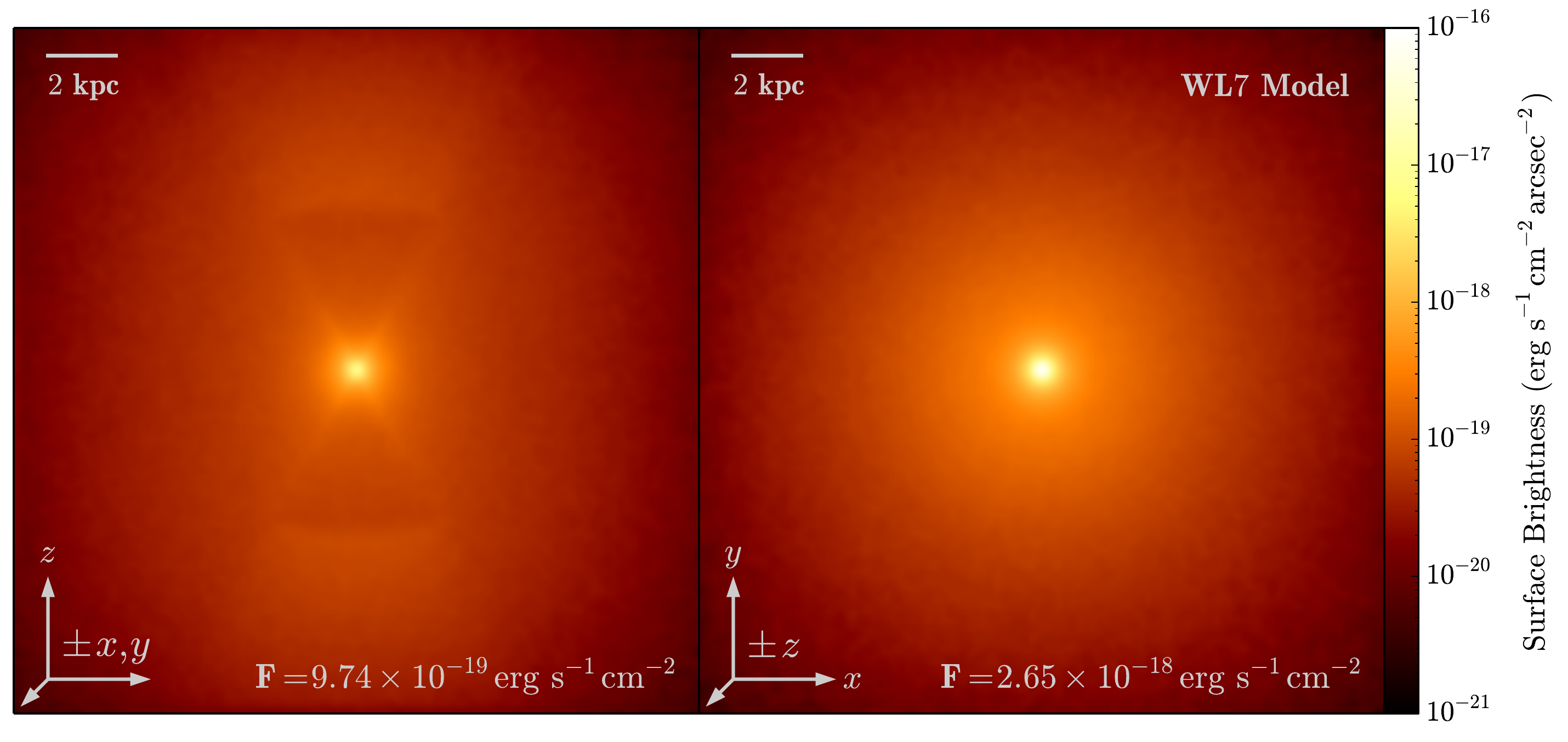}
    \vspace{-.2cm}
    \caption{Surface brightness profile for the WL7 idealized galaxy model at redshift $z=9$, which has a ``Wind'' velocity profile, a ``Late'' anisotropic ionization scenario, and total mass of hydrogen $M_{\rm H, tot} = 10^7~\Msun$ so that $M_{\rm vir} \sim 10^8~\Msun$. The central starburst has a Pop~III star formation efficiency of $\eta_\ast = 0.01$, which for this mass corresponds to a central starburst of $L_{\rm Ly\alpha} = 3.9 \times 10^8~\Lsun$ (see Table~\ref{tab:central_starburst} for more information). The galactic centre contains a Str\"{o}mgren sphere within the core radius $r_{\rm core} = 10$~pc and produces a biconic ionized cavity out to $r_{\rm edge} \approx 5$~kpc where the density drops to that of the background~$n_{\rm H, IGM}$. The simulation region is roughly 20~kpc across which allows the local IGM to play an important role. Indeed, a lighthouse effect is apparent as photons preferentially escape through the bipolar lobe aligned with the $z$-axis and are scattered when they hit the neutral IGM. The intrinsic bolometric flux~$F$ is given for each line of sight at the bottom of each panel.}
    \label{fig:halo_SB}
  \end{figure*}

  \subsection{Surface brightness construction}
    \label{sub:surface_brightness_construction}
    The first galaxies are positioned near the horizon of the currently observable Universe and therefore appear as very small and faint objects. The observability of individual galaxies depends on both the details of Ly$\alpha$ transfer and the sensitivity of the instruments. If a galaxy is completely unresolved we may only be able to measure a single integrated flux. However, if the feature of interest is spatially resolved we may also measure the surface brightness. Therefore, $\colt$ calculates the line of sight surface brightness using the next-event estimator method, similar to that of \citet{Tasitsiomi:2006} and \citet{Laursen:2008aq}. For each scattering we ask what is the probability that the photon would have been scattered toward a given line of sight and how the intervening medium would have attenuated the hypothetical signal. For anisotropic scattering the proper phase function~$W(\theta)$ from Equation~(\ref{eq:phaseW}) quantifies the probability of being scattered into the line of sight. Additionally, the optical depth integrated to the edge of the computational domain diminishes the photon's absolute weight by a factor of $e^{-\tau_{\rm esc}}$. If each photon originally has equal weight, i.e. $L_{\rm Ly\alpha}/N_{\rm ph}$ where $L_{\rm Ly\alpha}$ is again the total Ly$\alpha$ luminosity and \hlold{$N_{\rm ph} \sim 10^7$} is the number of photon packets, then a square CCD grid composed of pixels each subtending a solid angle~$\Omega_{\rm pix}$, observing a source at a luminosity distance~$d_{\rm L}$, receives a total binned surface brightness of
    \begin{equation} \label{eq:SB}
      \text{SB}_{\rm pix} \equiv \frac{\Delta E}{\Delta t \Delta A \Delta \Omega}
               = \frac{L_{\rm Ly\alpha}/N_{\rm ph}}{4 \pi d_{\rm L}^2 \Omega_{\rm pix}} \sum W(\theta) e^{-\tau_{\rm esc}-\tau_{\rm IGM}} .
    \end{equation}
    \hlold{Here we have approximated surface brightness by pixel quantities, i.e. energy~($\Delta E$), time~($\Delta t$), area~($\Delta A$), and solid angle~($\Delta \Omega$). For completeness, the luminosity distance in a flat universe is given by $d_{\rm L} = (1+z) \int_0^z c dz' / H(z')$ where $H(z)$ is the Hubble parameter at a given redshift.}
    The summation is over all scatterings of all photons within the pixel range. In Equation~(\ref{eq:SB}), the phase function~$W(\theta)$ is set to unity for isotropic scattering. At the relevant wavelengths the typical pixel size of \textit{JWST} instruments ranges from $\Omega_{\rm pix, NIRCam} \approx 10^{-3}$~arcsec$^2$ for photometry to $\Omega_{\rm pix, NIRSpec} \approx 0.1$~arcsec$^2$ for spectroscopy.

    The surface brightness may be calculated on the fly for any prescribed direction. In practice, however, we find it more efficient to ray trace along the six coordinate axes, yielding six orthogonal observers for the faces of a cube. Furthermore, due to the severe exponential damping with even moderate optical depths we only continue to ray trace as long as $\tau_{\rm esc} \lesssim 50$, which is conservatively large but still accelerates the process significantly. The line of sight flux is calculated with the same method but without the `per solid angle', i.e. $\Omega_{\rm pix}^{-1}$. $\colt$ produces spatial and frequency bins for $\sum W(\theta) e^{-\tau_{\rm esc}}$ which is multiplied by $L_{\rm Ly\alpha} / [4 \pi N_{\rm ph} (1+z)^4 206265^2 \ell_{\rm pix}^2]$ and integrated over frequency to obtain intensity\footnote{The units of $L_{\rm Ly\alpha}$ are erg~s$^{-1}$, the factor $\frac{180}{\pi} \times 60 \times 60 \approx 206265$ converts from radians to arcseconds, and $\ell_{\rm pix}$ is the physical size of a pixel in cm. The redshift and pixel size dependence originates from considering that $d_{\rm L} = (1+z)^2 d_{\rm A}$ and $\sqrt{\Omega_{\rm pix}} = \theta_{\rm pix} \approx \ell_{\rm pix}/d_{\rm A}$ where $d_{\rm A}$ is the angular diameter distance.}~$\rm{SB}$, expressed in units of erg~s$^{-1}$~cm$^{-2}$~arcsec$^{-2}$, or by $L_{\rm Ly\alpha} / (4 \pi N_{\rm ph} \Delta\lambda_{\rm bin, obs} d_{\rm L}^2)$ and integrated over the field of view to obtain flux density\footnote{The flux density is an observed quantity, therefore, we use the redshifted wavelength bin size, $\Delta\lambda_{\rm bin, obs} = (1+z) \Delta\lambda_{\rm bin}$.},~$f_\lambda$. Here $\Delta \lambda_{\rm bin, obs}$ corresponds to the observed wavelength bin size; throughout this paper we use a Doppler resolution of about $\Delta v \approx 10$~km~s$^{-1}$, corresponding to a spectral resolution of $R \equiv \lambda / \Delta \lambda \approx 30\,000$, which is achievable with next-generation large-aperture ground-based infrared observatories equipped with adaptive optics. Our results may be degraded by a factor of $\sim 30$ for comparison with the NIRSpec instrument aboard the \textit{JWST}.

    \begin{figure*}
      \centering
      \includegraphics[width=1.04599\columnwidth]{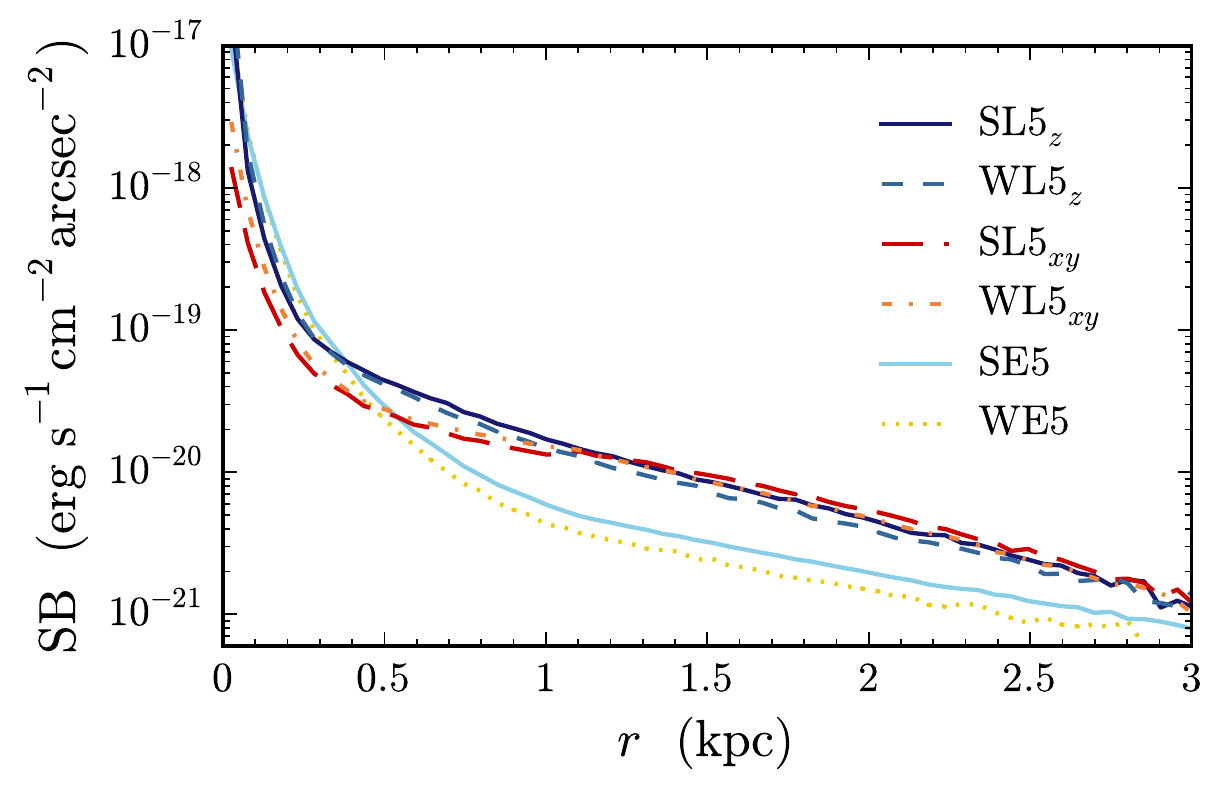}
      \includegraphics[width=1.04599\columnwidth]{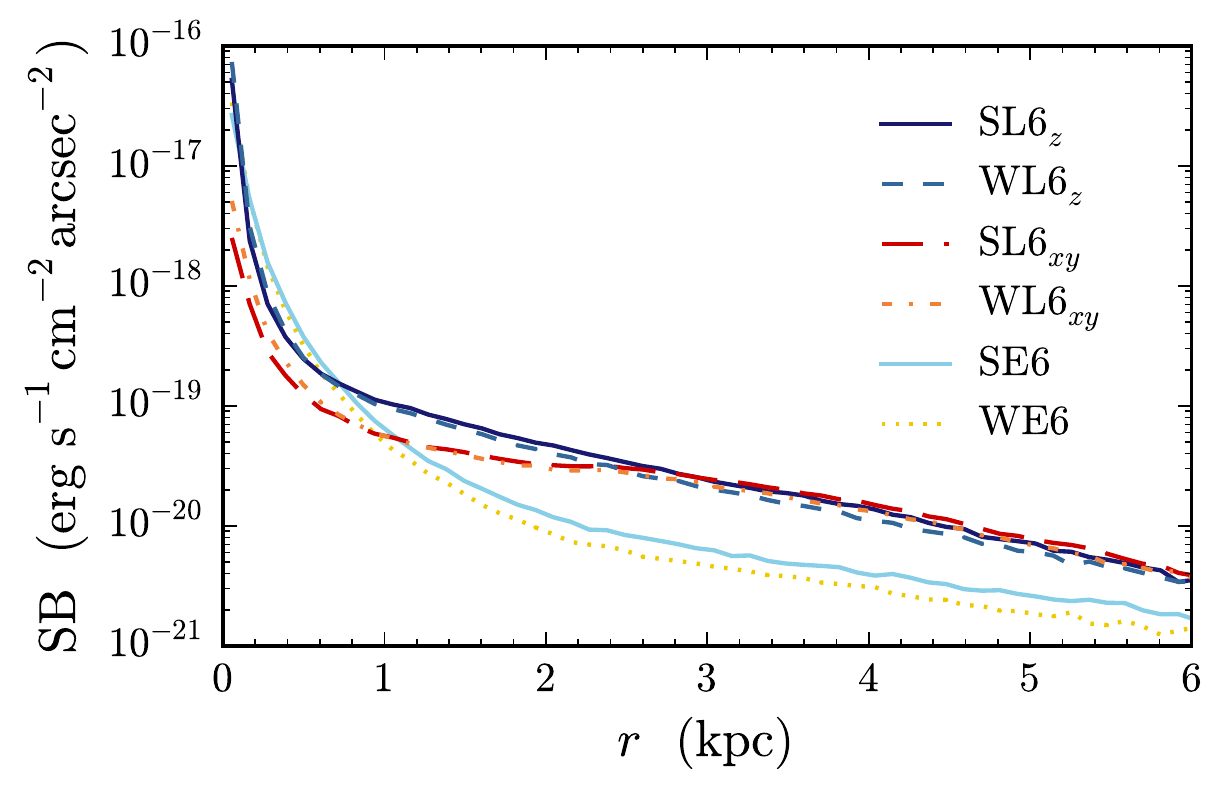}
      \includegraphics[width=1.04599\columnwidth]{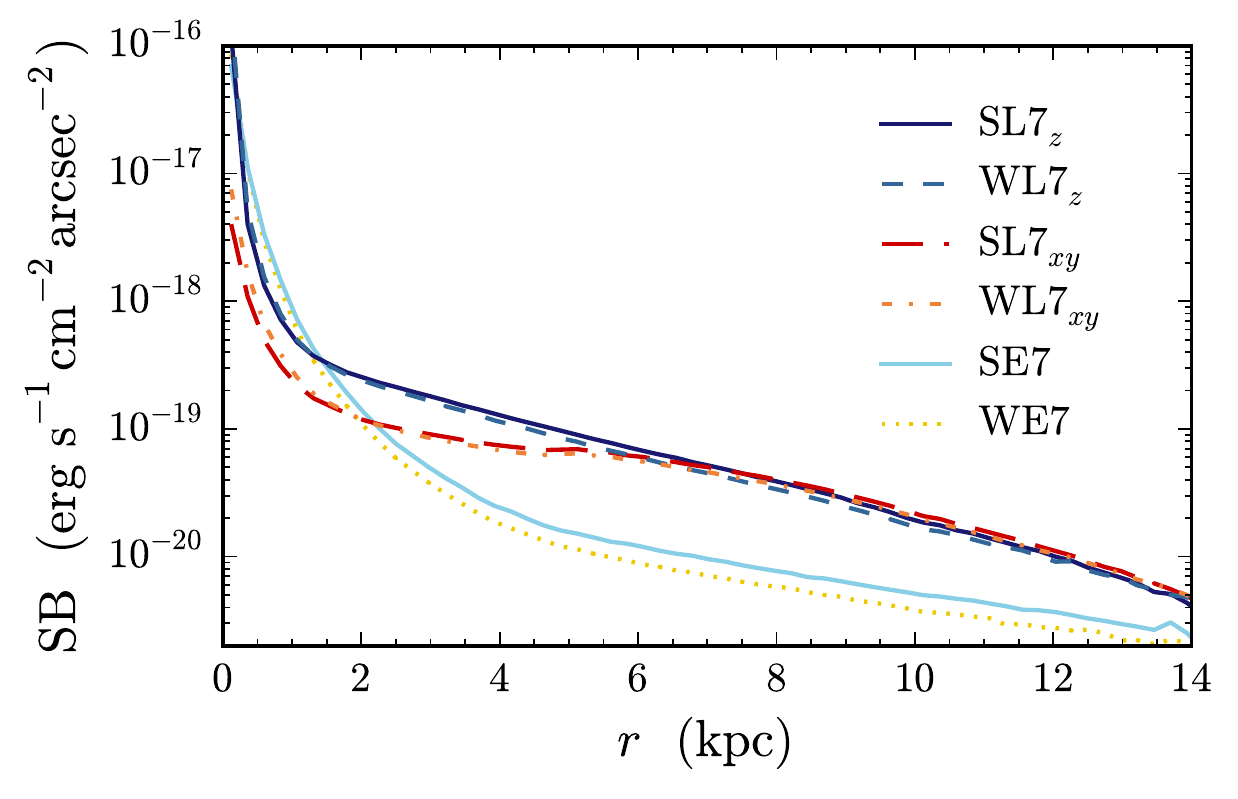}
      \includegraphics[width=1.04599\columnwidth]{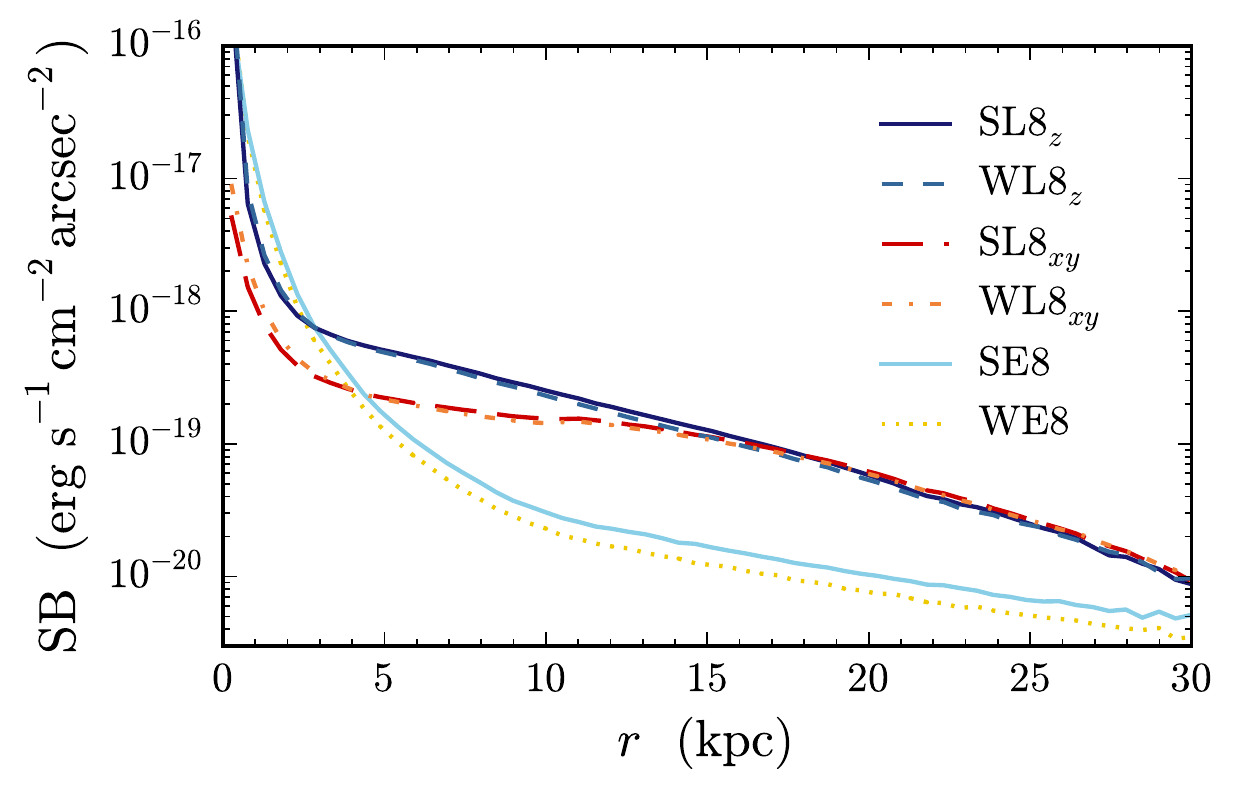}
      \vspace{-.4cm}
      \caption{Radial surface brightness profiles for each of the different mass models. The color scheme and intrinsic halo parameters are the same as that of Fig.~\ref{fig:halo_Flux} (i.e. $\eta_\ast = 0.01$ and $z=9$). For the anisotropic L--models there is a distinct feature at the edge radius corresponding to the hourglass ionization effect, where $r_{\rm edge} \approx 1.09$~kpc for the 5--models, $r_{\rm edge} \approx 2.35$~kpc for the 6--models, $r_{\rm edge} \approx 5.05$~kpc for the 7--models, and $r_{\rm edge} \approx 10.88$~kpc for the 8--models. The ``Late'' models are more extended than the bottled-up ``Early'' models. }
      \label{fig:halo_SB_r}
    \end{figure*}

    \begin{figure*}
      \centering
      \includegraphics[width=1.04599\columnwidth]{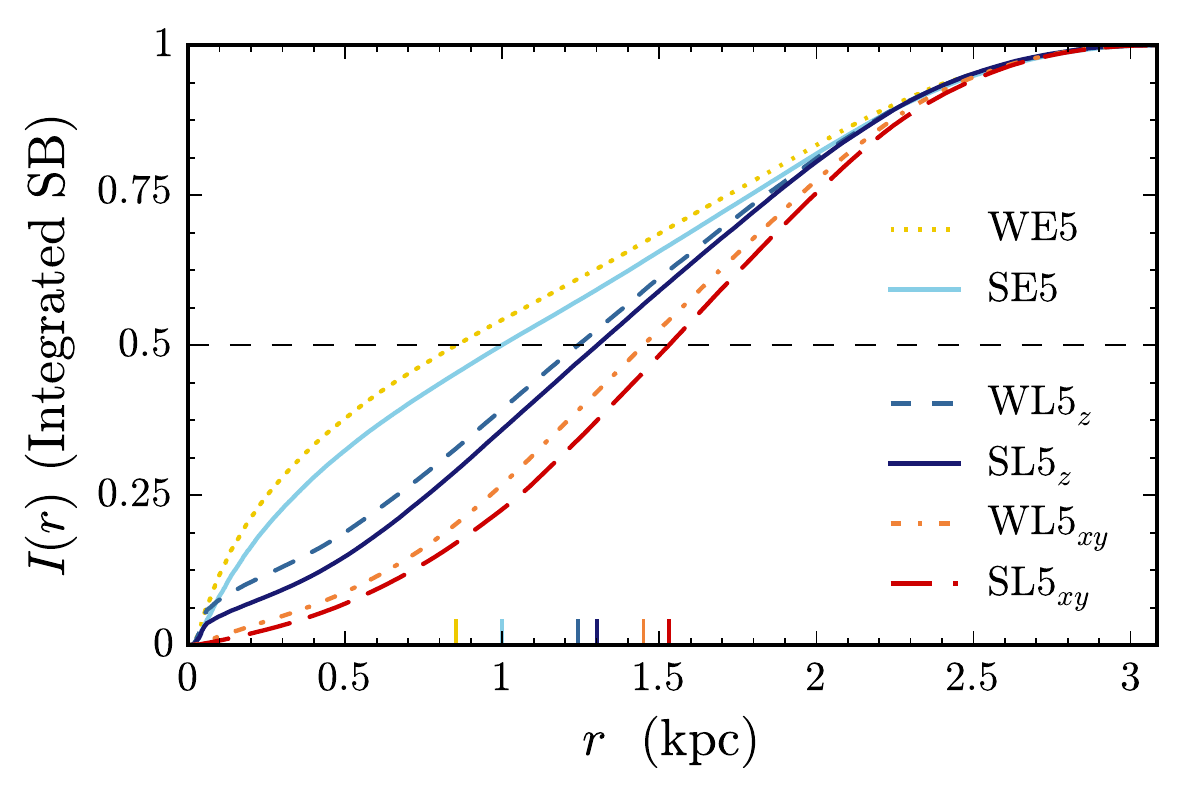}
      \includegraphics[width=1.04599\columnwidth]{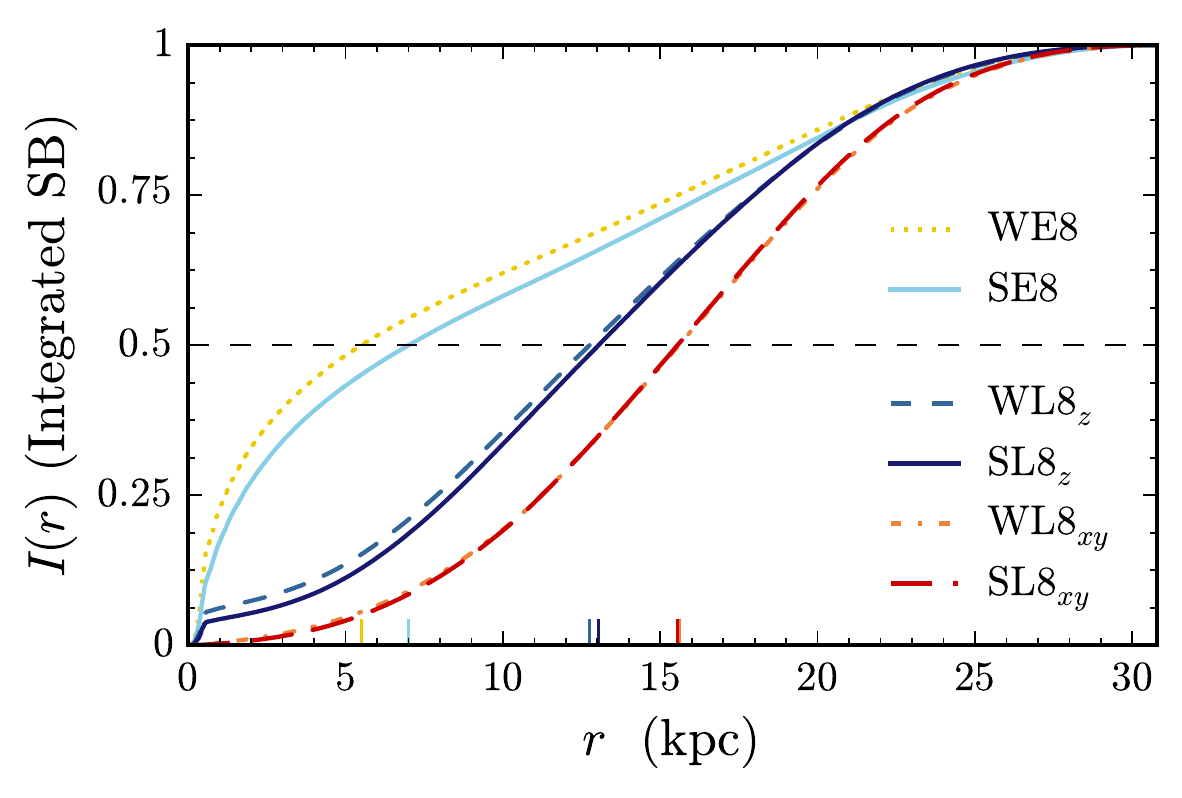}
      \vspace{-.4cm}
      \caption{Integrated light within a given radius, $I(r) \propto \int_0^r \text{SB}(r') r' dr'$, for the 5-- and 8--mass models, left and right respectively. The color scheme and intrinsic halo parameters are the same as Figures~\ref{fig:halo_Flux}~and~\ref{fig:halo_SB_r}. The anisotropic L--models are much more extended than the E-models as can be seen by comparing the half-light radii, $R_{1/2}$, shown as extended colored ticks on the radial axis. The curves have been normalized to unity to allow for direct comparison.}
      \label{fig:halo_SB_r_half_light}
    \end{figure*}

    Transmission through the IGM depends on the local environment and details of reionization. Our treatment follows that of \citet{Madau:2000} who examine the effect of local \HII\ bubbles on the red damping wing of the Gunn-Peterson~(GP) trough. In essence, the IGM opacity removes Ly$\alpha$ photons with a single scattering out of the line of sight, resulting in a spatially extended Ly$\alpha$ halo \citep{Loeb:1999er}. Normally the GP optical depth at line centre, $\tau_0(z) \approx 7 \times 10^5 \left[ (1+z) / 10 \right]^{3/2}$, is large enough to remove any flux blueward of the Ly$\alpha$ line. However, if the Ly$\alpha$ emitter resides within an ionized patch on the order of $\sim 0.1 - 1$ physical Mpc the radiation can redshift sufficiently far from resonance to avoid total suppression in the intervening IGM. In an Einstein--de~Sitter universe with a completely neutral medium outside the ionized bubbles the optical depth of the red damping wing is \citep{Madau:2000}
    \begin{equation} \label{eq:tauGP}
      \tau^{\rm red}_{\rm GP} = \frac{\tau_0(z_{\rm em})}{\pi R_\alpha^{-1}} \left[ \frac{\lambda_\alpha(1+z)}{\lambda_{\rm obs}} \right]^{3/2} \int_{x_{\rm rei}}^{x_i} \frac{dx \, x^{9/2}}{(1-x)^2 + R_\alpha^2 x^6} \, ,
    \end{equation}
    with dimensionless parameters $R_\alpha \equiv \Lambda \lambda_\alpha/(4 \pi c) \approx 2 \times 10^{-8}$, $x_{\rm rei} = (1+z_{\rm rei}) \lambda_\alpha / \lambda_{\rm obs}$, and $x_i = (1+z_i) \lambda_\alpha / \lambda_{\rm obs}$. The limits of integration are set by a cutoff at the redshift of reionization~$z_{\rm rei}$ and the redshift~$z_i$ to the edge of the \HII\ region around the source. \hlold{The line centre optical depth $\tau_0(z_{\rm em})$ is evaluated in terms of the emission redshift of the Ly$\alpha$ source.} For simplicity, we employ this prescription for the IGM opacity, i.e. $\tau_{\rm IGM} = \tau^{\rm red}_{\rm GP}$ in Equation~(\ref{eq:SB}).

  \subsection{Idealized models}
    \label{sub:idealized_models}
    The freedom to change one parameter at a time allows us to perform a direct comparison between various idealized first galaxy models. We discuss the line flux and surface brightness profiles, employing the S--W, E--L, and $M_{\rm H, tot}$ nomenclature, introduced above.

    Figure~\ref{fig:halo_Flux} demonstrates how line flux changes between the halo models. Some of the most apparent trends are:
    \begin{itemize}
      \item Static profiles are symmetric about the Ly$\alpha$ line centre whereas moderate radiation-driven winds generate considerably more red photons than blue ones. This scenario would facilitate Ly$\alpha$ escape and reduce the fraction of photons subject to Gunn-Peterson absorption. On the other hand cosmological inflow models might produce the opposite effect, creating more extended profiles.
      \item The ``Late'' ionization models lead to a distinctive sharp drop near line centre. Ionized pockets prove to be a very efficient mode of escape. Indeed, there would be a third peak at line centre if it were not for the neutral IGM surrounding these models.
      \item The bolometric flux for a given ``Late'' anisotropic model is larger when observed face-on~($z$) than edge-on~($xy$). See Table~\ref{tab:model_bolometric_flux} for the line of sight flux normalized for comparison against the angular averaged flux of the same model. The most dramatic difference is for the WL8 model, where the viewing angle can lead to a dynamic range of $\sim 3$ from maximum to minimum apparent brightness.
      \item The intrinsic ratio~$F_{\rm r}/F_{\rm b}$ of ``red'' photons to ``blue'' photons with respect to the Ly$\alpha$ line centre characterizes the relative efficiency of the wind compared to resonant scattering and escape facilitated by ionized regions. The highest fraction is for the edge-on~($xy$) view of the WL5 model with $F_{\rm r}/F_{\rm b} = 6.36$ and the lowest is for the face-on~($z$) view of the WL8 model with $F_{\rm r}/F_{\rm b} = 2.38$.
    \end{itemize}

    \begin{table}
      \caption{The Ly$\alpha$ bolometric flux, $F = \int F_\lambda d\lambda$ where $F_\lambda$ is the intrinsic flux density, along different lines of sight~($\vec{\ell}$) for the isothermal galaxy models. The flux has been normalized for comparison against the isotropic, angular-averaged flux~($\Omega$), which is $1.4 \times 10^{-[20-17]}$~erg~s$^{-1}$cm$^{-2}$ for a source at redshift $z = 9$ and star formation efficiency $\eta_\ast = 0.01$. The difference between the face-on~($z$) and edge-on~($xy$) integrated flux is more pronounced in massive models.
      The asymmetric ``Wind'' models in Fig.~\ref{fig:halo_Flux} produce greater flux redward of the line centre. The intrinsic red-to-blue component flux ratio~$F_{\rm r}/F_{\rm b}$ characterizes the relative efficiency of the wind compared to resonant scattering and ionization facilitated escape.}
      \label{tab:model_bolometric_flux}
      \begin{tabular}{@{} c cc @{}}
        \hline
        \;\;\;Model\;\;\; & \;\;\;$\vec{\ell}$\;\;\; & \;\;\;\;$F / F_\Omega$\;\;\;\; \\
        \hline
        SL5 &  $z$  & 1.40 \\ 
        WL5 &  $z$  & 1.59 \\ 
        SL5 &  $xy$ & 0.72 \\ 
        WL5 &  $xy$ & 0.80 \\ 
        \hline
        SL6 &  $z$  & 1.60 \\ 
        WL6 &  $z$  & 1.80 \\ 
        SL6 &  $xy$ & 0.72 \\ 
        WL6 &  $xy$ & 0.74 \\ 
        \hline
        SL7 &  $z$  & 1.69 \\ 
        WL7 &  $z$  & 1.91 \\ 
        SL7 &  $xy$ & 0.70 \\ 
        WL7 &  $xy$ & 0.70 \\ 
        \hline
        SL8 &  $z$  & 1.70 \\ 
        WL8 &  $z$  & 1.90 \\ 
        SL8 &  $xy$ & 0.69 \\ 
        WL8 &  $xy$ & 0.68 \\ 
        \hline
      \end{tabular}
      \quad
      \begin{tabular}{@{} c cc @{}}
        \hline
        \;\;\;Model\;\;\; & \;\;\;$\vec{\ell}$\;\;\; & \;\;\;\;$F_{\rm r}/F_{\rm b}$\;\;\;\; \\
        \hline
        WL5 &   $z$    & 3.72 \\ 
        WL5 & $\Omega$ & 4.37 \\ 
        WL5 &   $xy$   & 6.36 \\ 
        WE5 & $\Omega$ & 5.91 \\ 
        \hline
        WL6 &   $z$    & 2.96 \\ 
        WL6 & $\Omega$ & 3.49 \\ 
        WL6 &   $xy$   & 4.58 \\ 
        WE6 & $\Omega$ & 5.72 \\ 
        \hline
        WL7 &   $z$    & 2.57 \\ 
        WL7 & $\Omega$ & 2.99 \\ 
        WL7 &   $xy$   & 3.71 \\ 
        WE7 & $\Omega$ & 5.60 \\ 
        \hline
        WL8 &   $z$    & 2.38 \\ 
        WL8 & $\Omega$ & 2.74 \\ 
        WL8 &   $xy$   & 3.29 \\ 
        WE8 & $\Omega$ & 5.64 \\ 
        \hline
      \end{tabular}
    \end{table}

  \begin{figure}
    \centering
    \includegraphics[width=1.01\columnwidth]{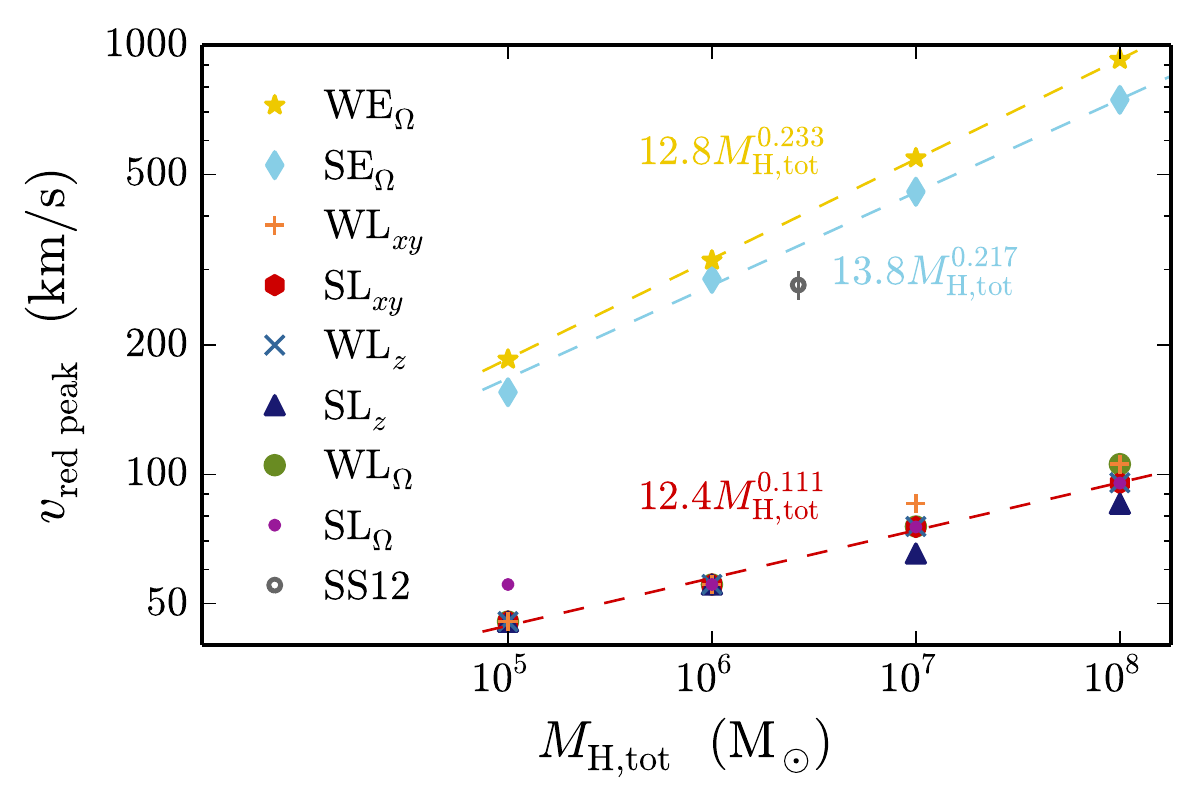}
    \vspace{-.4cm}
    \caption{Correlation between mass and the location of the red peak, $v_{\rm red~peak}$, in units of km~s$^{-1}$ for each halo model. Once again, there is a clear difference between the ``Early'' and ``Late'' models, where escape near line centre is possible for the latter because of anisotropic ionized bubbles. Note that the wavelength resolution of these simulations is $\sim 10$~km~s$^{-1}$ or $\sim 0.04~(1+z)$~\AA, which explains the deviation from the lower (red) least squares fit. The cosmological simulation of \citet{SafranekShrader:2012qn} is plotted as a gray circle with a vertical line to represent the uncertainty. SS12 is consistent with a relatively isotropic ionization scenario. To guide the eye we have included power law fits for selected models.}
    \label{fig:halo_V_red_correlations}
  \end{figure}

    \hlold{Figures}~\ref{fig:halo_SB}--\ref{fig:halo_SB_r_half_light}
    \hlold{illustrate various features of the spatial distribution of the emergent photons. The main qualitative difference may be seen in models with a ``Late'' ionization scenario, which when viewed from the edge-on direction display a prominent outline of the butterfly-shaped cavity. To demonstrate this effect we include Fig.}~\ref{fig:halo_SB}, \hlold{which shows the surface brightness profile for the WL7 idealized galaxy model from edge-on and face-on lines of sight.} Areas within the extended ionized regions are darker because the Ly$\alpha$ photons only scatter once they reach the neutral gas at the boundaries. Therefore, Ly$\alpha$ surface brightness images highlight sharp ionization fronts. \hlold{Recall that the galactic centre contains a Str\"{o}mgren sphere within the core radius $r_{\rm core} = 10$~pc and a biconic {\HII} region out to $r_{\rm edge} \approx 5$~kpc where the density drops to that of the background~$n_{\rm H, IGM}$. The simulation region itself is roughly 20~kpc across. This allows the local IGM to play an important role and capture the asymmetric morphology as photons preferentially escape through the lobes aligned with the $z$-axis.

    In order to compare the various models we present each of the radially averaged surface brightness profiles in Fig.}~\ref{fig:halo_SB_r}. \hlold{The four panels group models of the same mass. As expected, the ``Late'' edge-on ($xy$) profiles have an intensity deficit within the edge radius $r_{\rm edge}$ compared to the face-on ($z$) view. Furthermore, the ``Late'' models are generally more extended than the bottled-up ``Early'' models, which is increasingly true for more massive galaxies. Finally, in order to obtain a quantitative description of the relative spatial extension of the idealized models Fig.}~\ref{fig:halo_SB_r_half_light} \hlold{contains plots of the (normalized) integrated light within a given radius, $I(r) \propto \int_0^r \text{SB}(r') r' dr'$, for the 5-- and 8--mass models, respectively. A clear ordering of the observed Ly$\alpha$ size emerges as we compare the half-light radii, $R_{1/2}$, shown as extended coloured ticks on the radial axis. Large ionized regions and subsequent diffusion in the IGM may increase $R_{1/2}$ to many times the original ``Early'' size.}

  \begin{figure*}
    \centering
    \includegraphics[width=1.062\columnwidth]{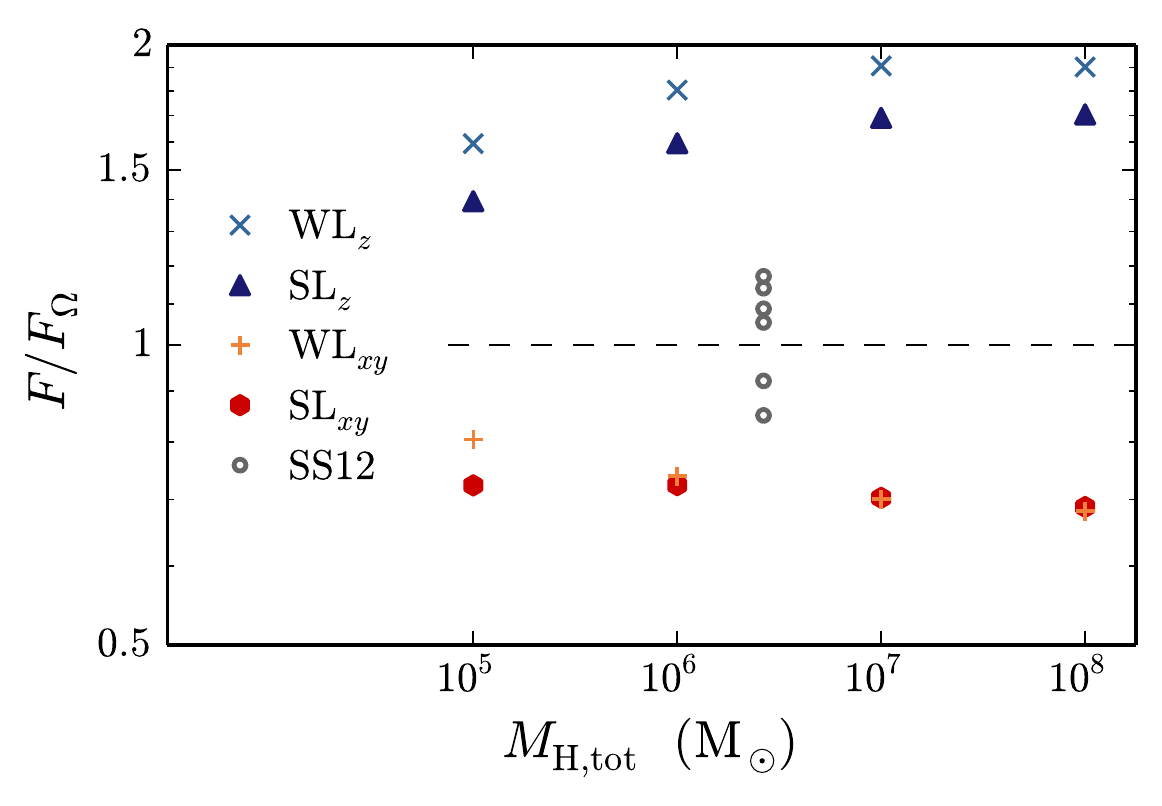}
    \includegraphics[width=1.03\columnwidth]{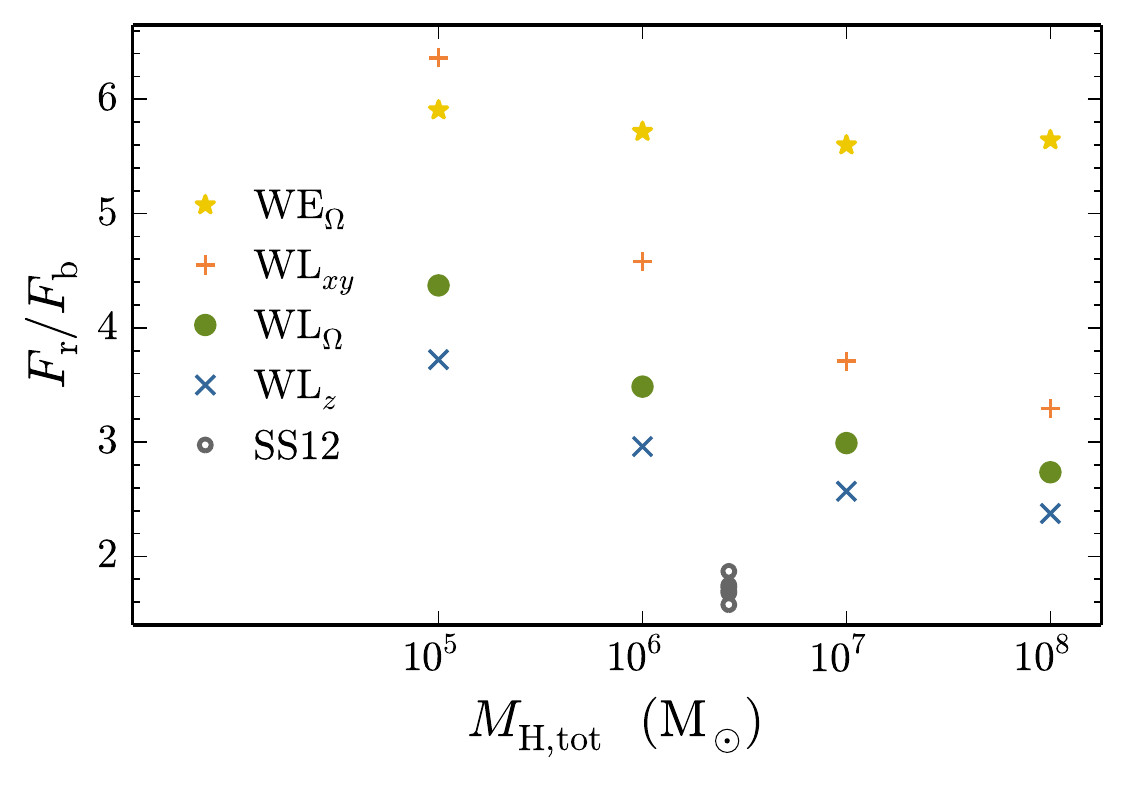}
    \vspace{-.4cm}
    \caption{Relationships between mass and flux properties for relevant idealized halo models. The left panel shows a clear difference in the bolometric flux between face-on~($z$) and edge-on~($xy$) views. The right panel illustrates trends for the ratio of `red' to `blue' flux. For the ``Early'' models $F_{\rm r} / F_{\rm b}$ only depends on the strength of the wind, however, for the ``Late'' models the ratio decreases for larger haloes. For reference, the cosmological simulation of \citet{SafranekShrader:2012qn} is plotted as gray circles for each of the six sightlines. The SS12 results indicate that the halo emits roughly isotropically -- perhaps due to the large (1~Mpc)$^3$ comoving box size -- and has a relatively small bulk velocity. For this model we plot the ratio of `blue' to `red' flux as the velocity is largely due to cosmological inflow. If additional feedback mechanisms are included the SS12 results may be different. See Table~\ref{tab:model_bolometric_flux} for a quantitative comparison. }
    \label{fig:halo_flux_correlations}
  \end{figure*}

  \begin{figure*}
    \centering
    \includegraphics[height=.70545\columnwidth]{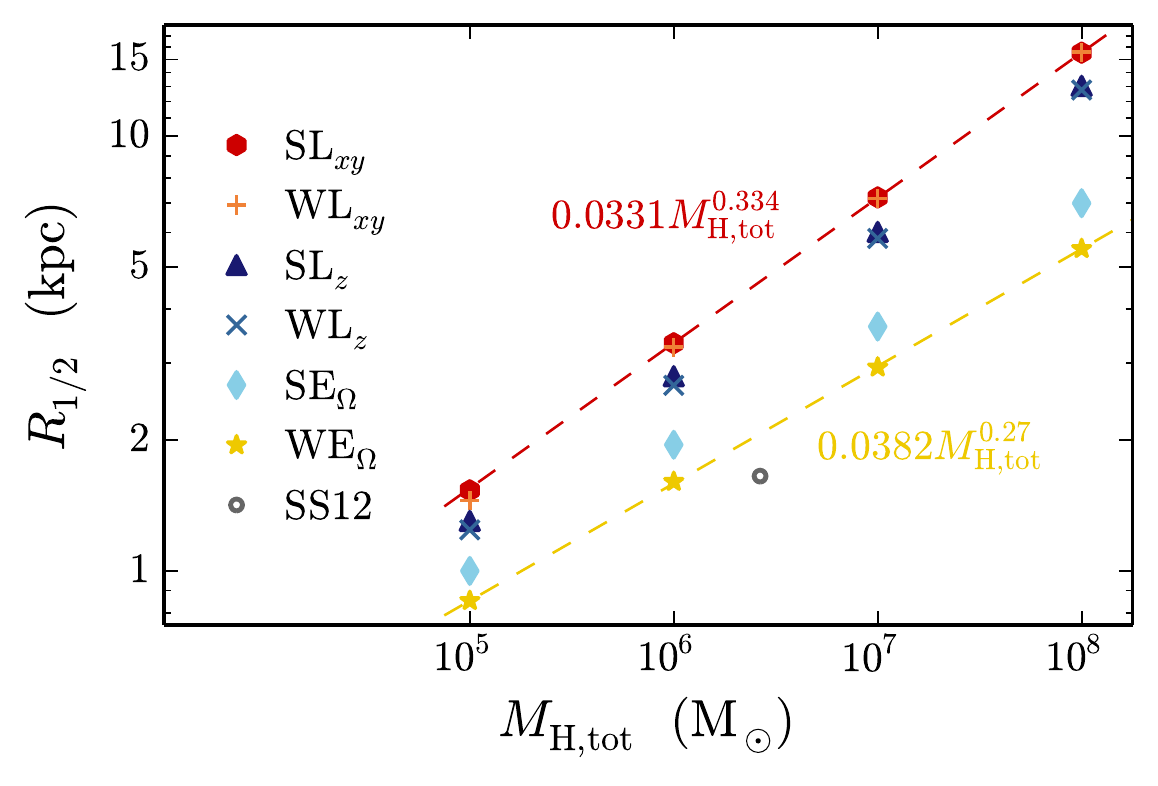}
    \includegraphics[height=.70545\columnwidth]{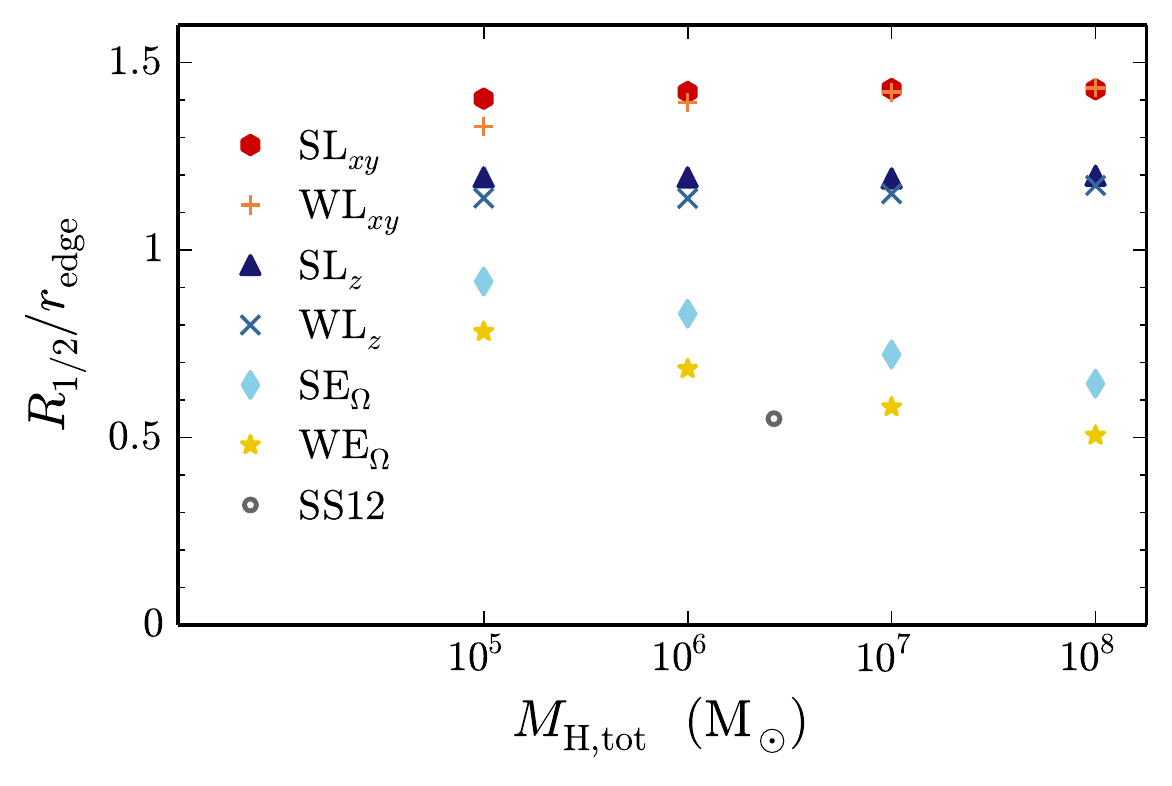}
    \vspace{-.4cm}
    \caption{Trends between mass and the half-light radius, $R_{1/2}$, for the idealized halo models. The relative shape of the integrated surface brightness and the corresponding value of $R_{1/2}$ indicate how extended the source appears. In the left panel there is a clear relationship of $R_{1/2} \propto M_{\rm H, tot}^{1/3}$ which shadows the established relationship of $\chi$ in Equation~(\ref{eq:chi}). In the right panel we have taken out the dependence on $r_{\rm edge}$ (or box size). The ``Early'' models still demonstrate a noticeable trend of becoming more singular with respect to the radius of the halo. To guide the eye we have included power law fits for selected models.}
    \label{fig:halo_half_light_correlations}
  \end{figure*}

    Finally, we illustrate some observable trends between these idealized first galaxy models. Figure~\ref{fig:halo_V_red_correlations} shows the location of the red peak,~$v_{\rm red~peak}$, in Doppler velocity units from line centre, i.e. $\Delta v = c \Delta \lambda / \lambda$, which increases as a function of mass. A power law fit of the data shows that $v_{\rm red~peak}$ is twice as sensitive to mass for ``Early'' models than for ``Late'' models. Figure~\ref{fig:halo_flux_correlations} shows the qualitative differences of the flux properties listed in Table~\ref{tab:model_bolometric_flux}. The anisotropic models appear more luminous when observed face-on~($z$) than edge-on~($xy$) by a factor of a few. Additionally, for ``Wind'' models the relative flux redward of line centre ($F_{\rm r} / F_{\rm b} \sim \text{a~few}$) is generally more exaggerated for the ``Early'' models. Finally, Figure~\ref{fig:halo_half_light_correlations} compares the half-light radius,~$R_{1/2}$, \hlold{for each of the galaxies. The various models} roughly follow the relation predicted by Equation~(\ref{eq:chi}) that $R_{1/2} \propto \chi \propto M_{\rm H, tot}^{1/3}$. \hlold{Recall that the integrated light within a given radius, $I(r) \propto \int_0^r \text{SB}(r') r' dr'$, may be normalized to unity at the largest radii so $I(R_{1/2}) = 1/2$. The value of the half-light radius indicates how extended the source appears. The difference between the bottled-up ``Early'' models and butterfly-shaped ``Late'' models is especially pronounced for more massive galaxies.} When appropriate we have included power law fits for selected models to guide the eye.

  \begin{figure*}
    \centering
    \includegraphics[height=.675\columnwidth]{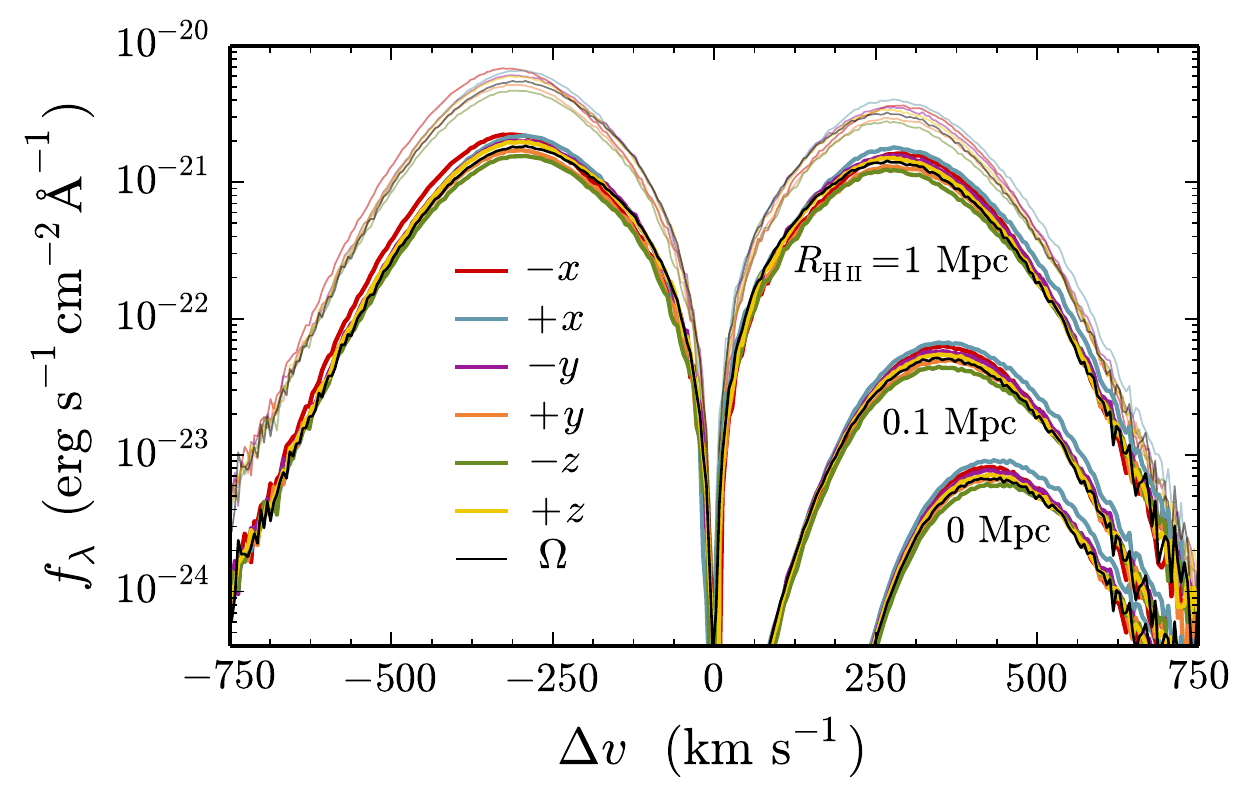} \hspace{-.265cm}
    \includegraphics[height=.675\columnwidth]{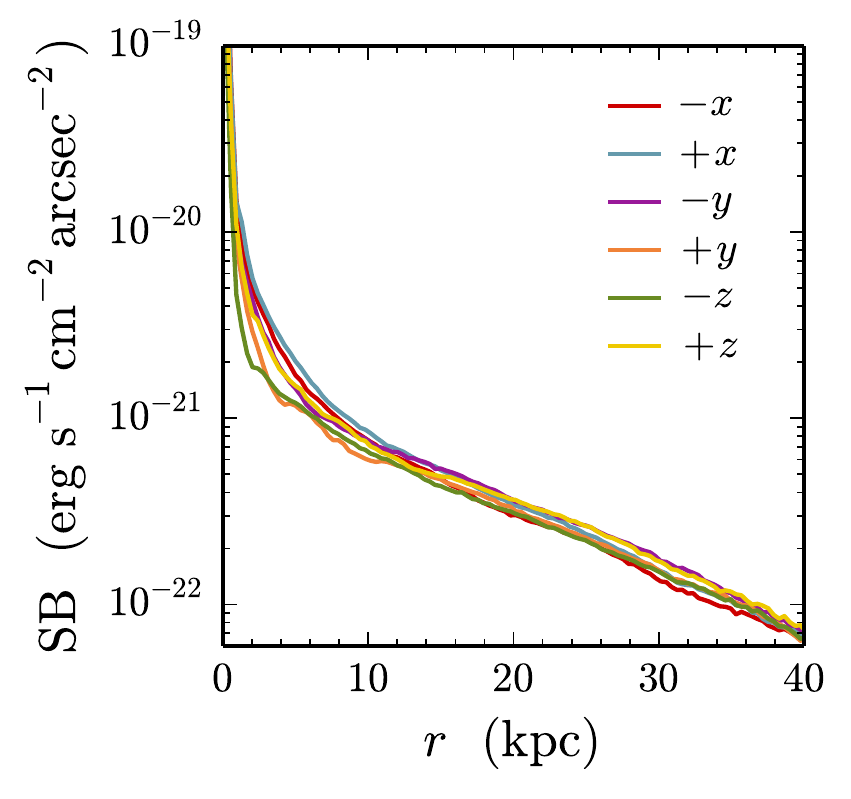} \hspace{-.32cm}
    \includegraphics[height=.6575\columnwidth,trim=0in -0.3in 0in 0in,clip=true]{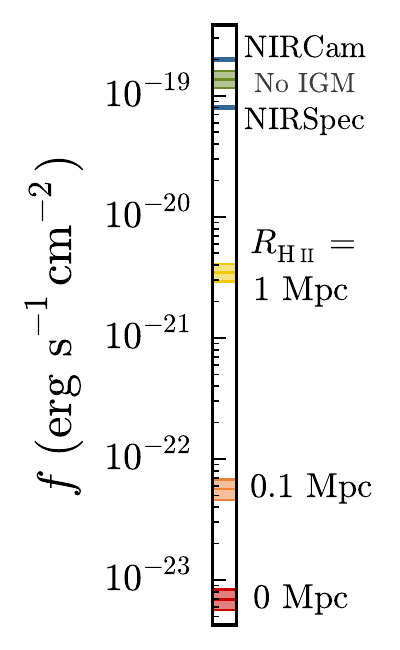}
    \vspace{-.4cm}
    \caption{Line of sight flux density (left), radial surface brightness profile (middle), and bolometric flux (right) for the six coordinate faces of the $(67.5$~kpc$)^3$ physical extraction region, assuming a $10^8~\Lsun$ source at $z = 13.8$. The specific flux in the left panel is calculated for a Doppler resolution of $\Delta v \approx 10$~km~s$^{-1}$, corresponding to a spectral resolution of $R \equiv \lambda / \Delta \lambda \approx 30\,000$, achievable with next-generation large-aperture ground-based infrared observatories with adaptive optics. The light-shaded curves are intrinsic to the galaxy whereas the other three sets of curves include suppression from IGM opacity, i.e. a frequency dependent factor of $\exp(-\tau^{\rm red}_{\rm GP})$ defined in Equation~(\ref{eq:tauGP}). The difference between the transmission models is the size of the local ionized bubble~$R_{\rm \HII}$ which has a strong effect on the observed flux. Although Figures~\ref{fig:extractor_density}~and~\ref{fig:extractor_SB_cube_2kpc} demonstrate many distinct inhomogeneous features, e.g. obscuration from clouds or anisotropic excess intensity, the spatially averaged flux and radial surface brightness are quite similar across different sightlines. The middle panel illustrates the singular nature of the intrinsic Ly$\alpha$ source and the transition to an exponentially damped halo, which in this case roughly coincides with $\text{SB} \approxprop \exp (-r/12.5~\text{kpc})$. The right panel compares the effect of $R_{\rm \HII}$ on the total observed flux for comparison with \textit{JWST} sensitivities. }
    \label{fig:extractor_32kpc}
  \end{figure*}

  \begin{figure*}
    \centering
    \includegraphics[width=1.625\columnwidth]{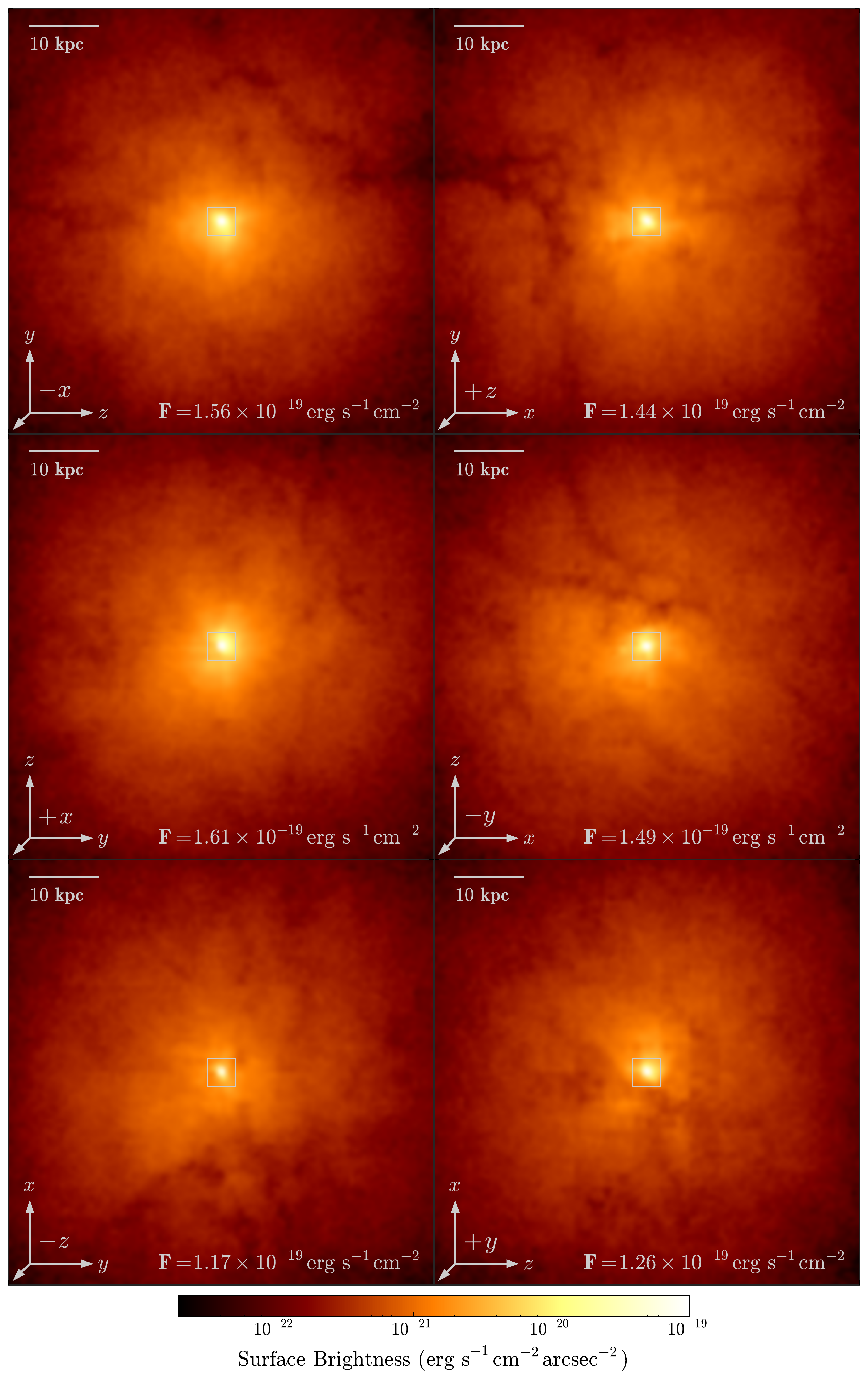}
    \vspace{-.4cm}
    \caption{Line of sight surface brightness profiles for the six coordinate faces of the entire $(1$~Mpc$)^3$ comoving volume or $(67.5$~kpc$)^3$ in physical units. \hlold{The central square corresponds to the size of Fig.}~\ref{fig:extractor_SB_cube_2kpc}. On larger scales the IGM tends to smooth out the profiles so the viewing angle differences are less severe.}
    \label{fig:extractor_SB_cube_32kpc}
  \end{figure*}

  \begin{figure*}
    \centering
    \includegraphics[width=1.625\columnwidth]{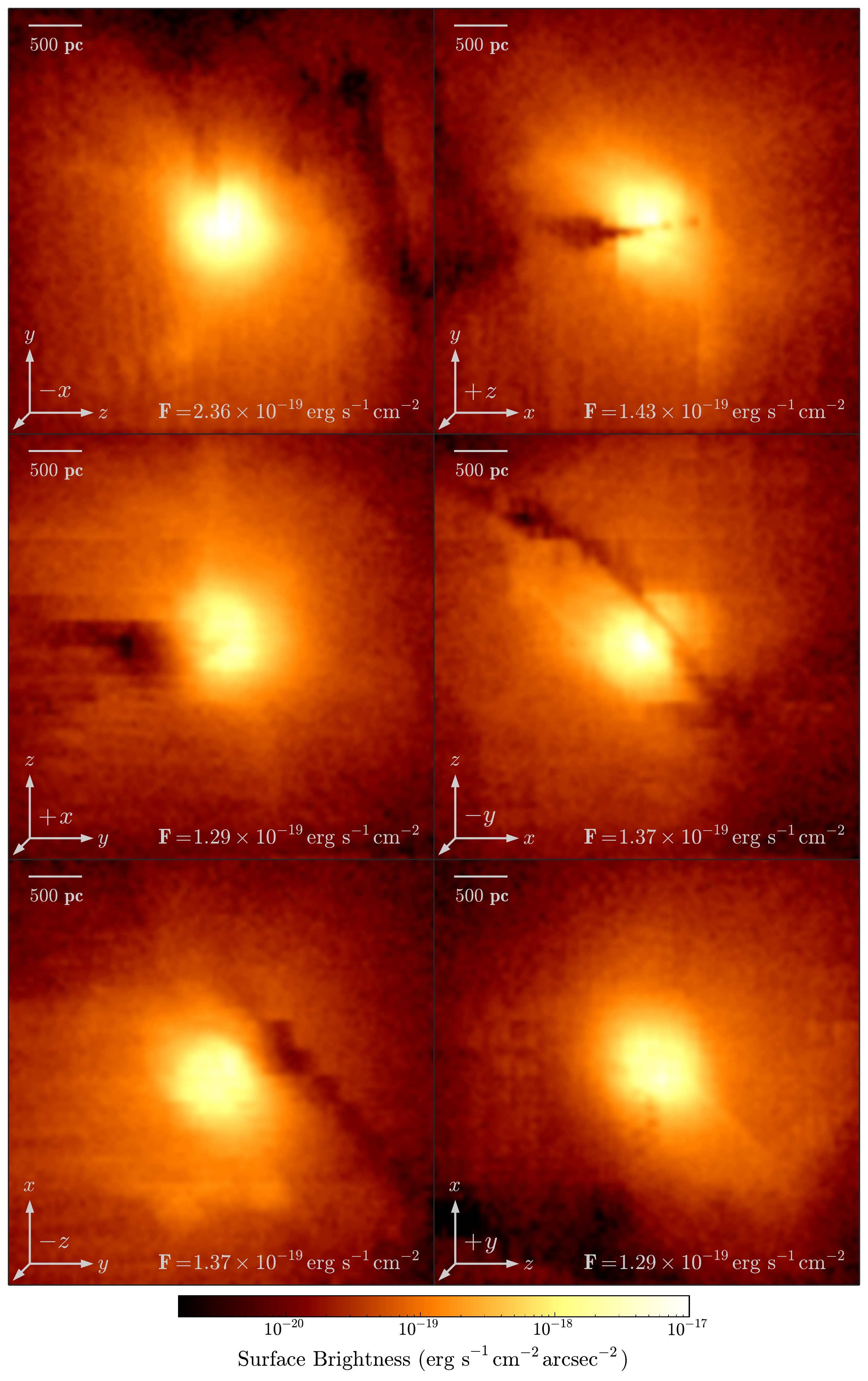}
    \vspace{-.4cm}
    \caption{Line of sight surface brightness profiles for the six coordinate faces of the $\sim (4$~kpc$)^3$ extraction region. The dark fluffy streaks are wisps or clouds of neutral hydrogen blocking the particular sightline. Artifacts of the next-event estimator method sometimes appear, which does not resolve intensity features on scales smaller than the intervening AMR grid structure. Although these features are smoothed out when considering larger volumes (cf.~Fig.~\ref{fig:extractor_SB_cube_32kpc}) and transmission through the IGM, such a halo could possible serve as an analog for resolved systems at lower redshifts.}
    \label{fig:extractor_SB_cube_2kpc}
  \end{figure*}

  \subsection{Realistic first galaxy}
    \label{sub:realistic_first_galaxy}
    We now present the $\colt$ output of the realistic cosmological simulation introduced in Section~\ref{sec:sim}. Figure~\ref{fig:extractor_32kpc} shows the line of sight flux and radial surface brightness profiles for the six coordinate faces of the $(67.5$~kpc$)^3$ extraction region in physical units. \hlold{The specific flux in the left panel is calculated for a Doppler resolution of $\Delta v \approx 10$~km~s$^{-1}$, corresponding to a spectral resolution of $R \equiv \lambda / \Delta \lambda \approx 30\,000$, achievable with next-generation large-aperture ground-based infrared observatories with adaptive optics. The light-shaded curves are intrinsic to the galaxy whereas the other three sets of curves denoted by different values of $R_{\rm \HII}$ are included to consider suppression from IGM opacity, however, we defer such discussion until Section}~\ref{sec:detectability}. \hlold{The middle panel illustrates the singular nature of the intrinsic Ly$\alpha$ source and the transition to an exponentially damped halo, which in this case roughly coincides with $\text{SB} \approxprop \exp (-r/12.5~\text{kpc})$. The right panel compares the observed bolometric flux under different $R_{\rm \HII}$ scenarios with \textit{JWST} sensitivities -- see Section}~\ref{subsub:JWST_sensitivity}.

    \hlold{For the most part the spatially averaged flux and radial surface brightness profiles are quite smooth and qualitatively similar across different sightlines.} Likewise, the actual intensity images (see Fig.~\ref{fig:extractor_SB_cube_32kpc}) also appear relatively isotropic and featureless despite the obvious inhomogeneous and anisotropic features illustrated by the column density (see Fig.~\ref{fig:extractor_density}) and surface brightness images (see Fig.~\ref{fig:extractor_SB_cube_2kpc}) captured in the immediate vicinity of the galaxy, i.e. a few virial radii away. The dark fluffy streaks are clouds of neutral hydrogen blocking particular sightlines. The main distinguishing characteristic is that certain faces of the cube are significantly brighter than others, especially in the central $\sim 1$~kpc region. This is due to the inhomogeneous medium which provides preferred channels of escape. It is apparent from Fig.~\ref{fig:halo_flux_correlations} that the deviation from isotropy for the SS12 model is not very pronounced compared to the idealized anisotropic models. In this particular case $| 1 - F/F_\Omega | \lesssim 0.2$, although smaller extraction cubes exhibit significant anisotropic variance in the emergent spectra, especially when comparing opposite lines of sight, i.e $\pm x$, $\pm y$, and $\pm z$ (see Appendix~\ref{appendix:box_sizes}). Such line-of-sight difference is likely due to the location of the central starburst within the dense galactic environment. However, much of the relative variation between sightlines may wash out as we account for the additional diffusion required to escape the vast neutral IGM. Furthermore, looking at the temperature structure of Fig.~\ref{fig:extractor_density} indicates that it may also be possible to experience similar \hlold{``thermal effects'' that shape the emergent spectra in a nontrivial way. Although there is no temperature dependence on the Lorentz wing optical depth it is apparent from Equations}~(\ref{eq:tau_wing})~and~(\ref{eq:atau0min}) \hlold{that the products $a\tau_0$ and $a\tau_{\rm wing}$ depend on temperature as $T^{-1}$ and $T^{-1/2}$, respectively. To demonstrate the thermal effects of the core-wing transition $x_{\rm cw}$ we include two test suites of uniform slabs with varying temperature but (i) constant optical depth and (ii) constant column density in Appendix}~\ref{appendix:temperature_tests}.

    \begin{table*}
      \caption{\textit{JWST} instrument sensitivity for a $5\sigma$ detection after $10^6$~seconds of exposure time based on sensitivities assuming the G235M grating with the F170LP filter for NIRSpec and the F150W filter for NIRCam. Flux densities are related by $f_\lambda \approx (c/\lambda^2)~f_\nu$ where $1~\text{Jy} = 10^{-23}$~erg~s$^{-1}$~cm$^{-2}$~Hz$^{-1}$. Entries denoted by G06 represent values taken from \citet{Gardner:2006ky}, scaled from the quoted sensitivity based on a $10\sigma$ detection after $10^4$~seconds.}
      \label{tab:JWST}
      \begin{tabular}{@{} c cccccccc @{}}
        \hline
        \;Instrument\; & $R$ & $\Delta \lambda$ & $\Delta \nu$ & $\Delta \Omega_{\rm pix}$
                   & $f$ & $f_\nu$ & $f_\lambda$ & SB \\
        \hline
        \vspace{.05cm}
        Units      & -- & \AA & Hz & arcsec$^2$
                   & erg~s$^{-1}$~cm$^{-2}$ & nJy
                   & erg~s$^{-1}$~cm$^{-2}$~\AA$^{-1}$ & erg~s$^{-1}$~cm$^{-2}$~arcsec$^{-2}$\; \\
        NIRSpec    & 1000 & 20 & $10^{11}$ & $0.1$
                   & $8 \times 10^{-20}$ (G06) & 50
                   & $4 \times 10^{-21}$ & $8 \times 10^{-19}$ \\
        NIRCam     & 4 & 4500 & $4 \times 10^{13}$ & $10^{-3}$
                   & $2 \times 10^{-19}$ & 0.56 (G06)
                   & $5 \times 10^{-23}$ & $2 \times 10^{-16}$ \\
        \hline
      \end{tabular}
    \end{table*}

  \subsection{Detectability of individual first galaxies}
    \label{sec:detectability}

    \subsubsection{IGM Transmission}
      \label{subsub:IGM_Trans}
      The observability of this particular galaxy model depends on the subsequent transmission through the IGM. In Fig.~\ref{fig:extractor_32kpc} we present the intrinsic flux density~$f_\lambda$ (shown as semi-transparent curves) and three scenarios that include suppression from the IGM. All signals have been corrected for redshift and assume a $10^8~\Lsun$ source. The lower three sets of curves include a frequency dependent factor of $\exp(-\tau^{\rm red}_{\rm GP})$ defined in Equation~(\ref{eq:tauGP}) using physical sizes for the local ionized bubble $R_{\rm \HII}$ of $1$~Mpc, $0.1$~Mpc, and $0$~Mpc, respectively. The $R_{\rm \HII} = 0$~Mpc curves represent the worst case scenario of no \HII\ region while the $R_{\rm \HII} = 1$~Mpc curves are likely a best case scenario, under the assumption that the IGM is fully neutral outside the ionized bubble. \hlold{To a certain degree, this ionization model may be inconsistent with the assumption in Equation}~(\ref{eq:L_Lya}) \hlold{that the fraction of ionizing photons escaping the central starburst region is small. In other words, the buildup of an ionized bubble requires $f^{\rm ion}_{\rm esc} > 0$ and consequently a lower Ly$\alpha$ luminosity. For simplicity we employ this model as a means of exploring the limiting cases between a fully ionized and a fully neutral IGM. Furthermore, such bubbles may be due to the exterior environment, e.g. neighboring galaxies or early patches of reionization.}

    \subsubsection{Direct detection in deep JWST surveys}
      \label{subsub:JWST_sensitivity}
      At $z = 13.8$ the Ly$\alpha$ line is redshifted to $1.8~\mu$m which will be detected by \textit{JWST} with NIRSpec at a (medium) spectral resolution of $R \sim 1000$, corresponding to a Doppler velocity resolution of $\sim 300$~km~s$^{-1}$. Therefore, if we assume a $5 \sigma$ signal after $10^6$~seconds of exposure time the expected flux detection limit\footnote{See \hlold{the predicted} \textit{JWST} sensitivity limits for detecting spatially unresolved line fluxes at \href{http://www.stsci.edu/jwst/science/sensitivity}{www.stsci.edu/jwst/science/sensitivity}.} for observations with the NIRSpec is $f_{\rm \lambda, NIRSpec} \approx 4 \times 10^{-21}$~erg~s$^{-1}$~cm$^{-2}$~\AA$^{-1}$, or $f_{\rm \nu, NIRSpec} \approx 50$~nJy \citep*[see Table~\ref{tab:JWST}; Figure~\ref{fig:extractor_32kpc};][]{Gardner:2006ky,Johnson:2009,Pawlik:2011}. For the most part only optimistic \HII\ scenarios allow a significant detection of the Ly$\alpha$ line in the first galaxies. However, if the strength of the source is increased and redshift is decreased then possibly even the $R_{\rm \HII} = 0$~Mpc scenario may be observable. The NIRSpec instrument will have an integrated flux sensitivity of $f_{\rm NIRSpec} \approx 8 \times 10^{-20}$~erg~s$^{-1}$~cm$^{-2}$ and therefore, a surface brightness sensitivity of $\text{SB}_{\rm NIRSpec} \approx 8 \times 10^{-19}$~erg~s$^{-1}$~cm$^{-2}$~arcsec$^{-2}$. The NIRCam instrument is capable of $f_{\rm \nu, NIRCam} = 0.56$~nJy photometry, or $f_{\rm \lambda, NIRCam} \approx 5 \times 10^{-23}$~erg~s$^{-1}$~cm$^{-2}$~\AA$^{-1}$, over $10^{-3}$~arcsec$^2$ pixels, providing an equivalent sensitivity of $\text{SB}_{\rm NIRCam} \approx 2 \times 10^{-16}$~erg~s$^{-1}$~cm$^{-2}$~arcsec$^{-2}$ and $f_{\rm NIRCam} \approx 2 \times 10^{-19}$~erg~s$^{-1}$~cm$^{-2}$ for the $5~\mu$m range of the F150W filter. See Table~\ref{tab:JWST} for a summary of detection limits for NIRSpec and NIRCam aboard the \textit{JWST}. Neither instrument is sensitive enough to detect Ly$\alpha$ emission from the SS12 first galaxy model without an additional boost from gravitational lensing.

\begin{figure}
        \centering
        \includegraphics[width=\columnwidth,trim=0.095in 0in 0.09in 0in,clip=true]{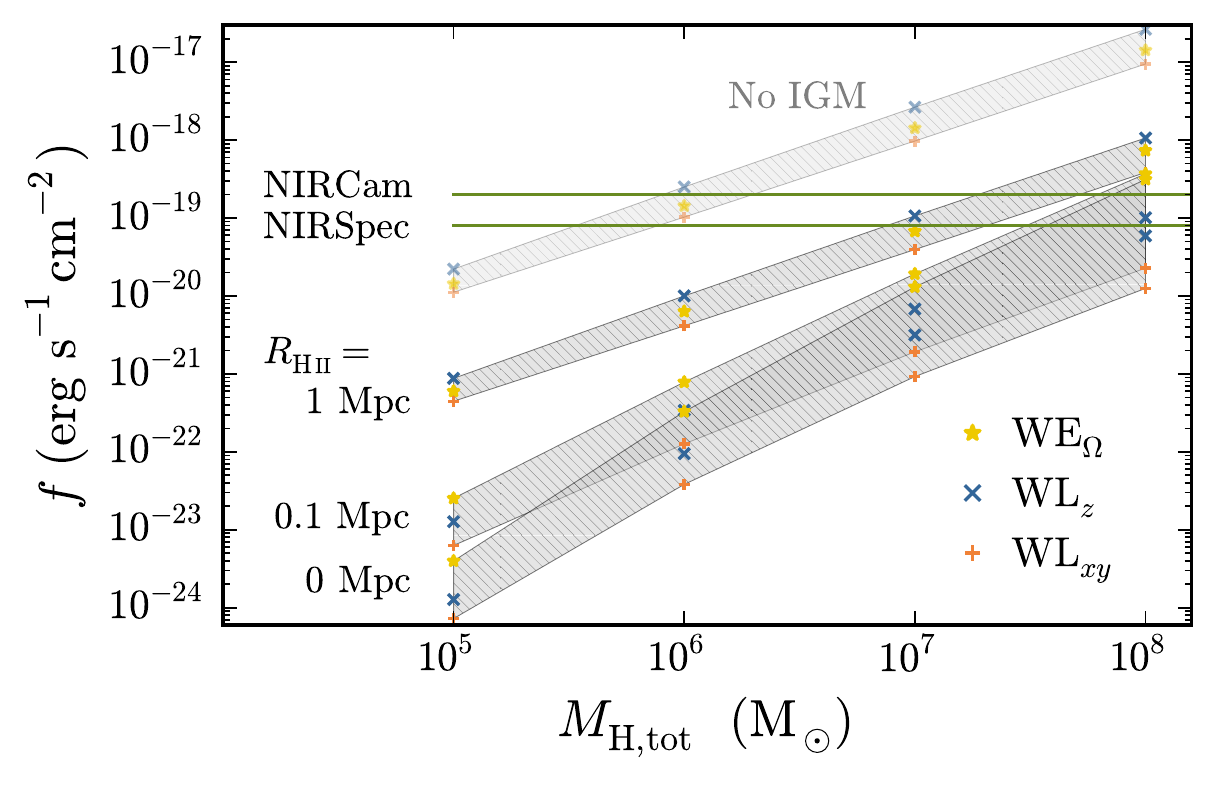}
        \vspace{-.4cm}
        \caption{\hlold{The total observed flux for ``Wind'' models after considering different IGM ionization scenarios, denoted by the size of the local bubble $R_{\rm \HII}$. A roughly linear relation exists for most galaxies due to the assumed constant Pop~III star formation efficiency of $\eta_\ast = 0.01$ at redshift $z = 9$. Smaller bubbles have significant variance in bolometric flux based on the ionization morphology and line of sight. On the other hand, local model differences are minimized for the largest bubble sizes.}}
        \label{fig:halo_BF_IGM}
      \end{figure}

      \hlold{Although we have focused on the SS12 model we may also explore the detectability of the suite of idealized models. The advantage of this approach is that a number of morphologies and masses may be explored, although the physical setups assume a number of simplifications as discussed in Section}~\ref{sec:toy}. \hlold{The bolometric flux predicted for the idealized models, after considering the Ly$\alpha$ radiative transfer and IGM ionization, is shown if Fig.}~\ref{fig:halo_BF_IGM}. \hlold{We focus on ``Wind'' models to avoid overpopulating the figure. This also provides a slight boost in flux compared to the static models. We again use the same IGM ionization scenarios denoted by the bubble size $R_{\rm \HII}$. A roughly linear relation exists for most galaxies due to the assumed constant Pop~III star formation efficiency of $\eta_\ast = 0.01$ at redshift $z = 9$, but when $R_{\rm \HII} = 0$~Mpc the power-law slope increases to $f \approxprop M_{\rm H, tot}^{1.5}$. Smaller bubbles have significant variance in flux based on the ionization morphology and line of sight. On the other hand, local model differences are minimized for the largest bubble sizes. Figure}~\ref{fig:halo_BF_IGM} \hlold{is consistent with the previous results extrapolated from the SS12 galaxy. For completeness, we note that there may still be significant theoretical uncertainties due to choices in modeling, scattering in the IGM, or additional sources of feedback in and around the host galaxy.}

    \subsubsection{Gravitational lensing to boost the Ly$\alpha$ luminosity}
      \label{subsub:lensing}
      Currently, the \textit{Hubble Space Telescope}~(\textit{HST}) is carrying out the Frontier Fields programme, which uses high-magnification foreground galaxy clusters to produce the deepest lensing observations to date. This method has the potential to sufficiently boost observed Ly$\alpha$ luminosities to detect high-$z$ target galaxies. \citet{Zackrisson:2012} explore the prospects of detecting Pop~III galaxies behind the $z=0.546$ galaxy cluster MACS\,J0717.5+3745. With a magnification $\mu \gtrsim 10$, the cluster is an ideal candidate for an even deeper \textit{JWST} Frontier Field programme. The authors conclude that if $\gtrsim 0.1$~per cent of the available baryons are converted into Pop~III stars, then one expects a statistically significant number of lensed Pop~III galaxy images in a single \textit{JWST}/NIRCam field. Therefore, even conservative Ly$\alpha$ galaxies with virial mass $M_{\rm vir} \gtrsim 10^8~\Msun$ and redshift $8 < z < 15$ may be observable with the \textit{JWST}. However, current estimates are based on semi-analytic modeling of the transition from dark matter halo mass to total stellar luminosity \citep[e.g.][]{SafranekShrader:2012qn}. More precise Ly$\alpha$ fluxes from additional simulations would help provide input for the upcoming deep lensing searches.

\section{Summary and Conclusions}
  \label{sec:conc}
  Lyman-$\alpha$ emitting sources provide observational clues about the formation and evolution of distant galaxies. Future observatories, such as the \textit{JWST} and large-aperture ground-based facilities, will help focus and extend our view into the high-$z$ Universe. As we better understand the properties of Ly$\alpha$ radiative transfer we can more fully assess the potential of this probe of the cosmic dark ages. The modeling of both individual galaxies and the background emission from all Ly$\alpha$ sources is highly complementary at these redshifts. Here we have carried out an exploratory survey of Ly$\alpha$ radiative transfer in two classes of first galaxy models. The first was an idealized analytic model and the second was an atomic cooling halo extracted from a high-resolution cosmological simulation for post-processing. We have found that the diffusion in space and frequency is sensitive to the mass density distribution, the velocity profile, and the ionization structure. The specific line-of-sight flux and surface brightness profiles show unique aspects of Ly$\alpha$ transfer for both the idealized models and the cosmological simulation.

  The intervening IGM has a significant effect on the Ly$\alpha$ line flux prior to and during the epoch of reionization. We expect the Gunn-Peterson effect to eliminate the blue peak entirely and significantly destroy the signal out to at least $\Delta v \sim 500$~km~s$^{-1}$, which corresponds to $\Delta \lambda_{\rm obs} \sim 20~[(1+z)/10]$~\AA. The idealized models with ``Late'' type ionization are intrinsically peaked close to line centre; therefore, Ly$\alpha$ sources from $z \gtrsim z_{\rm rei}$ associated with a highly anisotropic ionization scenario from the host galaxy may be nearly impossible to detect. However, the ``Early'' galaxy models with virial mass $M_{\rm vir} \gtrsim 10^8~\Msun$ have resonantly scattered far enough into the wings to possibly survive IGM transmission. This may be inferred from Fig.~\ref{fig:extractor_32kpc} for the post-processing results of the cosmological simulation described above (see also Fig.~\ref{fig:halo_Flux} for the idealized models). We note that our treatment of Ly$\alpha$ transmission through the IGM could be extended. The analytic prescription can hardly capture the details of the epoch of reionization~(EoR). Indeed, the EoR was not instantaneous and inhomogeneous reionization boosts the Ly$\alpha$ visibility, especially if local \HII\ patches are large enough for photons to redshift out of resonance. The IGM model we considered does not include line-of-sight overdensities (e.g. Damped Lyman-$\alpha$ systems, etc.), gravitational lensing, or realistic prescriptions for the ionizing background. The sum total of all such effects may produce a large variance in Ly$\alpha$ observations across different sightlines.

  The specific flux detected from high-$z$ Ly$\alpha$ sources depends on the spectral resolution and sensitivity of the instrument. Throughout this study we have presented numerical calculations of $f_\lambda$ with a resolution of $R \equiv \lambda / \Delta \lambda \approx 30\,000$, achievable with next-generation large-aperture ground-based infrared observatories with adaptive optics. The NIRSpec instrument aboard the \textit{JWST} is capable of obtaining $R \approx 1000$, so many of the Ly$\alpha$ profiles here marginally span $\sim 10$ wavelength bins. Furthermore, at $z = 9$ a physical size of $4$~kpc corresponds to 1~arcsec, i.e. $\sim 30$ NIRCam pixels or $\sim 3$ NIRSpec pixels, thus the \textit{JWST} also has sufficient angular resolution to consider surface brightness measurements and spatially varying spectral features. We anticipate ongoing and future deep field surveys which take advantage of Ly$\alpha$ selection for further spectroscopic follow-up. As seen from Fig.~\ref{fig:extractor_32kpc} the atomic cooling halo from SS12 with $M_{\rm vir} = 2 \times 10^7~\Msun$ at $z = 13.8$, with a Pop~III star formation efficiency of $\eta_\ast = 0.01$, residing in a super bubble with $R_{\rm \HII} = 100$~kpc, and a boost from gravitational lensing is still a factor of 100 below the \textit{JWST} detection limits for a $5\sigma$ signal after $10^6$ seconds of exposure time. Thus, extrapolation from our result implies that haloes with $M_{\rm vir} < 10^9~\Msun$ are generally too faint to be amenable to the detection of Ly$\alpha$ emission from stellar sources. More massive haloes, on the other hand, should be within reach for the \textit{JWST}. Their observability is further boosted by the expected broader spectral profiles which are less susceptible to the opacity of the IGM.

  With post-processing results from additional cosmological simulations of more evolved haloes we will be better equipped to discuss the observability of the Ly$\alpha$ signature of the first galaxies. Furthermore, additional processes not considered in this study may have an important effect on Ly$\alpha$ observations. For example, diffuse emission may account for a significant source of radiation and numerical methods should be developed to compute this directly from the conditions of the ambient gas.

  Finally, we have not included dust in these models. The presence of high amounts of dust in quasars at $z > 6$ constrains the production timescale to $\la 100$~Myr \citep[e.g.][]{Bertoldi:2003}. Therefore, the origin of high-redshift dust may be almost exclusively due to $\sim 8 - 40~\Msun$ core-collapse supernovae \citep[SNe;][]{Gall:2011vy}. Current models based on chemical kinematics of $\la 1000$~day old SNe ejecta predict the formation of a significant amount of silicate dust along with other metals \citep*{Dwek:2010rq}. However, it is unclear how much dust actually survives in these hostile environments \citep{Gall:2014}. Dust grain destruction may be caused by shock-heating from the SN UV flash, hot gas in the reverse shock $\sim 10^4$~years after the explosion, or lower order effects such as radioactivity. Still, there is empirical evidence for resilient dust production in SN ejecta, e.g. observations of remnants with yields of $0.1-1~\Msun$ \citep{Matsuura:2011,Gomez:2012}. This coincides with numerical simulations demonstrating rapid metal enrichment in young galaxies \citep[e.g.][]{Greif:2010,Wise:2012b}. We expect to be able to model dust accurately by post-processing cosmological simulations that include models for metal enrichment. This may give additional insight and can be compared with models that assume mixed or clumpy distributions based on an intrinsic dust to gas mass ratio.

\section*{Acknowledgements}
  AS thanks T.~Chonis, S.~Finkelstein, B.~Tsang, J.~Ritter, Y.~Yang,
  and J.~Hummel for technical advice and stimulating discussion.
  AS was supported in part by the McDonald Observatory and Dept.
  of Astronomy's Board of Visitors Scholarship at UT Austin and
  the NSF Graduate Research Fellowship Program (GRFP).
  CSS is grateful for generous support provided by the NASA Earth
  and Space Science Fellowship (NESSF) programme.
  The authors acknowledge the Texas Advanced Computing Center~(TACC)
  at The University of Texas at Austin for providing HPC resources
  under XSEDE allocation TG-AST120024.
  This work was supported by NSF grants AST-1009928 and AST-1413501.
  \hlold{We thank the anonymous referee for an exceptionally 
  helpful review of the 
  text.}

\appendix

\section{Additional details}
  \label{appendix:details}
  \subsection{Calculation of \texorpdfstring{$\bmath{H(a,x)}$}{H(a,x)}}
    \label{appendix:H_approx}
    $\colt$ uses the following approximation for $H(a,x)$:
    \begin{align} \label{eq:AB}
      &H_{\rm approx}(a,z) = \notag \\
      &\begin{cases}
        \displaystyle e^{-z} \left[1 - a\left( A_0 + \frac{A_1}{z - A_2 + \displaystyle \frac{A_3}{ \displaystyle z - A_4 + \frac{A_5}{z - A_6} } } \right) \right] \;\; \text{for} \;\; z \leq 3 \vspace{.1cm} \\
        \displaystyle e^{-z} + a \left( B_0 + \displaystyle \frac{B_1}{ \displaystyle z - B_2 + \frac{B_3}{ \displaystyle z + B_4 + \frac{B_5}{ \displaystyle z - B_6 + \frac{B_7}{z-B_8} } } } \right) \begin{matrix} \text{for} \\ 3 < z < 25 \end{matrix} \vspace{.1cm} \\
        \displaystyle \frac{a/\sqrt{\pi}}{ \displaystyle z - 1.5 - \frac{1.5}{ \displaystyle z - 3.5 - \frac{5}{z-5.5} } } \quad \text{for} \;\; z \geq 25
      \end{cases}
    \end{align}
    where $z = x^2$ and the constants $A_i$ and $B_i$ are given in Table~\ref{tab:AB}.

    \begin{table}
      \caption{Coefficients for the rational function approximation of the central ($x^2 \leq 3$) and intermediate ($3<x^2<25$) regions in Equation~(\ref{eq:AB}).}
      \label{tab:AB}
      \begin{tabular}{@{} c cc  @{}}
      \hline
      \;\;$i$\;\; & $A_i$ & $B_i$ \, \\
      \hline
      $0$ & $15.75328153963877$ & $0.0003300469163682737$ \\
      $1$ & $286.9341762324778$ & $0.5403095364583999$ \\
      $2$ & $19.05706700907019$ & $2.676724102580895$ \\
      $3$ & $28.22644017233441$ & $12.82026082606220$ \\
      $4$ & $9.526399802414186$ & $3.21166435627278$ \\
      $5$ & $35.29217026286130$ & $32.032981933420$ \\
      $6$ & $0.8681020834678775$ & $9.0328158696$ \\
      $7$ & -- & $23.7489999060$ \\
      $8$ & -- & $1.82106170570$ \\
      \hline
      \end{tabular}
    \end{table}

  \subsection{Tests for \texorpdfstring{$\bmath{x_{\rm crit}}$}{x\_crit}}
    \label{appendix:x_crit}
\begin{figure}
      \centering
      \includegraphics[width=\columnwidth]{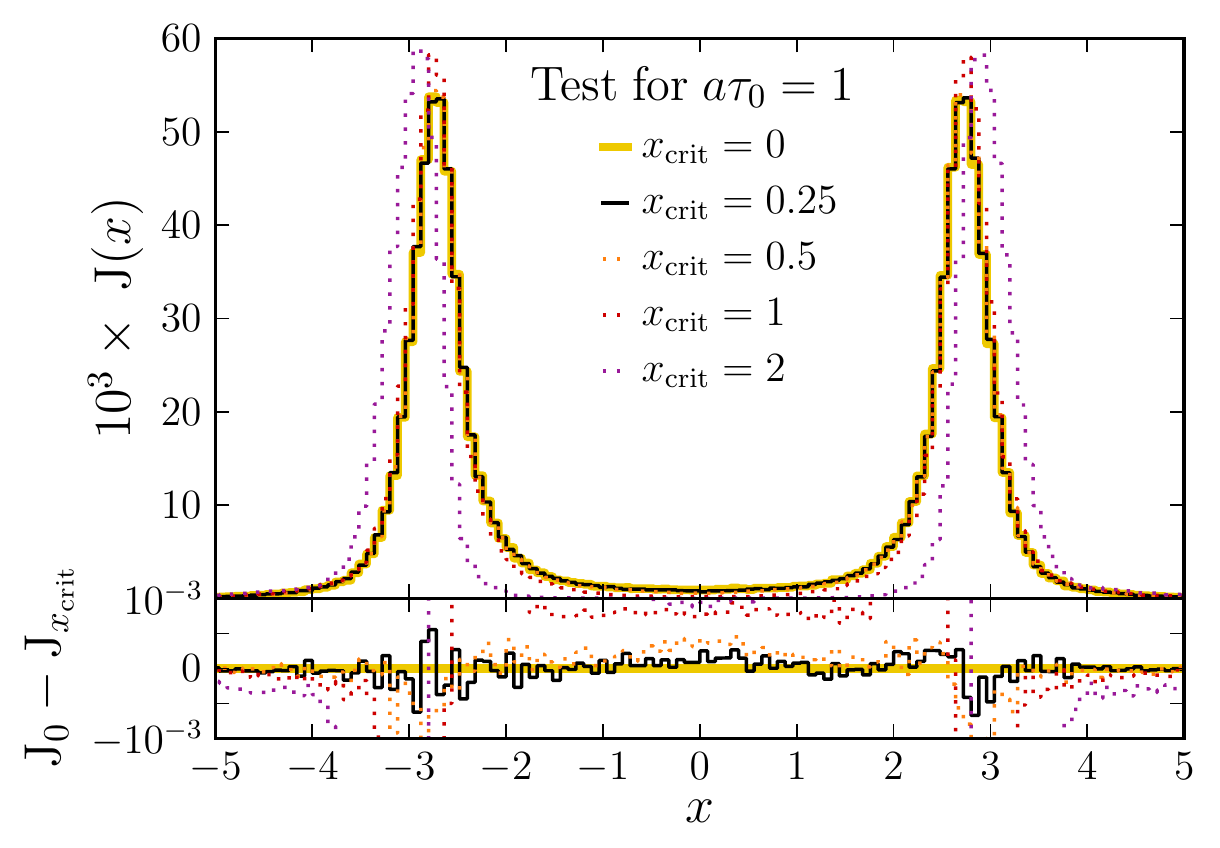}
      \includegraphics[width=\columnwidth]{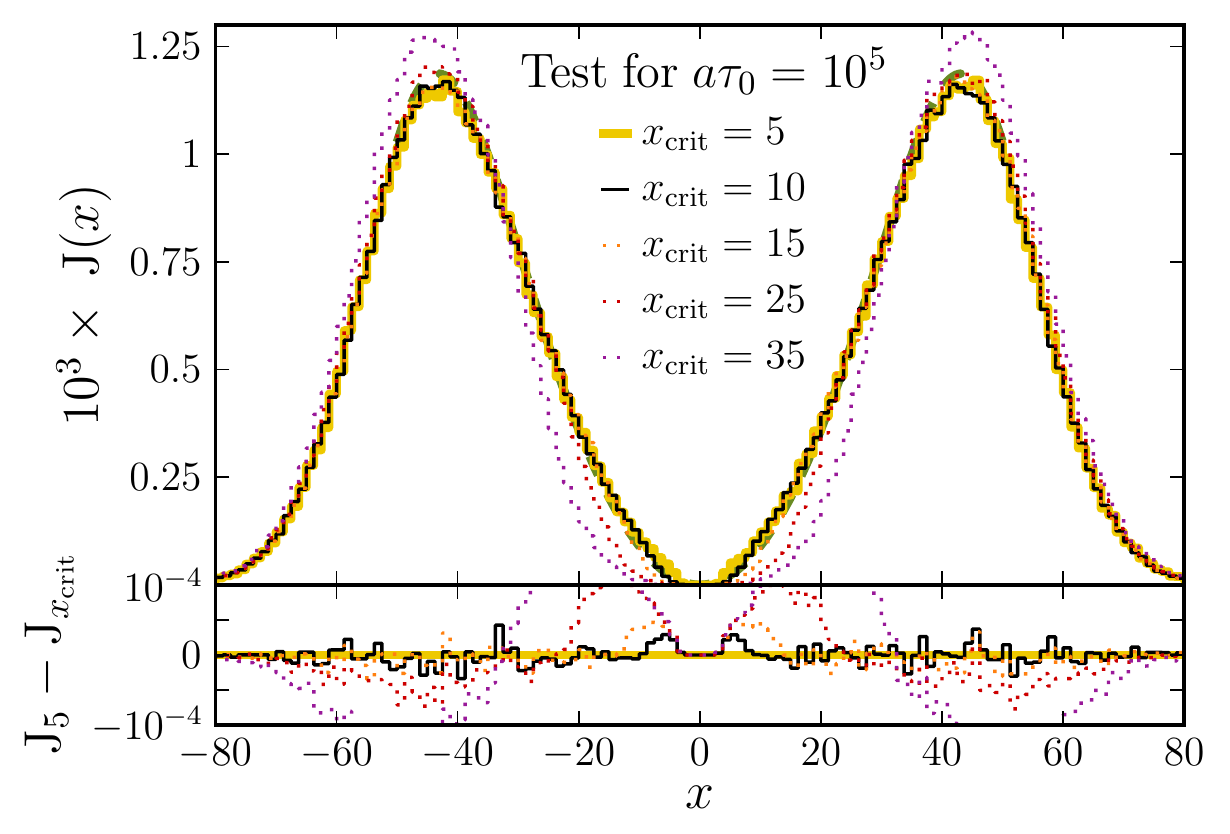}
      \vspace{-.25cm}
      \caption{Top panel: A test for $a\tau_0 = 1$ designed to compare different values of $x_{\rm crit}$. The converged solution (yellow histogram) is given by $x_{\rm crit} = 0$ while a low value of $x_{\rm crit} = 0.25$ provides excellent agreement and is shown in black. A sample of values which are too high and affect the emergent spectrum are $x_{\rm crit} = \{0.5,1,2\}$ and are respectively given by orange, red, and purple dotted histograms. Bottom panel: Same as the top panel except for $a\tau_0 = 10^5$. A value of $x_{\rm crit} = 5$ is sufficiently converged for our purposes. An acceptable value of $x_{\rm crit} = 10$ is given by a black line while non-converged values of $x_{\rm crit} = \{15,25,35\}$ are again given by the orange, red, and purple dotted histograms. Here $a\tau_0$ is large enough that the analytical solution of Equ.~(\ref{eq:sphere}) is accurate, so it is included as the green dashed line in the background. Both tests used $\sim500,000$ photon packets.}
      \label{fig:xcrit_test}
    \end{figure}

    The expression $x_{\rm crit} \propto (a\tau_0)^{1/3}$ is based on comparing an expanding of the analytical solution for a static uniform sphere to the height of its peak -- see Equation~(\ref{eq:sphere}). However, the constant of proportionality must be found by empirical tests. The tests were run on many different values of $a\tau_0$, however, we only show two to demonstrate the validity across the parameter space. The first is for $a\tau_0 = 1$, for which Equation~(\ref{eq:xcrit}) gives $x_{\rm crit} = 0.2$, while the second is for $a\tau_0 = 10^5$, where $x_{\rm crit} = 9.3$. As can be seen from Fig.~\ref{fig:xcrit_test}, both values produce excellent results.

  \subsection{Extraction size for the cosmological simulation}
    \label{appendix:box_sizes}
\begin{figure}
      \centering
      \includegraphics[width=\columnwidth]{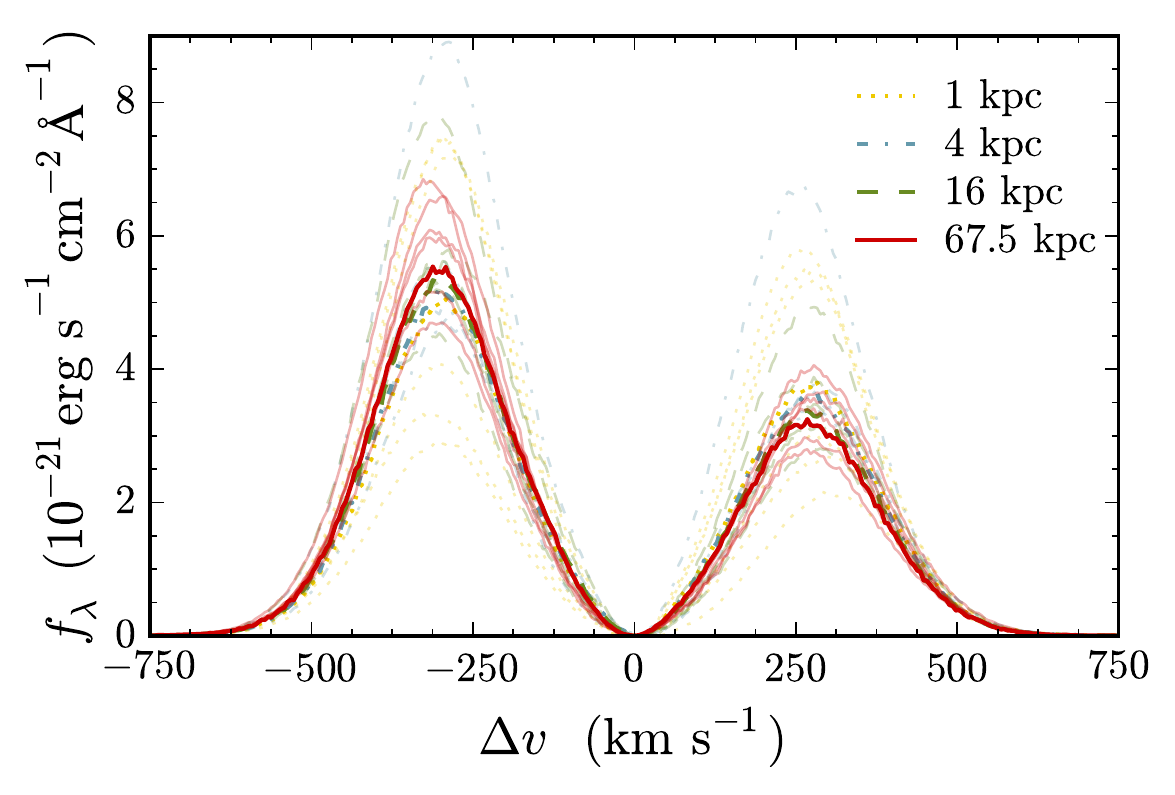}
      \vspace{-.25cm}
      \caption{A test to examine the effect of extraction size for the post-processing conditions of \citet{SafranekShrader:2012qn}. Solid curves represent the angular averaged spectra while the transparent curves show the observed flux as viewed along each each of the six coordinate axes. The red, blue, green and yellow curves represent the results from extraction cubes with a physical edge size of $1$~kpc, $4$~kpc, $16$~kpc, and $67.5$~kpc, respectively. The normalization is set by a $10^8~\Lsun$ source at $z = 13.8$. }
      \label{fig:box_size}
    \end{figure}

    In order to test the grid structure of the cosmological simulation for edge effects and sensitivity to the extraction size we examine the emergent spectra for cubes with a centre to edge distance of $500$~pc, $2$~kpc, $8$~kpc, and $32$~kpc. The largest size represents the radiative transfer through the entire $(1$~Mpc$)^3$ comoving volume. Figure~\ref{fig:box_size} demonstrates the convergence of the flux density~$f_\lambda$ toward that of the largest extraction, although there is still a significant variance across different lines of sight due to the inhomogeneous nature of the cosmic structure.

  \subsection{Thermal effects on Ly$\bmath{\alpha}$ spectra}
    \label{appendix:temperature_tests}
\begin{figure}
      \centering
      \includegraphics[width=\columnwidth]{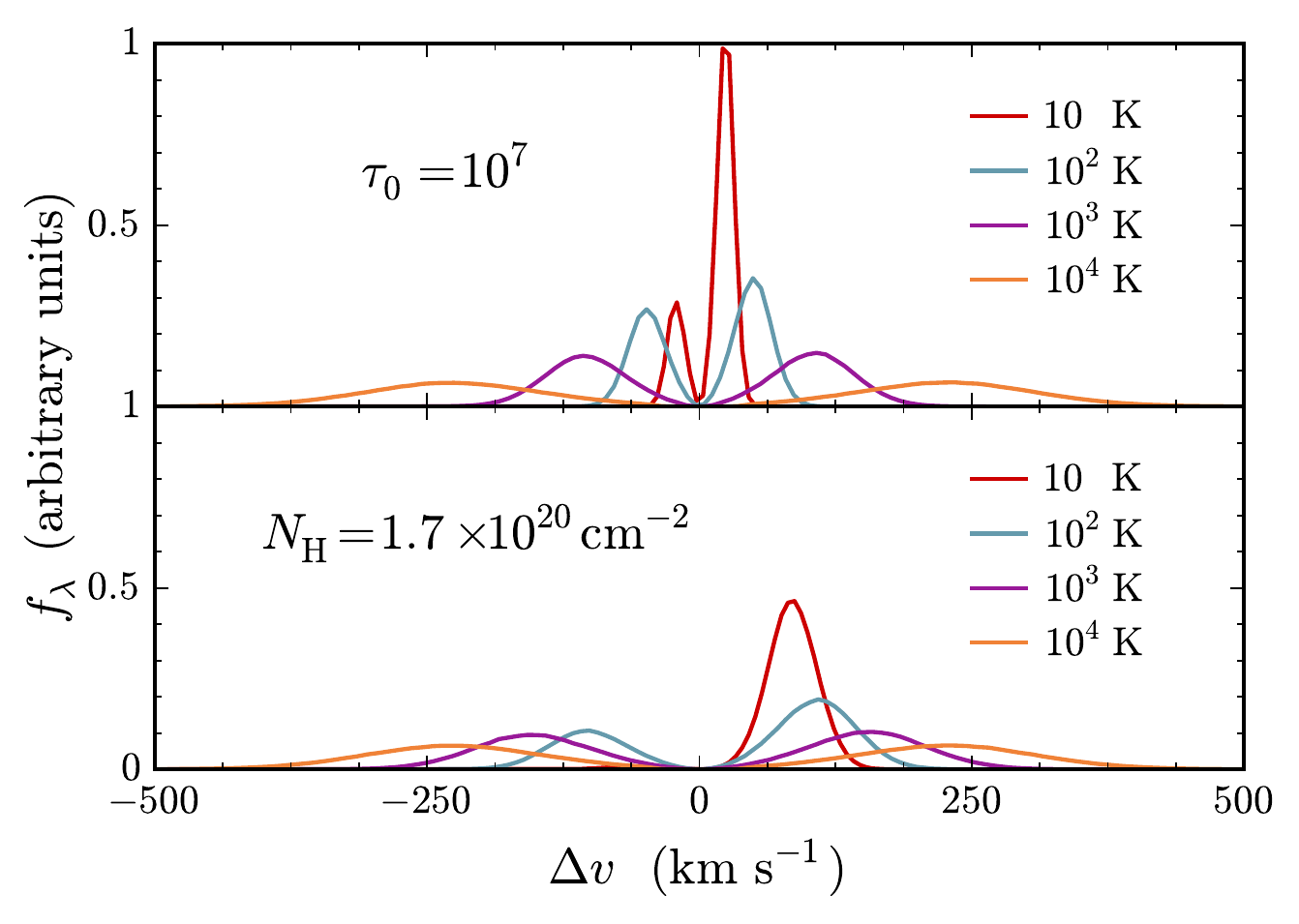}
      \vspace{-.25cm}
      \caption{\hlold{Two test suites to explore the effect of temperature on Ly$\alpha$ radiative transfer. Both cases represent the angular averaged intensity as a function of Doppler velocity $\Delta v = c \Delta \lambda / \lambda$ for uniform slabs of varying temperature. For direct comparison between the models the normalization is arbitrary but consistent in all cases. The top panel maintains a constant optical depth at line centre of $\tau_0 = 10^7$ while the bottom panel ensures a fixed column density of $N_{\rm H} = 1.7 \times 10^{20}$~cm$^{-2}$, 
      corresponding to $\tau_0 = 10^7$ at $T = 10^4$~K. Even though the optical depth or column density are fixed, the product $a \tau$ depends on temperature and the emergent spectra is affected. The number of photon packets used in these simulations is $N_{\rm ph} = 10^6$.}}
      \label{fig:temperature_tests}
    \end{figure}

    \hlold{In order to demonstrate the ``thermal effects'' on Ly$\alpha$ radiative transfer, e.g. via the Doppler width~$\Delta \nu_{\rm D}$ and core-wing transition $x_{\rm cw}$, we include two test suites of uniform slabs with varying temperature. The first maintains a constant optical depth at line centre of $\tau_0 = 10^7$ while the second ensures a fixed column density of $N_{\rm H} = 1.7 \times 10^{20}$~cm$^{-2}$, which corresponds to $\tau_0 = 10^7$ at $T = 10$~K.} \hl{To be clear, a fixed $\tau_0$ requires the column density to vary according to $N_{\rm H} \propto T_4^{1/2}$ while a fixed $N_{\rm H}$ requires $\tau_0 \propto T_4^{-1/2}$.} \hlold{The angular averaged intensity for each model is plotted in Fig.}~\ref{fig:temperature_tests}. \hlold{Even though the optical depth or column density is fixed, the product $a \tau$ depends on temperature and the emergent spectra is affected. Furthermore, we note that although the Lorentz wing optical depth is always independent of temperature, $a \tau_0$ and $a \tau_{\rm wing}$ are proportional to $T^{-1}$ and $T^{-1/2}$, respectively.

    The prominent asymmetric profile at lower temperatures is due to energy loss from recoil at each scattering event. Equations}~(\ref{eq:x_f})~\hlold{and}~(\ref{eq:recoil_parameter}) \hlold{demonstrate that for recoil shifting to appreciably affect the Ly$\alpha$ profile it must be comparable to the typical redistribution at each scattering event. For example, if $g \approx 2.5 \times 10^{-4} T_4^{-1/2}$ then under thermally-dominated redistribution we typically have $u_{\rm atom} \gg g$. However, Lorentzian-dominated redistribution may become a reasonable approximation at low temperatures. In this case, the scattering atom's parallel velocity component is roughly $u_{\rm atom} \approx x_i \pm (c/v_{\rm th}) (\Delta \nu_{\rm L}/\nu_0)$ and the ratio to recoil is significantly reduced for core photons, i.e. $u_{\rm atom} / g \approx (\Delta \nu_{\rm L} / \nu_0) (m_{\rm H} c^2 / h \nu_0) \approx 3.7$. The total relative contribution over all scattering events induces a runaway reddening effect in cold, optically thick environments.}

\label{lastpage}
\end{document}